\documentclass[twocolumn]{aa} 

\usepackage{graphicx}
\usepackage{txfonts}
\usepackage{natbib}
\usepackage{hyperref}
\usepackage{geometry}
\usepackage{multirow}
\usepackage{threeparttable}
\usepackage{subfigure}
\usepackage{booktabs}
\usepackage{lscape}
\usepackage{longtable}
\usepackage{multicol}
\usepackage{CJK}
\usepackage{placeins}

\usepackage{color}

\begin{document}

   \title{Cyanopolyyne line survey towards high-mass star-forming regions with TMRT}

   \subtitle{}
   \author{\protect\begin{CJK*}{UTF8}{gbsn}Y. X. Wang (汪友鑫)\protect\end{CJK*}\inst{1},
          \protect\begin{CJK*}{UTF8}{gbsn}J. S. Zhang (张江水)\protect\end{CJK*}\inst{1,}\thanks{\email jszhang@gzhu.edu.cn},
          \protect\begin{CJK*}{UTF8}{gbsn}Y. T. Yan (闫耀庭)\protect\end{CJK*}\inst{2,}\thanks{Member of the International Max Planck Research School (IMPRS) for Astronomy and Astrophysics at the universities of Bonn and Cologne.},
          \protect\begin{CJK*}{UTF8}{gbsn}J. J. Qiu (邱建杰)\protect\end{CJK*}\inst{3},
          \protect\begin{CJK*}{UTF8}{gbsn}J. L. Chen(陈家梁)\protect\end{CJK*}\inst{1}, 
          \protect\begin{CJK*}{UTF8}{gbsn}J. Y. Zhao(赵洁瑜)\protect\end{CJK*}\inst{1},
          \protect\begin{CJK*}{UTF8}{gbsn}Y. P. Zou(邹益鹏)\protect\end{CJK*}\inst{1},
          \protect\begin{CJK*}{UTF8}{gbsn}X. C. Wu(吴贤聪)\protect\end{CJK*}\inst{1},
          \protect\begin{CJK*}{UTF8}{gbsn}X. L. He(何晓玲)\protect\end{CJK*}\inst{1},
          \protect\begin{CJK*}{UTF8}{gbsn}Y. B. Gong(龚宇彬)\protect\end{CJK*}\inst{1},
          \protect\begin{CJK*}{UTF8}{gbsn}J. H. Cai(蔡嘉华)\protect\end{CJK*}\inst{1}
          }

   \institute{Centre For Astrophysics, Guangzhou University, Guangzhou 510006, PR China\\
         \and
             Max-Planck-Institut f{\"u}r Radioastronomie, Auf dem H{\"u}gel 69, D-53121 Bonn, Germany\\
         \and
             School of Physics and Astronomy, Sun Yat-sen University, Guangzhou 510275, PR China\\}
%
   \date{}


  \abstract
   {Cyanopolyynes (HC$_{2n+1}$N, n = 1,2,3), which are the linear carbon chain molecules, are precursors for the prebiotic synthesis of simple amino acids. They are important for understanding prebiotic chemistry and may be good tracers of the star formation sequence.}
   {We aim to search for cyanopolyynes in high-mass star-forming regions (HMSFRs) at possibly different evolutionary stages, investigate the evolution of HC$_{3}$N and its relation with shock tracers, and detect the existence of HC$_{5}$N and HC$_{7}$N in HMSFRs with a formed protostar.}
   {We carried out a cyanopolyyne line survey towards a large sample of HMSFRs using the Shanghai Tian Ma 65m Radio Telescope (TMRT). Our sample consisted of 123 targets taken from the TMRT C band line survey. It included three kinds of sources, namely those with detection of the 6.7 GHz CH$_{3}$OH maser alone, with detection of the radio recombination line (RRL) alone, and with detection of both (hereafter referred to as Maser-only, RRL-only, and Maser-RRL sources, respectively). For our sample with detection of cyanopolyynes, their column densities were derived using the rotational temperature measured from the NH$_{3}$ lines. We constructed and fitted the far-infrared (FIR) spectral energy distributions (SED; obtained from the Herschel FIR data and the Atacama Pathfinder Experiment data at 870 $\mu$m) of our HC$_{3}$N sources. Moreover, by analysing the relation between HC$_{3}$N and other shock tracers, we also investigate whether HC$_{3}$N is a good tracer of shocks.}
   {We detected HC$_{3}$N in 38 sources, HC$_{5}$N in 11 sources, and HC$_{7}$N in G24.790+0.084, with the highest detection rate being found for Maser-RRL sources and a very low detection rate found for RRL-only sources. The mean column density of HC$_{3}$N was found to be (1.75$\pm$0.42)$\times$10$^{13}$, (2.84$\pm$0.47)$\times$10$^{13}$, and (0.82$\pm$0.15)$\times$10$^{13}$ cm$^{-2}$ for Maser-only, Maser-RRL, and RRL-only sources, respectively. Based on a fit of the FIR SED, we derive their dust temperatures, H$_{2}$ column densities, and abundances of cyanopolyynes relative to H$_{2}$. The mean relative abundance of HC$_{3}$N was found to be (1.22$\pm$0.52)$\times$10$^{-10}$ for Maser-only, (5.40$\pm$1.45)$\times$10$^{-10}$ for Maser-RRL, and (1.65$\pm$1.50)$\times$10$^{-10}$ for RRL-only sources, respectively.}
  {The detection rate, the column density, and the relative abundance of HC$_{3}$N increase from Maser-only to Maser-RRL sources and decrease from Maser-RRL to RRL-only sources. This trend is consistent with the proposed evolutionary trend of HC$_{3}$N under the assumption that our Maser-only, Maser-RRL, and RRL-only sources correspond to massive young stellar objects, ultra-compact H{\sc ii} regions, and normal classical H{\sc ii} regions, respectively. Our detections enlarge the sample of HC$_{3}$N in HMSFRs and support the idea that unsaturated complex organic molecules can exist in HMSFRs with a formed protostar. Furthermore, a statistical analysis of the integrated line intensity and column density of HC$_{3}$N and shock-tracing molecules (SiO, H$_{2}$CO) enabled us to find positive correlations between them. This suggests that HC$_{3}$N may be another tracer of shocks, and should therefore be the subject of further observations and corresponding chemical simulations. Our results indirectly support the idea that the neutral--neutral reaction between C$_{2}$H$_{2}$ and CN is the dominant formation pathway of HC$_{3}$N.}

   \titlerunning{Cyanopolyynes line survey towards high-mass star-forming regions with TMRT 65m}
   \authorrunning{Youxin Wang et al.}
   \maketitle
%

\section{Introduction}

Molecules have been detected in space in various environments, including star-forming regions, the envelopes around evolved stars, well-developed protoplanetary discs around stars, and even distant galaxies. They serve as a useful probe of the physical conditions of their environments, and can be used to derive the ages of their natal molecular cloud \citep{2009ARA&A..47..427H,2020ARA&A..58..727J,2014A&A...563A..97G,2021A&A...652A..71S,2021A&A...648A..66G}. So far, around 250 molecules have been detected in the interstellar medium and circumstellar shells\footnote{\href{https://cdms.astro.uni-koeln.de/classic/molecules}{https://cdms.astro.uni-koeln.de/classic/molecules}}. Among them, carbon chain molecules contribute a very large portion of the known chemical complexity of the interstellar medium \citep{2001AcSpe..57..757T,2016MNRAS.463.4175L,2018ApJ...863...88L}. As typical linear carbon chain molecules, cyanopolyynes (HC$_{2n+1}$N, n = 1,2,3,...) have attracted great interest from astronomers in both observational studies and laboratory theoretical works \citep[e.g.][]{1996ApJ...469L..65T,1997ApJ...483L..61B,1998ApJ...494L.231M,2016MNRAS.463.4175L,2020A&A...642L...8C}. Due to their nitrile bond (-C$\equiv$N), cyanopolyynes are considered to be precursors for the prebiotic synthesis of simple amino acids, which is thought to be important for understanding prebiotic chemistry \citep{2018A&A...617A..95C}.

Cyanoacetylene (HC$_{3}$N), which is the shortest cyanopolyyne molecule, normally exists in warm and active star-forming regions \citep[e.g.][]{2019MNRAS.489.4497Y,2021PASJ...73..467F}. The University of Manchester Institute of Science and Technology (UMIST) database for astrochemistry\footnote{\href{http://udfa.ajmarkwick.net/}{http://udfa.ajmarkwick.net/}} suggests that HC$_{3}$N has four main formation pathways, including the neutral--neutral reaction between C$_{2}$H$_{2}$ and CN, the neutral--neutral reaction between C$_{2}$H and HNC, the ion--molecule reaction between C$_{3}$H$_{n}^{+}$ (n = 3-5) and nitrogen atoms, and the ion--molecule reaction between C$_{2}$H$_{2}^{+}$ and HCN \citep{2013A&A...550A..36M}. Up to now, it is not clear which formation pathway is the dominant one. Some theoretical models suggest that C$_{2}$H$_{2}$ is released from the grain mantle and produces HC$_{3}$N by the neutral--neutral reaction between C$_{2}$H$_{2}$ and CN \citep{2008ApJ...681.1385H,2009MNRAS.394..221C,2019ApJ...881...57T}. Using the isotopic fractionation method \citep[e.g.][]{1998A&A...329.1156T,2007ApJ...663.1174S}, the neutral--neutral reaction between C$_{2}$H$_{2}$ and CN was found to be the dominant formation pathway of HC$_{3}$N \citep[e.g.][]{1998A&A...329.1156T,2016ApJ...833..291A,2016ApJ...830..106T}. A number of chemical models \citep{2019ApJ...881...57T} suggested that there is enhancement of the key reactions of cyanopolyyne formation upon sublimation of CH$_{4}$ and C$_{2n}$H$_{2}$ from dust grains. As shocks heat and compress the surrounding gas, thereby releasing molecules (such as CH$_{4}$ and C$_{2n}$H$_{2}$) trapped in the dust mantles into the gas phase \citep[e.g.][]{2008ApJ...681L..21A}, the released CH$_{4}$ and C$_{2}$H$_{2}$ would produce HC$_{3}$N in the gas phase. Therefore, shocks may be associated with the formation process of HC$_{3}$N. Investigation of the correlation between HC$_{3}$N and shock tracers may therefore help us to constrain the formation pathway of HC$_{3}$N.

Unsaturated interstellar complex organic molecules \citep[iCOMs, e.g.][]{2009ARA&A..47..427H,2017ApJ...850..176C}\footnote{Unsaturated iCOMs are carbon-bearing molecules that have at least six atoms and are hydrogen poor \citep[e.g.][]{2011A&A...536A..33R,2011ApJ...741..120R}}, such as HC$_{5}$N and HC$_{7}$N, are generally abundant in quiescent infrared dark clouds (quiescent) \citep{1992ApJ...392..551S,2009ApJ...699..585H}. During the earliest phases of the star formation process, unsaturated iCOMS become deficient and saturated iCOMs (such as CH$_{3}$OH and CH$_{3}$CN) become abundant instead in more evolved sources \citep[e.g.][]{2006A&A...457..927G,2009ARA&A..47..427H}. Such chemistry is known as `hot corino' chemistry in low-mass star-forming regions (LMSFRs) and `hot core' chemistry in high-mass star-forming regions (HMSFRs). However, various unsaturated iCOMs (e.g. C$_{4}$H$_{2}$ and HC$_{5}$N) have been detected in many LMSFRs \citep{2008ApJ...678.1049S,2009ApJ...702.1025S,2016ApJ...819..140G,2018ApJ...863...88L,2021A&A...648A..83Z}. The theory of warm carbon-chain chemistry (WCCC) has been proposed, which is based on the gas-grain model \citep{2009ApJ...702.1025S}, and can explain the formation of cyanopolyynes in LMSFRs with a formed protostar. As compared to the hot corino chemistry, in WCCC, unsaturated iCOMs can be formed in the gas phase when CH$_{4}$  evaporates from grain mantle \citep{2008ApJ...674..984A, 2008ApJ...681.1385H} and are even more abundant in warm regions \citep[20-30 K,][]{2010ApJ...722.1633S}.

The fact that unsaturated iCOMs can form in LMSFRs with a formed protostar naturally poses the question of whether unsaturated iCOMs can also form in HMSFRs in advanced evolutionary stages. There have been a few successful observations of unsaturated iCOMs towards HMSFRs with a formed protostar \citep{2013A&A...559A..51E,2013A&A...559A..47B,2014MNRAS.443.2252G,2015A&A...581A..71F,2018ApJ...854..133T}. However, the sample with detection of unsaturated iCOMs (especially HC$_{7}$N) is still small. Several models involving chemical networks suggest that unsaturated iCOMs could be formed in HMSFRs with a formed protostar \citep[e.g.][]{2009MNRAS.394..221C,2019ApJ...881...57T}. These theoretical results were obtained by simulating certain sources (e.g. G305.2+0.2 in Chapman et al. 2009 and G28.28-0.36 in Taniguchi et al. 2019a) characterised by their specific properties. Therefore, unlike LMSFRs, there is no sufficient understanding of the formation of unsaturated iCOMs in HMSFRs with a formed protostar. Hence, more studies of unsaturated iCOMs in HMSFRs are required in order to obtain more details of their primary formation path.

We performed a cyanopolyynes line survey in Ku band using the Shanghai Tian Ma 65m Radio Telescope (TMRT) to investigate the evolution of HC$_{3}$N in HMSFRs and the existence of HC$_{5}$N and HC$_{7}$N in HMSFRs with a formed protostar. In Sect. 2, we introduce our methods of sample selection and observations. We summarise our  observational results and method of analysis in Sect. 3. We discuss the evolution of HC$_{3}$N in HMSFRs in Sect. 4.1. In Sect. 4.2, the relation between HC$_{3}$N and shock-tracing molecules (H$_{2}$CO and SiO) is studied. In Sect. 4.3, we discuss unsaturated iCOMs in HMSFRs with a formed protostar. A short summary is presented in Sect. 5.

\section{Sample selection and observations}
\subsection{Sample selection and distance}
Successful surveys of unsaturated iCOMs \citep[HC$_{5}$N, HC$_{7}$N and HC$_{9}$N etc.,][]{2016ApJ...824..136L,2017A&A...606A..74Z} in the C and Ku band were performed using TMRT. A systematic survey in C-band using TMRT was performed recently on a galactic radio recombination line (RRL) and a 6.7 GHz methanol maser towards a sample of 3348 sources obtained from the all-sky Wide-field Infrared Survey Explorer (WISE) point-source catalogue, applying colour criteria \citep{2017ApJ...846..160Y,2019ApJS..241...18Y,2020ApJS..248....3C}. The 6.7 GHz CH$_{3}$OH maser and RRL were detected in 240 sources and 517 sources, respectively. A HMSFRs sample consisting
of 654 sources was built by consolidating the 6.7 GHz CH$_{3}$OH maser sample and the RRL sample \citep{2020ApJS..248....3C}. This incorporates three subsamples of sources, namely 137 sources with detection of the 6.7 GHz CH$_{3}$OH maser alone (hereafter Maser-only), 414 sources with detection of the RRL alone (RRL-only), and 103 sources with detection of both (Maser-RRL). It was suggested that these three types of sources correspond to different star-forming stages, taking into consideration that the methanol maser should appear before the formation of the H{\sc ii} region \citep{2020ApJS..248....3C}.

The sources with the strong 6.7 GHz CH$_{3}$OH maser (flux density \textgreater 0.4 Jy) or RRL emission ($T_\mathrm{peak}$ \textgreater 0.02 K) that also have a galactocentric distance distribution (ranging from 3 to 12 kpc) similar to that of the original HMSFRs sample \citep[3-13 kpc,][]{2020ApJS..248....3C} were chosen as our targets in order to improve the detection rate of cyanopolyynes. We then chose 50 targets from each class.
 
The selected sources have accurate heliocentric distances, which were determined from trigonometric parallax measurements \citep{2014ApJ...783..130R,2019ApJ...885..131R} or the latest parallax-based distance calculator V2\footnote{\href{http://bessel.vlbi-astrometry.org/node/378}{http://bessel.vlbi-astrometry.org/node/378}}. The galactocentric distances of our sources were then determined using the heliocentric distance according to the following formula \citep{2009ApJ...699.1153R}:
\begin{equation}
   \centering
   D_{\rm GC}= \sqrt{{[R_0cos{\left(l\right)-Dcos\left(b\right)}}]^2+\ R_0^2{sin}^2{(l)}},
\end{equation}
where \textit{R$_{0}$} is the galactocentric distance of the sun, which is equal to 8.122 $\pm$ 0.031 kpc \citep{2018A&A...615L..15G}, \textit{l} and \textit{b} are the Galactic longitude and Galactic latitude, respectively, and \textit{D} is the heliocentric distance (Table \ref{tab2}).

\subsection{Observations}

We carried out a cyanopolyyne line survey in August, September, and October 2018, April, May, November, and December 2019, and January 2020 using the 65m TMRT located in Shanghai, China. Due to limited observation time, we observed 123 sources comprising 29 Maser-only, 50 Maser-RRL, and 44 RRL-only ones, while the remaining 27  sources, which are relatively weak, were not observed. In view of our detection rate, several detections among these 27 relatively weak sources could be missed. However, this should not introduce significant error in our results. Using a cryogenically cooled Ku band receiver with the frequency coverage of 12-18 GHz and the DIBAS (Digital Backend System) being a Field-Programmable Gate Array (FPGA)-based spectrometer \citep{2012AAS...21944610B}, we were able to receive and record the signals. Mode 22, which has 16 spectral windows, was used in our observations. Each window has 16384 channels with bandwidths of 23.4375 MHz and a frequency resolution of 1.431 KHz. The centre frequency setups for our observations of cyanopolyynes are listed in Table $\ref{tab1}$. The half power beam widths (HPBWs) of TMRT at the transitions of HC$_{3}$N (J = 2-1), HC$_{5}$N (J = 6-5), and HC$_{7}$N (J = 15-14) were about 52" at 18195.31 MHz, 60" at 15975.93 MHz, and 56" at 16919.98 MHz, respectively. The typical root mean square (rms) was about 10 mK. The original velocity resolution was 0.02 km s$^{-1}$ for HC$_{3}$N (J = 2-1), and 0.03 km s$^{-1}$ for HC$_{5}$N (J = 6-5) and HC$_{7}$N (J = 15-14). The system temperatures were about 30-80 K in the Ku band. A noise diode was used for calibration. The uncertainly on the antenna temperature ($T_{\rm A}^{\ast}$) was about 20$\%$. The main beam brightness temperature ($T_{\rm mb}$) was obtained from the antenna temperature as $T_{\rm mb}$ = $T_{\rm A}^{\ast}$/$\eta _{\rm mb}$, where the main beam efficiency ( $\eta _{mb}$) of the TMRT was about 0.6 in the Ku band \citep{2017AcASn..58...37W}. During our observations, we adopted a position-switching mode with the off-position at (-30',0) or (30',0) offset in azimuth and elevation from the source. Each cycle adopted 2 mins for both ON position and OFF position and each source was observed for 15 ON-OFF cycles. For a few weak objects, more scans were adopted to improve the signal-to-noise ratio (S/N) of the molecular lines.

\section{Results and analyses}

The GILDAS/CLASS\footnote{\href{https://www.iram.fr/IRAMFR/GILDAS/}{https://www.iram.fr/IRAMFR/GILDAS/}} software was used for the data reduction. Poor scans (system temperature \textgreater 80 K) were removed and the rest of the scans were averaged to improve the S/N. Linear or multi-order (\textless 4) polynomial baselines were subtracted for the spectra. To further improve the S/N, the spectra were smoothed with a velocity resolution of $\sim$ 0.4 km s$^{-1}$. Among six hyperfine structure (HFS) lines of HC$_{3}$N (J = 2-1) \citep[e.g.][]{2010A&A...524A..32L,2016ApJ...824..136L}, only blended lines of the J = 2-1, F = 3-2 and J = 2-1, F = 2-1 transitions were detected towards our targets, except for G30.810-0.050. In this latter source, two extra groups of HFS components, that is, F=1-1, and unresolved F=2-2 and F=1-0 were also detected. Due to the presence of the spectra with unresolved HFS components, the Gaussian method was used to fit the spectra  instead of the HFS fitting method.

\begin{figure}
    \centering
    \includegraphics[width=0.3\textwidth]{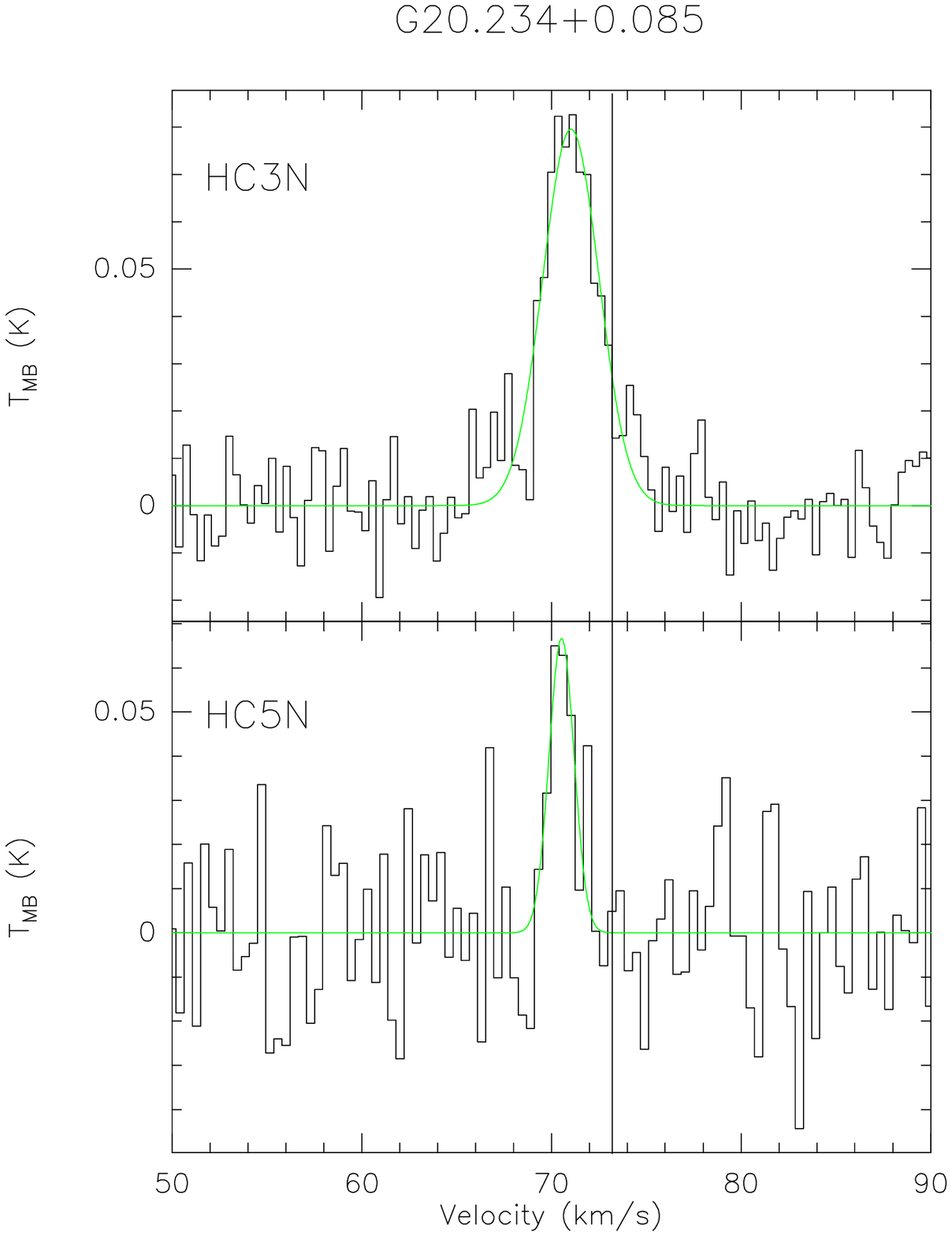}
    \centering
    \includegraphics[width=0.3\textwidth]{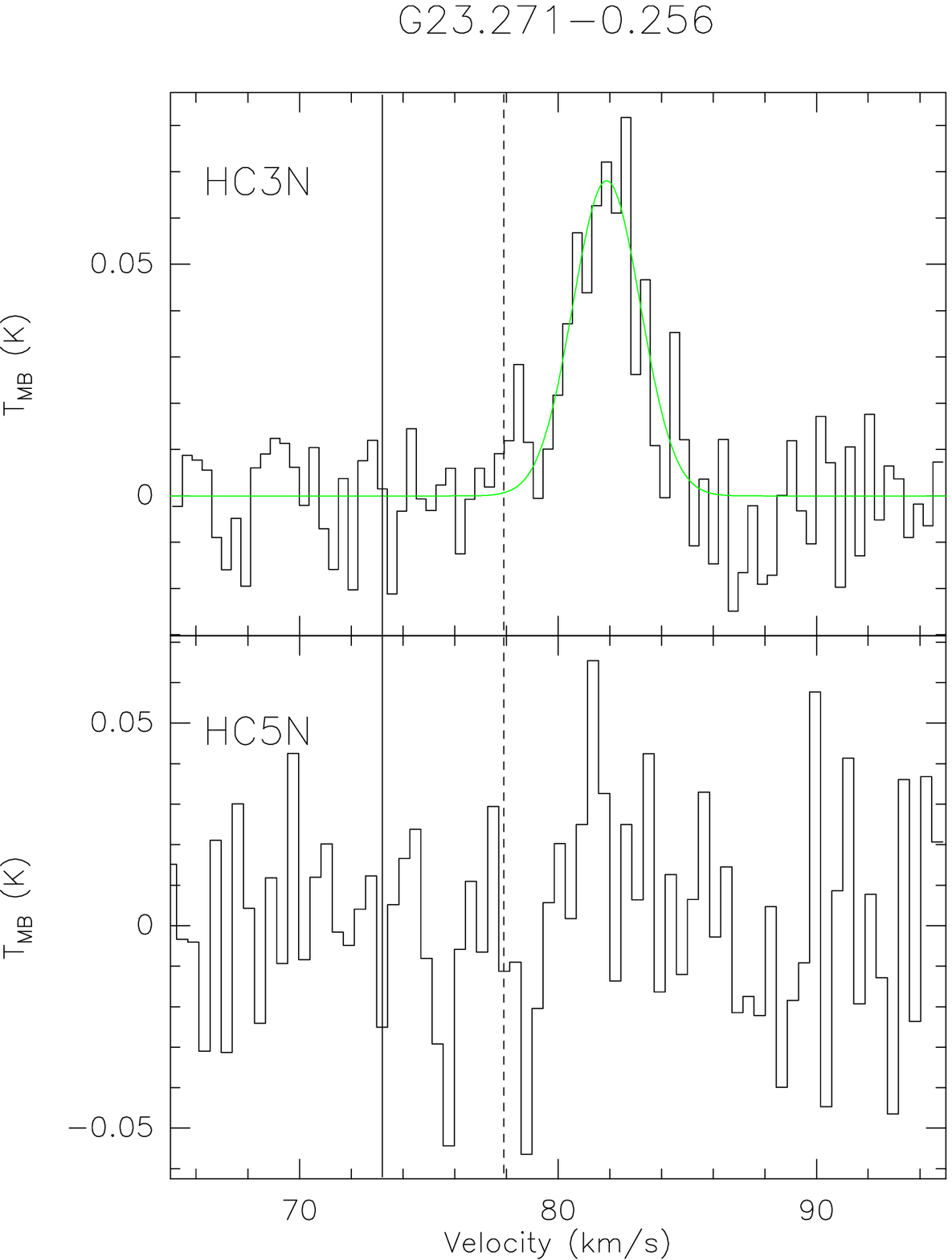}
    \centering
    \includegraphics[width=0.3\textwidth]{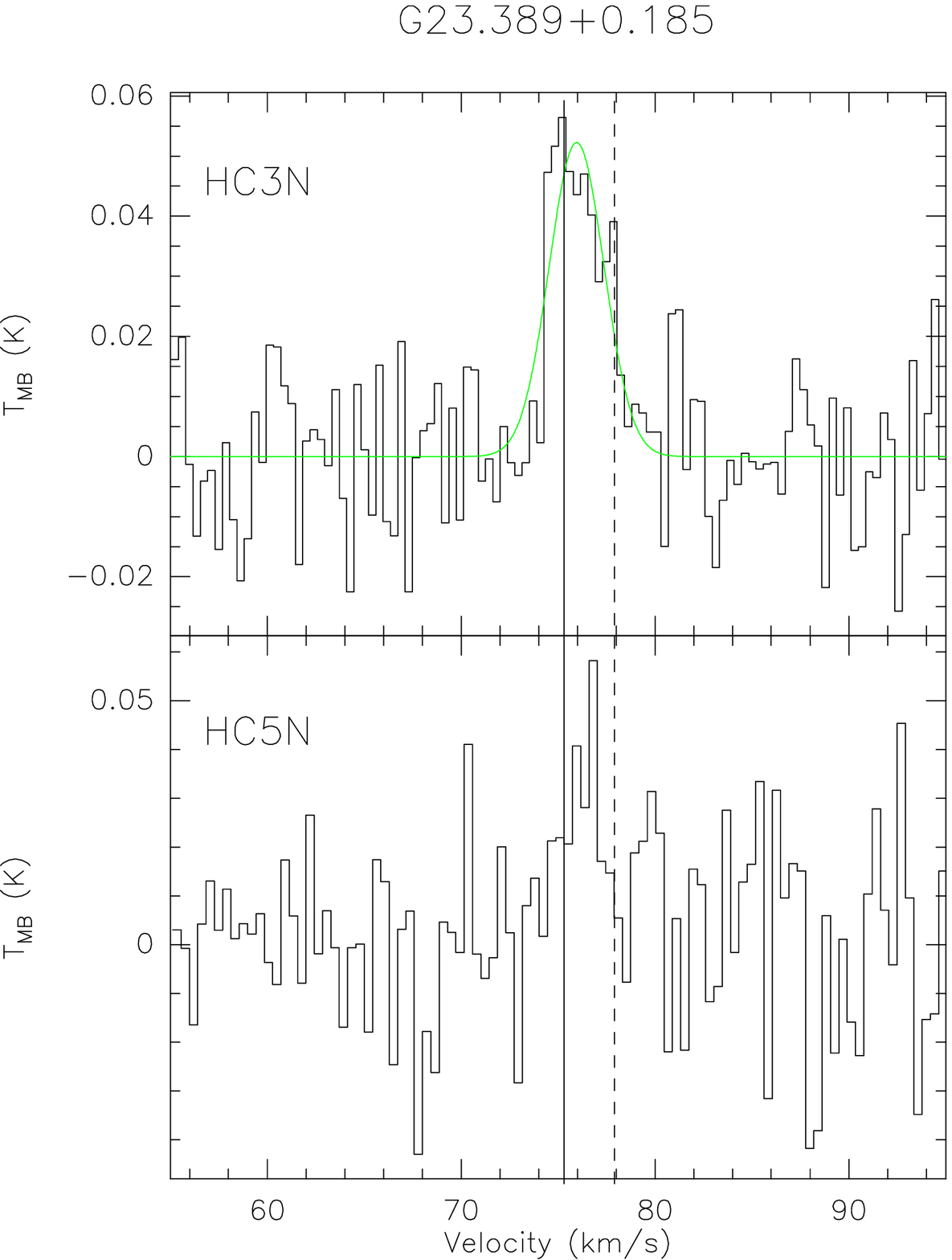}
    \caption{Examples for HC$_{3}$N, HC$_{5}$N, and HC$_{7}$N with TMRT telescope. Solid and dashed lines show the LSR velocity values of 6.7 GHz CH$_{3}$OH maser \citep{2017ApJ...846..160Y,2019ApJS..241...18Y}, and of RRL \citep{2020ApJS..248....3C}, respectively. The full spectra are shown in Appendix Fig. \ref{figa1}.}
    \label{fig1}
\end{figure}

We successfully detected the HC$_{3}$N (J = 2-1) emission line in 38 HMSFRs (S/N \textgreater 3) towards 123 sources, which results in a detection rate of 31$\%$. The sources with HC$_{3}$N detection include 10 Maser-only, 22 Maser-RRL, and 6 RRL-only sources with detection rates of 34$\%$, 44$\%,$ and 14$\%$, respectively. As far as we know, among our 38 HC$_{3}$N (J = 2-1) sources, 34 are newly detected \citep[except G24.328+0.144, G24.790+0.084, G49.466-0.408 and G111.532+0.759,][]{2017ApJS..232....3W}. Among the 38 sources with HC$_{3}$N detection, the HC$_{5}$N (J = 6-5) line was also detected in 11 sources with a detection rate of $\sim$29$\%$, including 2 Maser-only and 9 Maser-RRL sources with the detection rates of $\sim$20$\%$ and $\sim$41$\%$, respectively. Apart from G24.790+0.084 \citep{2014MNRAS.443.2252G}, another 10 sources were new HC$_{5}$N detections. The HC$_{7}$N (J = 15-14) transition was first detected in G24.790+0.084. All spectra for HC$_{3}$N, HC$_{5}$N, and HC$_{7}$N are shown in Fig. \ref{fig1}, where solid and dashed lines show the local standard of rest (LSR) velocities of 6.7 GHz CH$_{3}$OH maser \citep{2017ApJ...846..160Y,2019ApJS..241...18Y} and RRL \citep{2020ApJS..248....3C}, respectively. The small differences in LSR velocity for these three molecules may be due to the fact that the molecules have different excitation mechanisms and exist in HMSFRs with  different characteristics. The spectral line parameters of HC$_{3}$N and unsaturated iCOMs (HC$_{5}$N and HC$_{7}$N) are listed in Tables $\ref{tab2}$ and $\ref{tab3}$, respectively.

\subsection{Column density of cyanopolyynes}
\begin{figure}
    \centering
    \includegraphics[width=0.8\columnwidth]{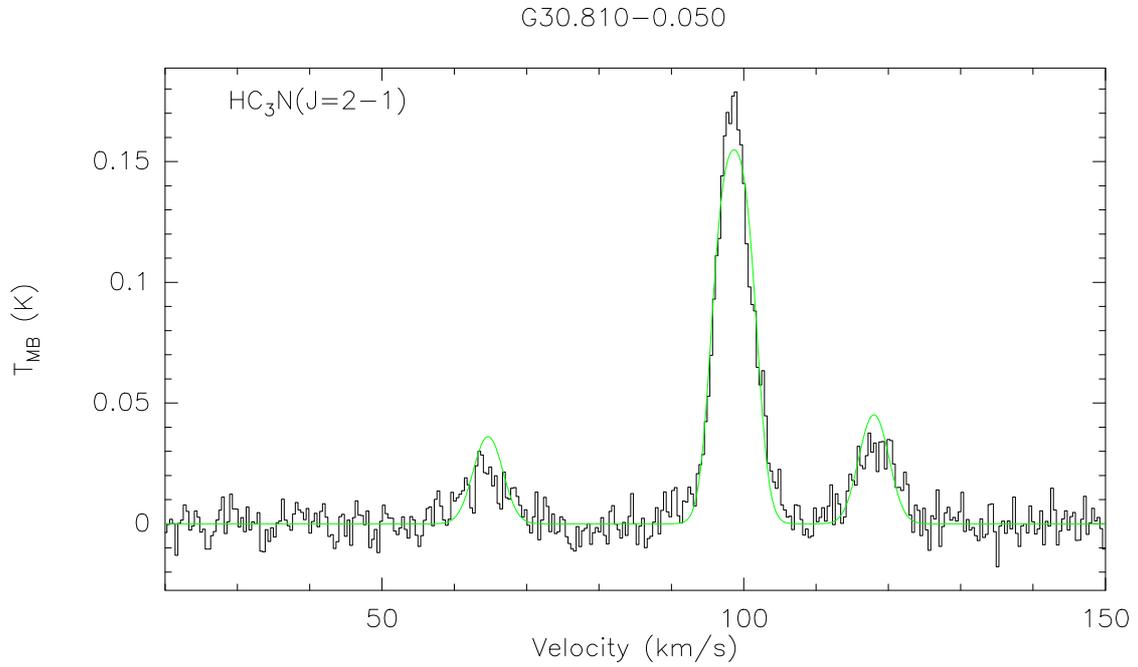}
    \caption{HFS fitting of G30.810-0.050.}
    \label{fig2}
\end{figure}

The HC$_{3}$N (J = 2-1) line is generally considered to be optically thin \citep[e.g.][]{1975ApJ...202...76M,2015A&A...581A..48G,2016ApJ...824..136L}. Among our detected sources, the strongest, G30.810-0.050, was found to have three groups of HFS structure components of HC$_{3}$N (J = 2-1). We can determine the optical depth of its HC$_{3}$N (J = 2-1) line using the HFS fitting method implemented in CLASS software (`Method' command) as shown in Fig. \ref{fig2}. A low optical depth value of \textless 0.1 is obtained, which supports the idea that the optical depth effect is insignificant and cannot be used for further analysis.

The volume density of H$_{2}$ can be estimated as $n_{\rm H_2} = N_{\rm H_2}/(2DR_{\rm eff})$\citep{2014A&A...568A..41U,2017A&A...599A.139K}, where $N_{\rm H_{2}}$ and $R_{\rm eff}$ are the column density of H$_{2}$ and the effective radius of the clump in rad (Table $\ref{tab5}$), respectively, and \textit{D} is the heliocentric distance (Table $\ref{tab2}$). The lowest value of the volume density for our sources is about 2.1$\times$10$^{3}$ cm$^{-3}$ (G24.328+0.144), which is larger than the critical density\footnote{The critical density is estimated from $n_{\rm crit} = A_{\rm ij}/C_{\rm ij}$ for optically thin emission, where $A_{\rm ij}$ and $C_{\rm ij}$ is the Einstein coefficient for spontaneous emission and the collision rate, respectively \citep{2015PASP..127..299S}.} of HC$_{3}$N (J = 2-1) \citep[$n_{\rm crit}$ = 9.7$\times$10$^{2}$ cm$^{-3}$, with a kinetic temperature of 10 K,][]{2015PASP..127..299S}. This suggests that the HC$_{3}$N molecule is under local thermodynamic equilibrium (LTE). HC$_{5}$N and HC$_{7}$N are generally assumed to be in LTE as well \citep[e.g.][]{2016ApJ...824..136L,2018ApJ...866...32T,2019MNRAS.488..495W,2021A&A...648A..83Z}. Therefore, assuming LTE conditions and a negligible optical depth value, the column density can be estimated by the following equation\citep{1986ApJS...60..819C}:
\begin{equation}
    N = \frac{3k_{B}\textit{W}}{8\pi ^{3}\nu S \mu ^{2}} Q(T_{\rm rot}) \frac{ T_{\rm rot} }{ T_{\rm rot} - T_{\rm bg}} exp(E_{\rm u}/k_{B}T_{\rm rot}),
\end{equation}
where \textit{k$_{B}$} is the Boltzmann constant, \textit{W} is the integrated line intensity, $\nu$ is the rest frequency of the transition, and $T_{\rm rot}$ and $T_{\rm bg}$ are the rotational temperature and the background temperature (2.73 K), respectively. Also, $E_{\rm u}/k_{B}$ and $S\mu^{2}$ (taken from Cologne Database for Molecular Spectroscopy, CDMS\footnote{\href{https://cdms.astro.uni-koeln.de/cdms/portal/}{https://cdms.astro.uni-koeln.de/cdms/portal/}}) are the upper level energy in K and the product of the line strength and the square of the electric dipole moment, respectively (see Table $\ref{tab1}$). The partition functions, Q($T_{rot}$), of each molecule are taken from CDMS:
\begin{align}
&Q (\rm HC_{3}N) =  4.582 T_{\rm rot} + 0.283 \\
&Q (\rm HC_{5}N) =  7.595{\times}10{^{-7}}T_{\rm rot}^{4} - 3.760 {\times} 10^{-4}T_{\rm rot}^{3}\nonumber \\
&       \qquad\qquad     + 5.177{\times} 10^{-2}T_{\rm rot}^{2}+13.38T_{\rm rot}+23.15\\
&Q (\rm HC_{7}N) =  1.633 {\times}10{^{-6}}T_{\rm rot}^{4} - 7.706 {\times} 10^{-4}T_{\rm rot}^{3}\nonumber \\
&   \qquad\qquad     - 9.174{\times} 10^{-2}T_{\rm rot}^{2}+119T_{\rm rot}-96.46.
\end{align}
For molecular transitions under LTE conditions, the rotation temperature is usually determined by the rotation diagram method \citep[e.g.][]{1999ApJ...517..209G}. Application of this method requires at least two transition lines of each molecule to be available. However, our observations provide only one transition line for each molecule. 

Here, $T_{\rm rot}$ was taken from the rotational temperature of NH$_{3}$, which could be obtained from the NH$_{3}$ inversion lines and was widely used as a good tracer of temperature in the molecular clouds \citep{2004A&A...416..191T,2015PASP..127..266M,2020MNRAS.499.4432W}. Notably, NH$_{3}$ and cyanopolyynes may come from different regions in the clumps and therefore lead to error in calculations of the column density of cyanopolyynes. However, HC$_{3}$N and NH$_{3}$ have similar critical densities, when the kinetic temperature is in the range of 10 to 50 K \citep{2015PASP..127..299S}. Therefore, the use of $T_{\rm rot}$ of NH$_{3}$ should not cause any significant bias in our calculations of the column density of cyanopolyynes. The rotational temperature of NH$_{3}$ was reported in 23 sources among the total of 38 sources with HC$_{3}$N detection \citep{2012A&A...544A.146W,2011MNRAS.418.1689U,2016AJ....152...92L,2013A&A...552A..40C}. For another 6 sources (G20.234+0.085, G24.328+0.144, G24.528+0.337, G24.790+0.084, G28.287-0.348, and G192.600-0.048), only the optical depth ($\tau$) and the main beam temperatures ($T_{\rm mb}$) of NH$_{3}$ lines were reported \citep{2013ApJ...764...61C,2016ApJ...822...59S,Yangtian2021}. For these sources, $T_{\rm rot}$ of NH$_{3}$ was derived as follows \citep[e.g.][]{2011ApJ...736..163R}:
\begin{align}
T_{\rm rot}=-41.5\left[ ln (\frac{-0.283}{\tau (1,1,m)}ln[1-\frac{T_{\rm mb}(2,2,m)}{T_{\rm mb}(1,1,m)}(1-e^{-\tau (1,1,m)})])\right ]^{-1},
\end{align}
where $m$ represents the main hyperfine component. Therefore, we obtained the $T_{\rm rot}$ values of NH$_{3}$ for 29 sources from our sample with HC$_{3}$N detection, including 9 Maser-only, 19 Maser-RRL, and 1 RRL-only sources. For 9 sources without $T_{\rm rot}$ (1 Maser-only, 3 Maser-RRL, and 5 RRL-only), the average $T_{\rm rot}$ value of each type was used as their $T_{\rm rot}$ values (Table $\ref{tab4}$). The column densities of cyanopolyynes molecules
were obtained using the values of $T_{\rm rot}$ of NH$_{3}$ and Eq. (2)  (Table $\ref{tab4}$). We made a comparison with the published HC$_{3}$N column densities towards our sample. These were found to be 2.00$\times$10$^{13}$ cm$^{-2}$ for G28.287-0.348 \citep{2018ApJ...866..150T} and 3.60$\times$10$^{13}$ cm$^{-2}$ for G81.752-0.691 \citep[DR21,][]{1991JKAS...24..217C}, which are comparable to our results (1.10$\times$10$^{13}$ cm$^{-2}$ for G28.287-0.348 and 3.12$\times$10$^{13}$ cm$^{-2}$ for G81.752-0.691, respectively). The mean column density of HC$_{3}$N was found to be (1.75$\pm$0.42)$\times$10$^{13}$ cm$^{-2}$ for Maser-only, (2.84$\pm$0.47)$\times$10$^{13}$ cm$^{-2}$ for Maser-RRL, and (0.82$\pm$0.15)$\times$10$^{13}$ cm$^{-2}$ for RRL-only sources. More details about the differences in the column density between these three types of sources is discussed in Sect. 4.1.

\subsection{Far-infrared spectral energy distributions}
\begin{figure}
    \centering
    \subfigure{
    \centering
    \includegraphics[width=0.25\textwidth]{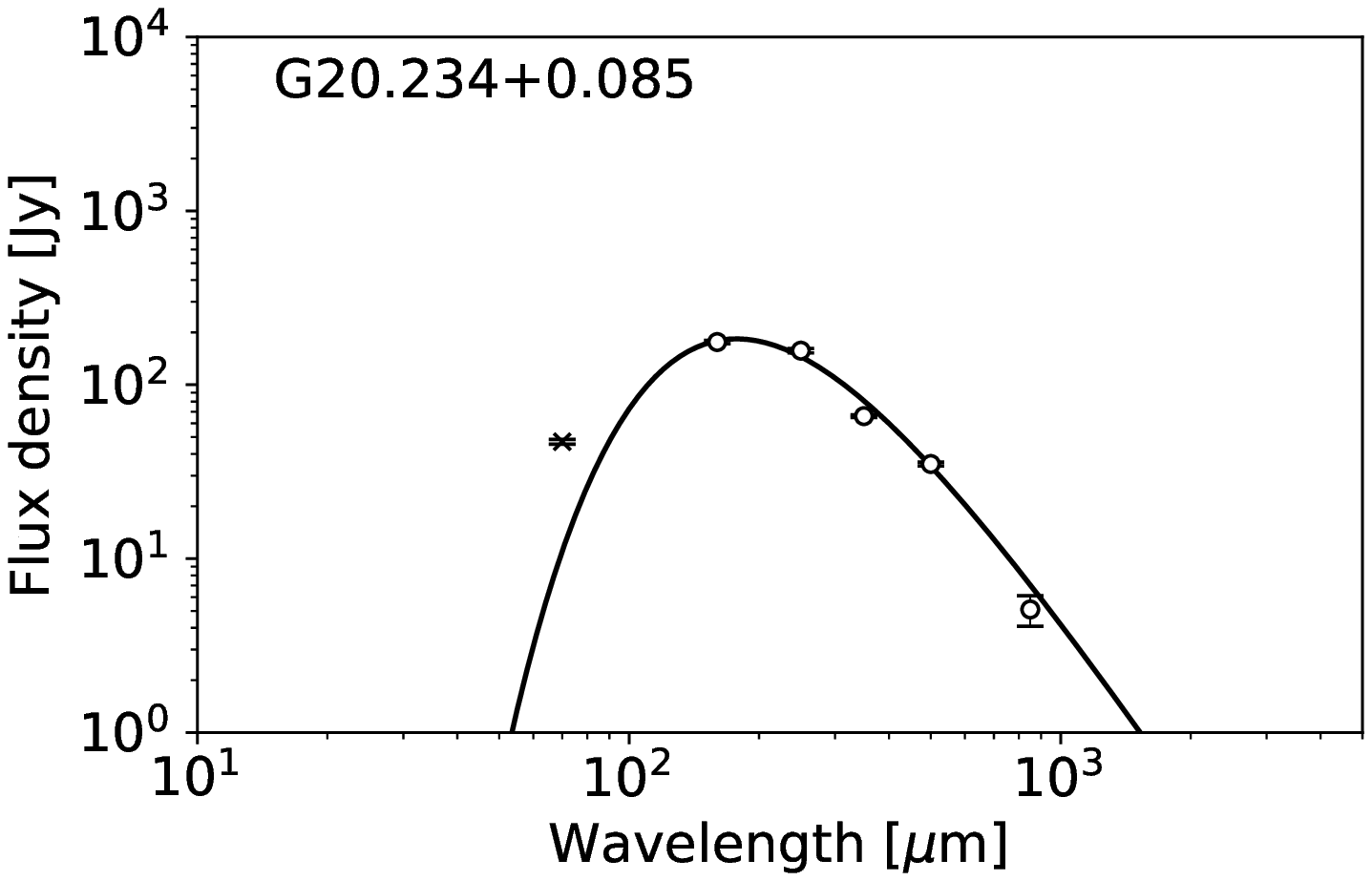}}
    \subfigure{
    \centering
    \includegraphics[width=0.25\textwidth]{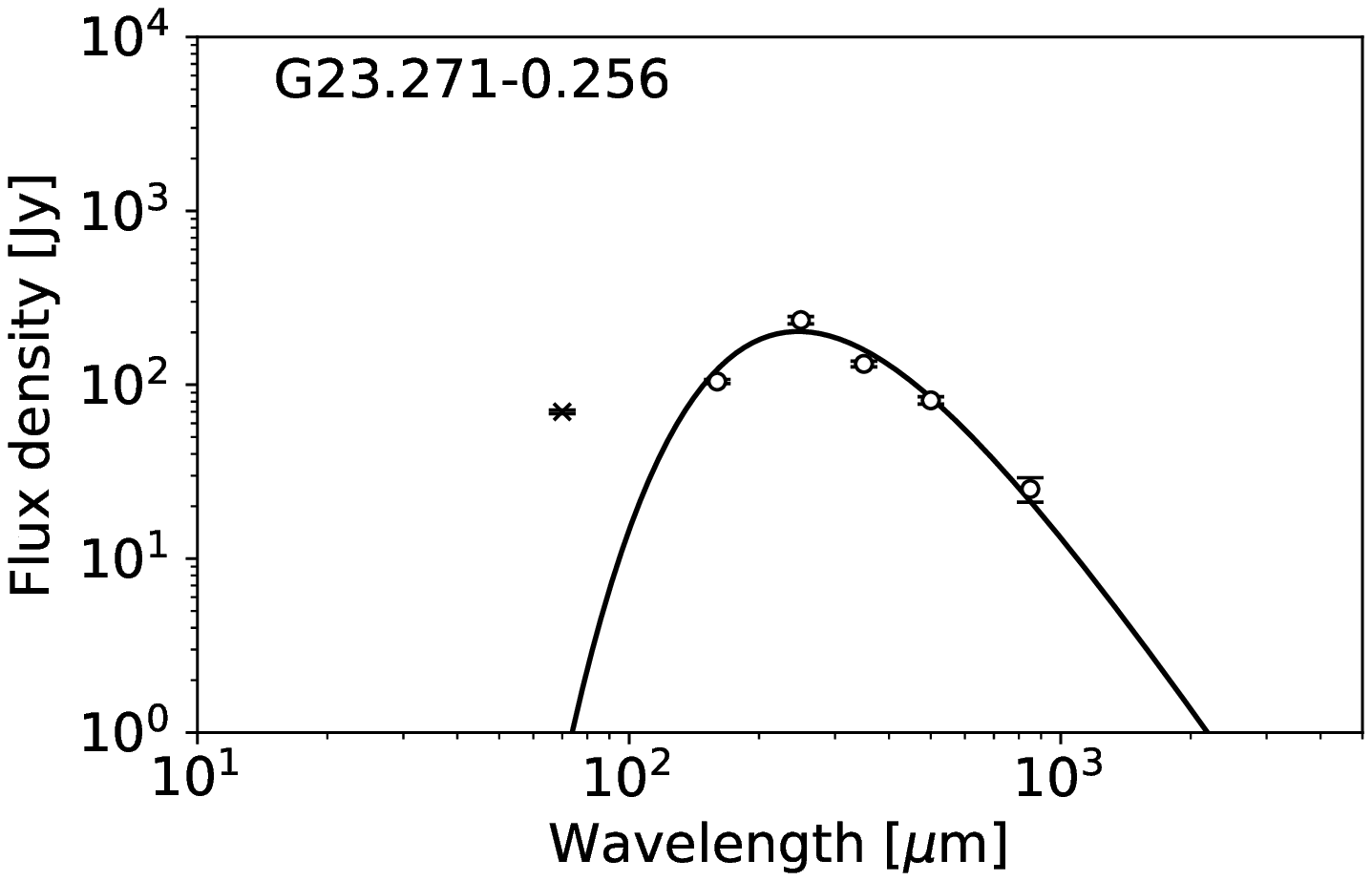}}
    \subfigure{
    \centering
    \includegraphics[width=0.25\textwidth]{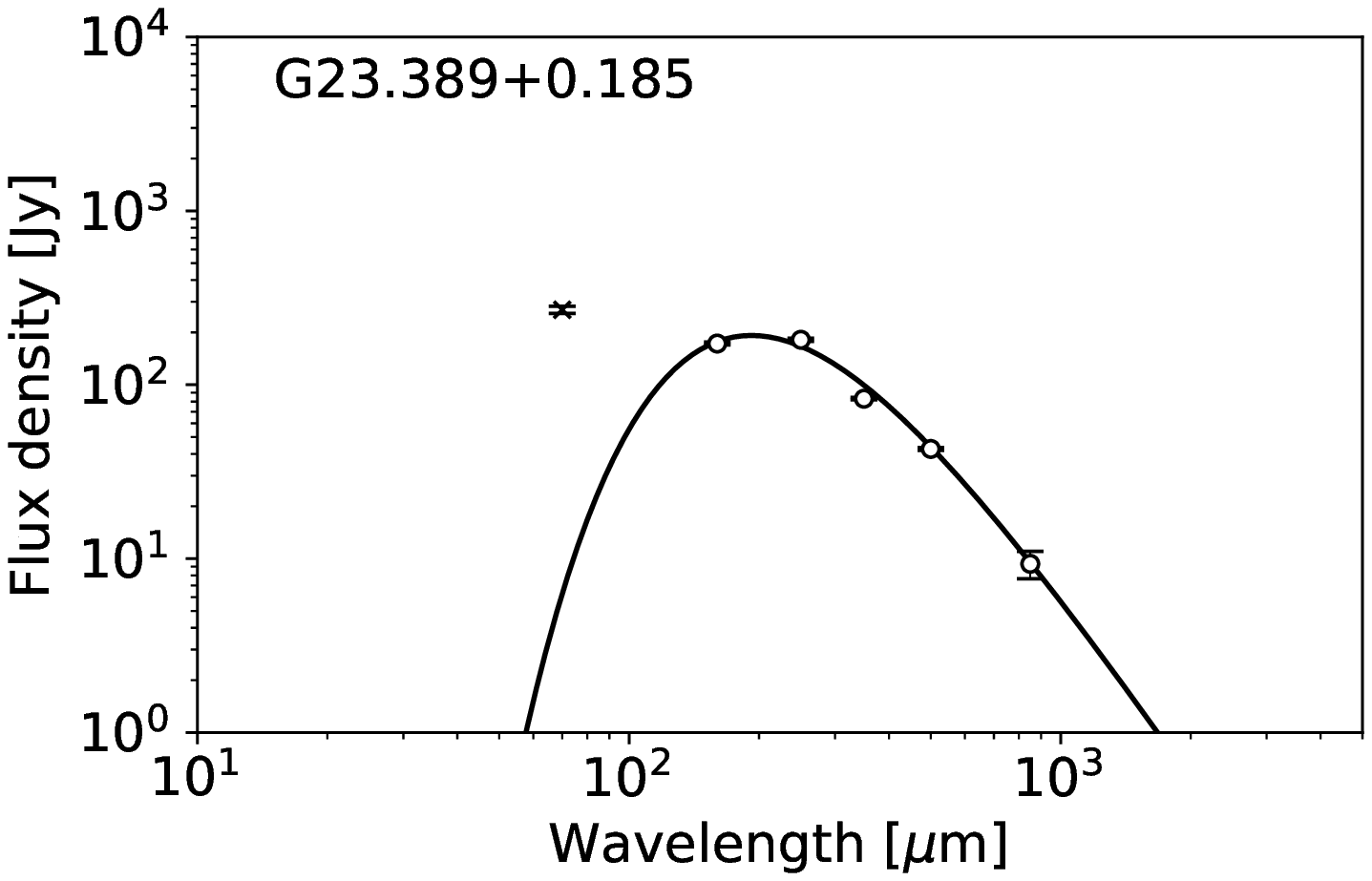}}
    \centering
    \subfigure{
    \centering
    \includegraphics[width=0.25\textwidth]{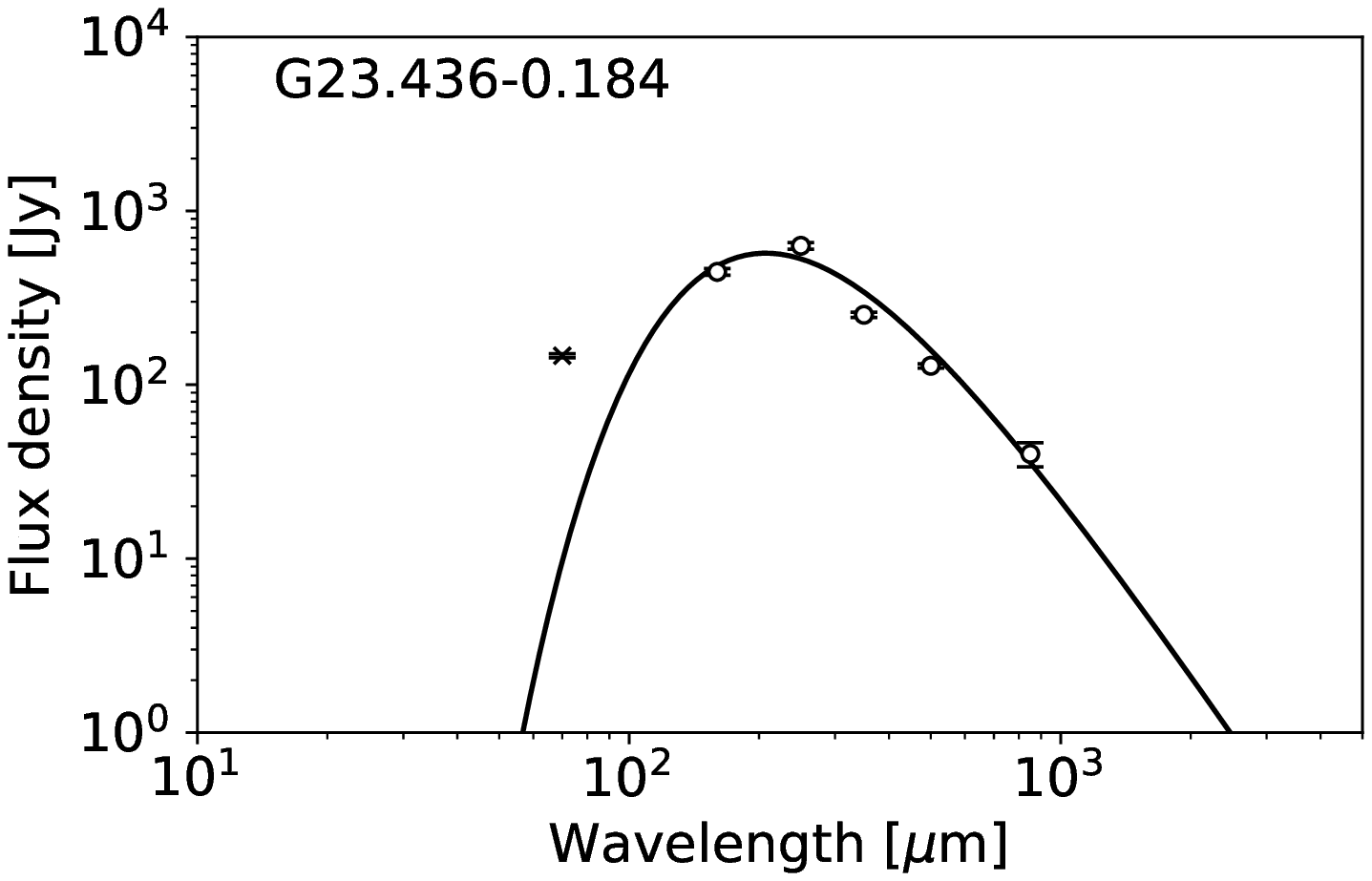}}
    \subfigure{
    \centering
    \includegraphics[width=0.25\textwidth]{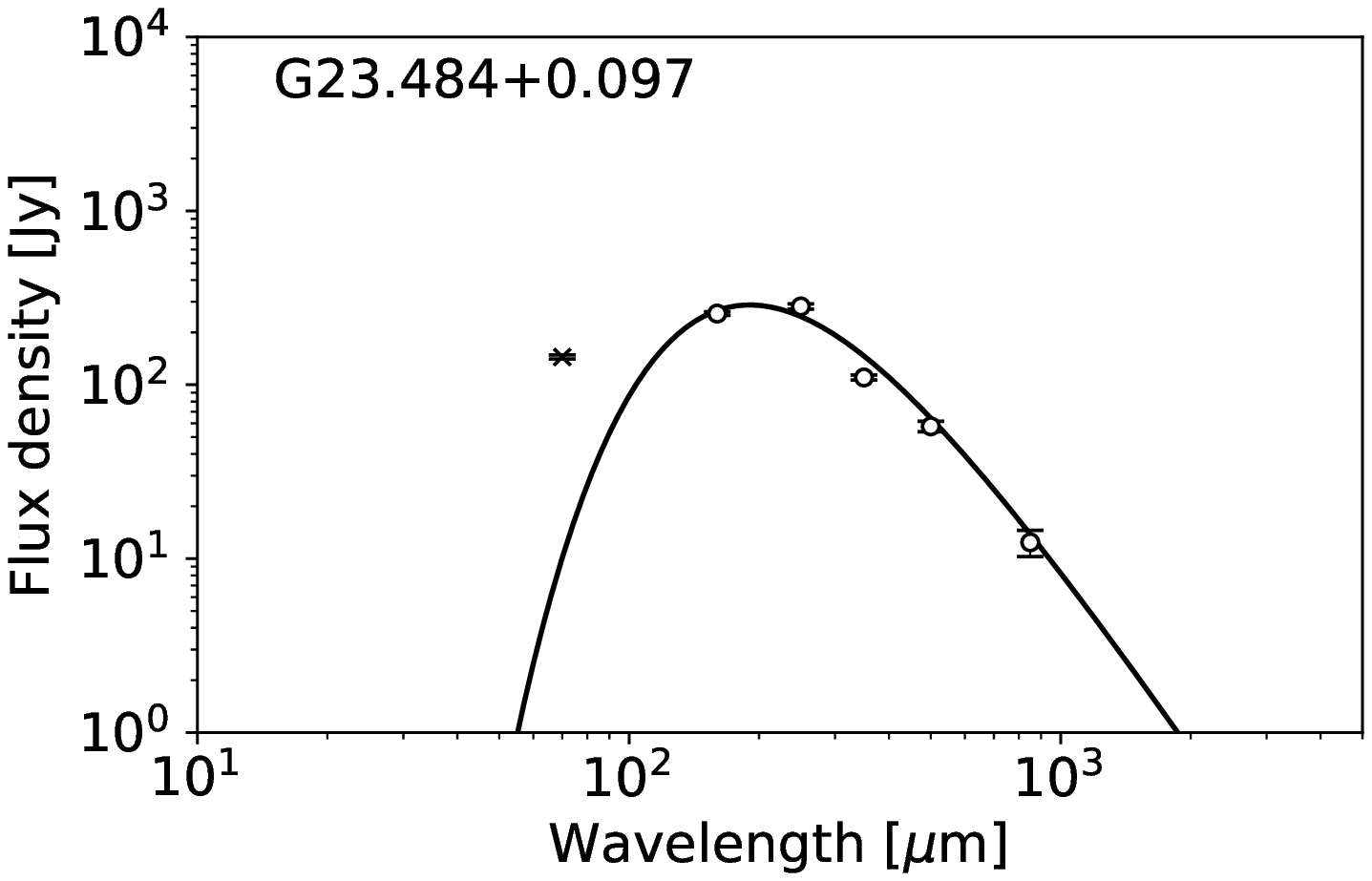}}
    \subfigure{
    \centering
    \includegraphics[width=0.25\textwidth]{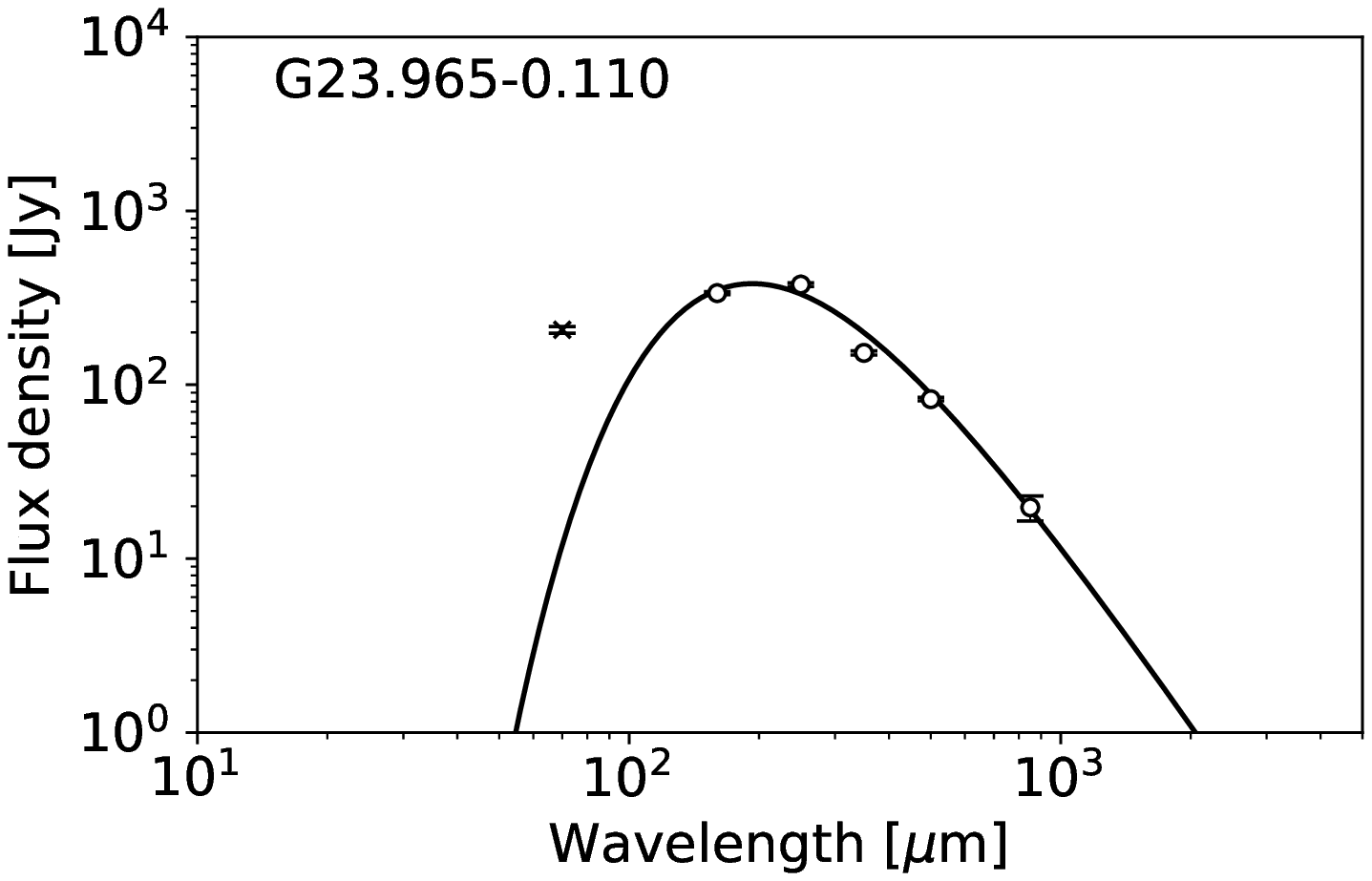}}
    \centering
    \subfigure{
    \centering
    \includegraphics[width=0.25\textwidth]{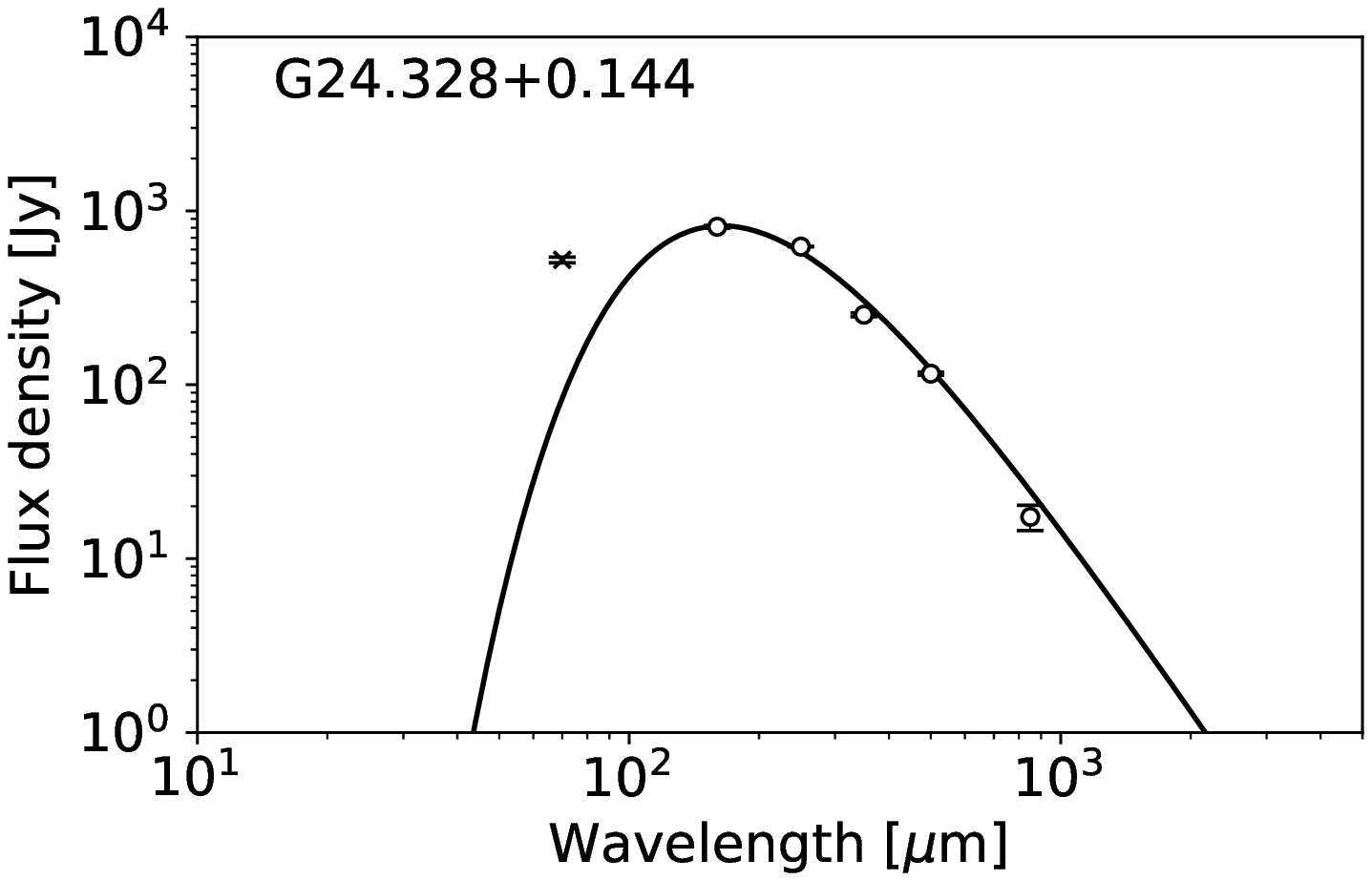}}
    \subfigure{
    \centering
    \includegraphics[width=0.25\textwidth]{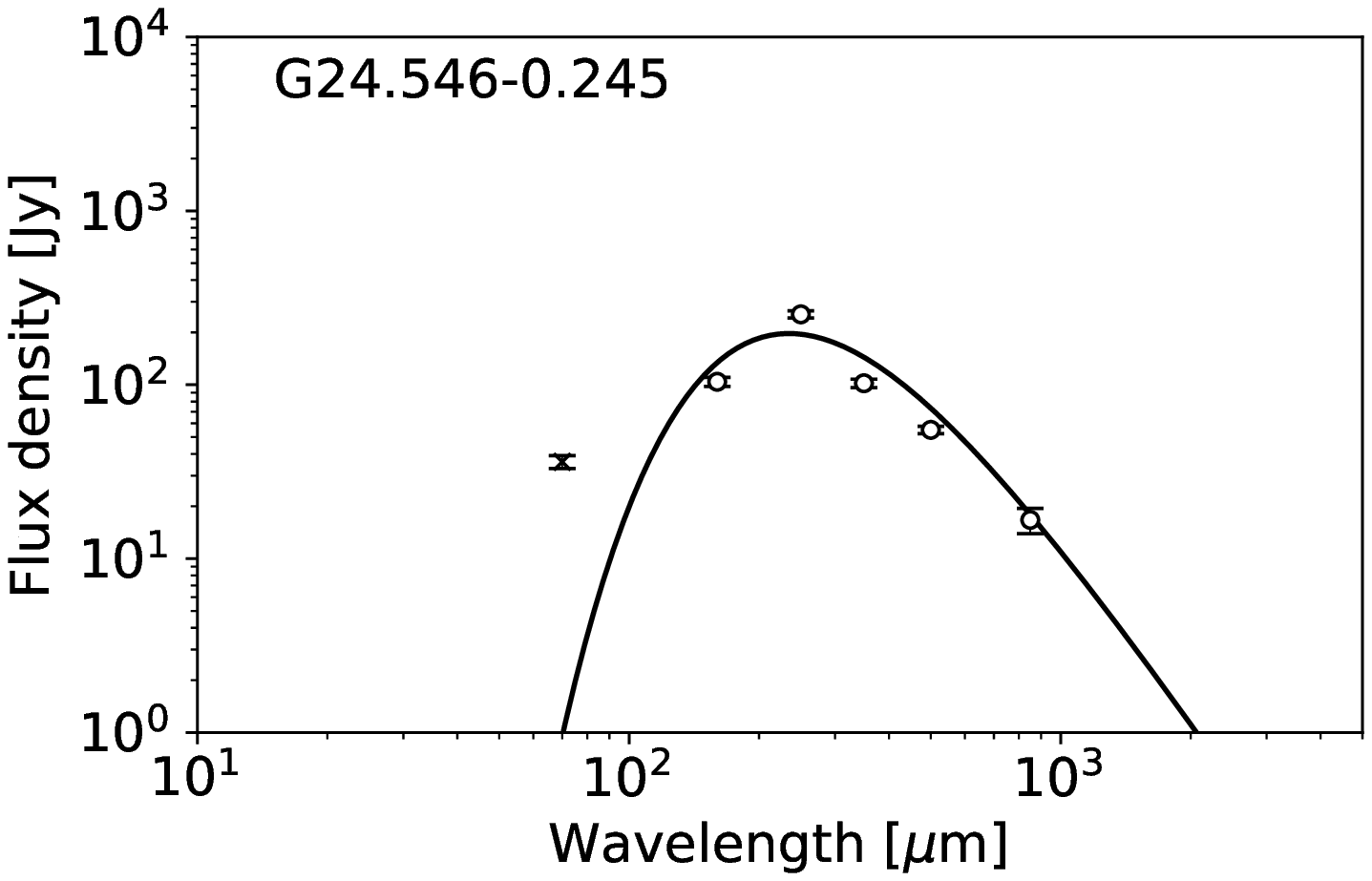}}
    \subfigure{
    \centering
    \includegraphics[width=0.25\textwidth]{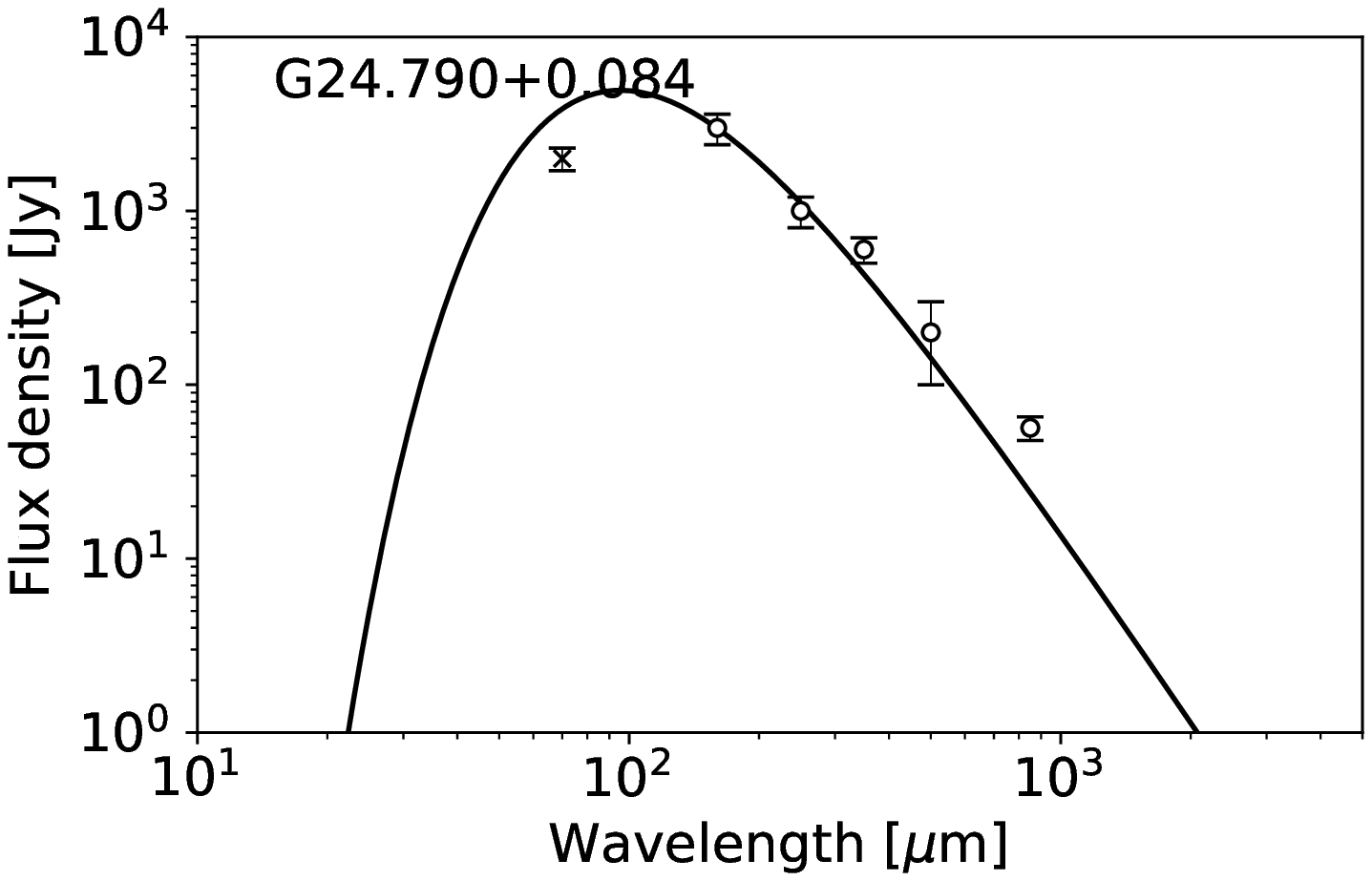}}
    \centering
    \includegraphics[width=0.25\textwidth]{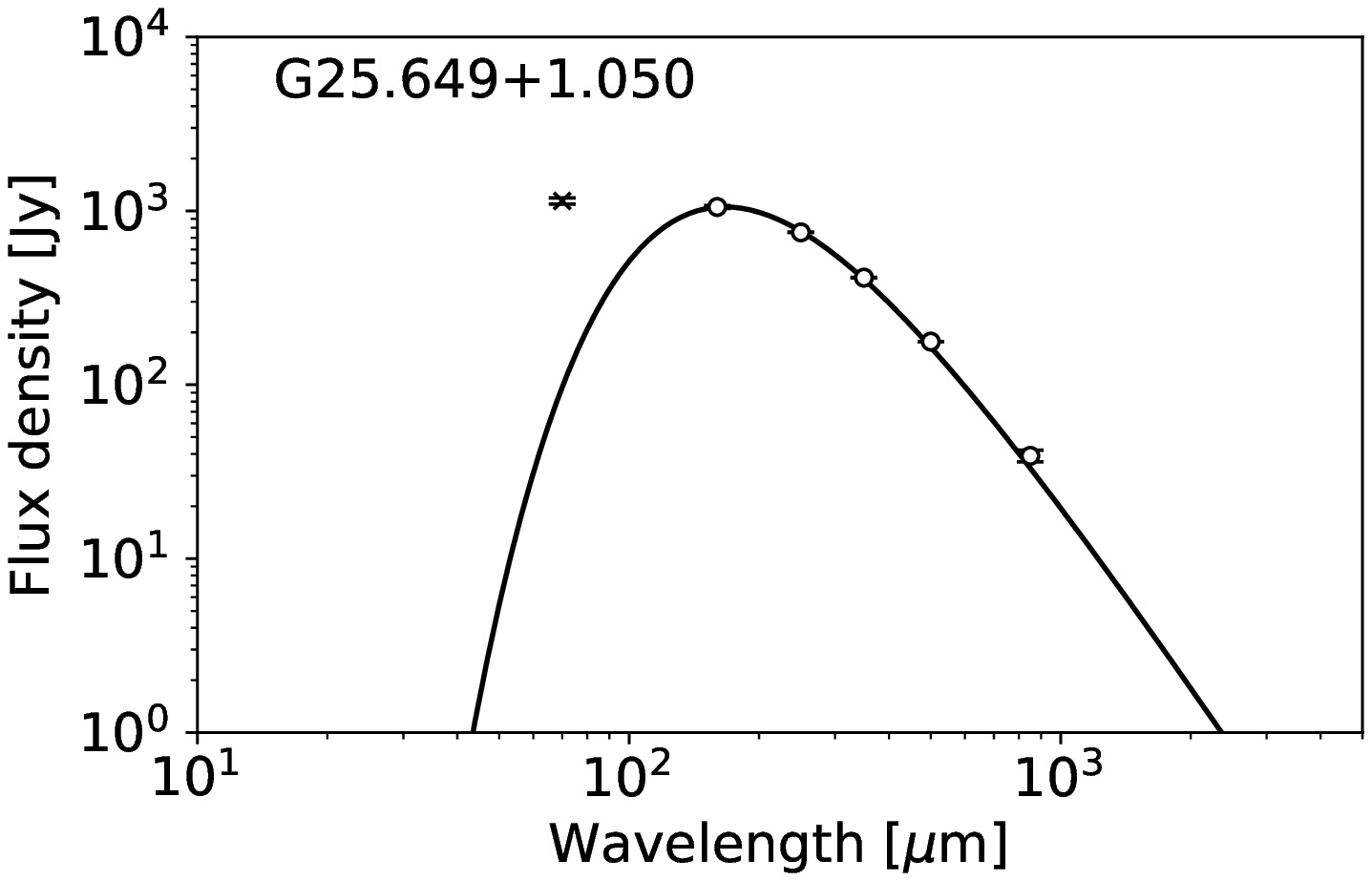}
    \centering
    \includegraphics[width=0.25\textwidth]{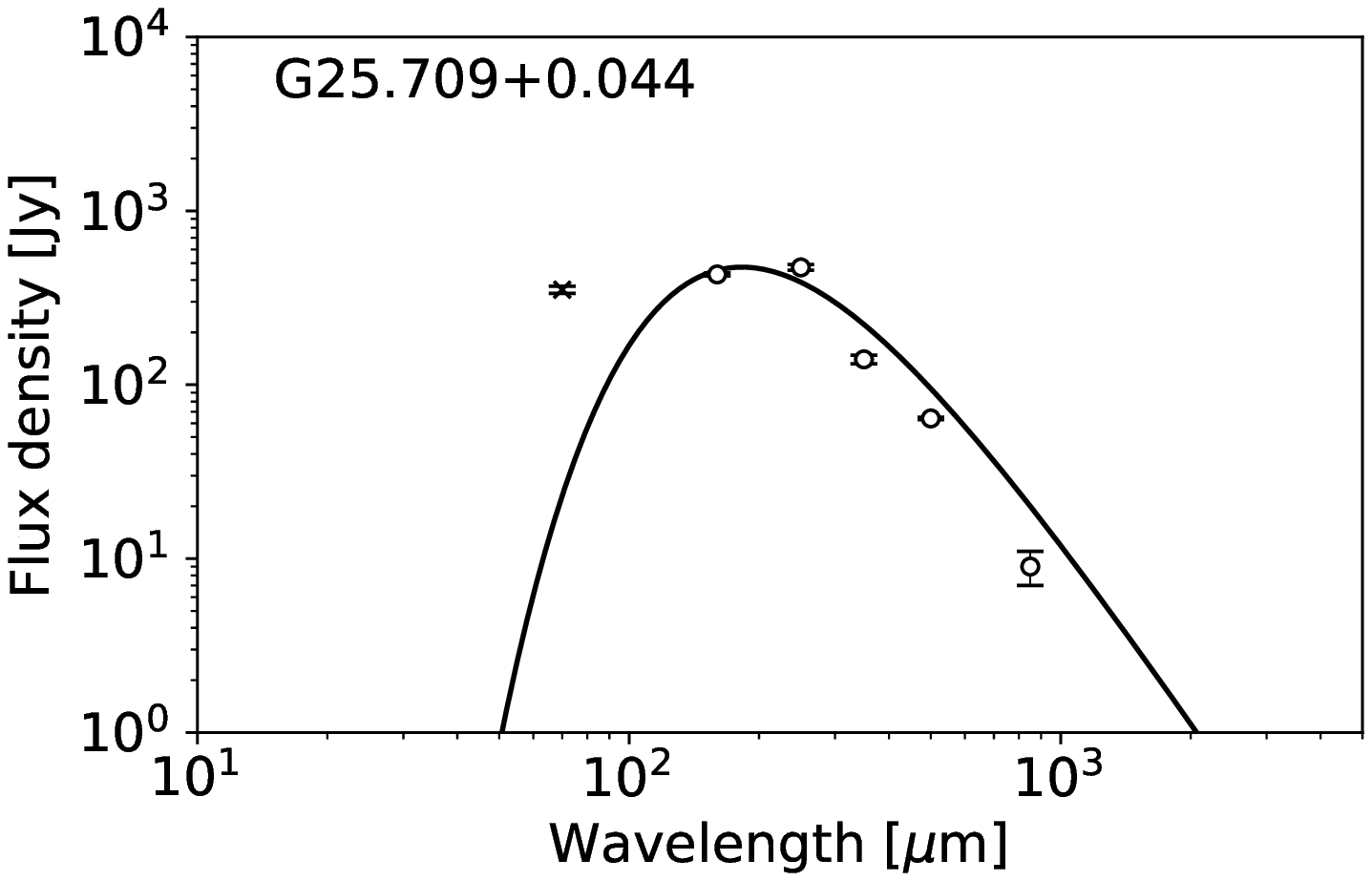}
    \centering
    \includegraphics[width=0.25\textwidth]{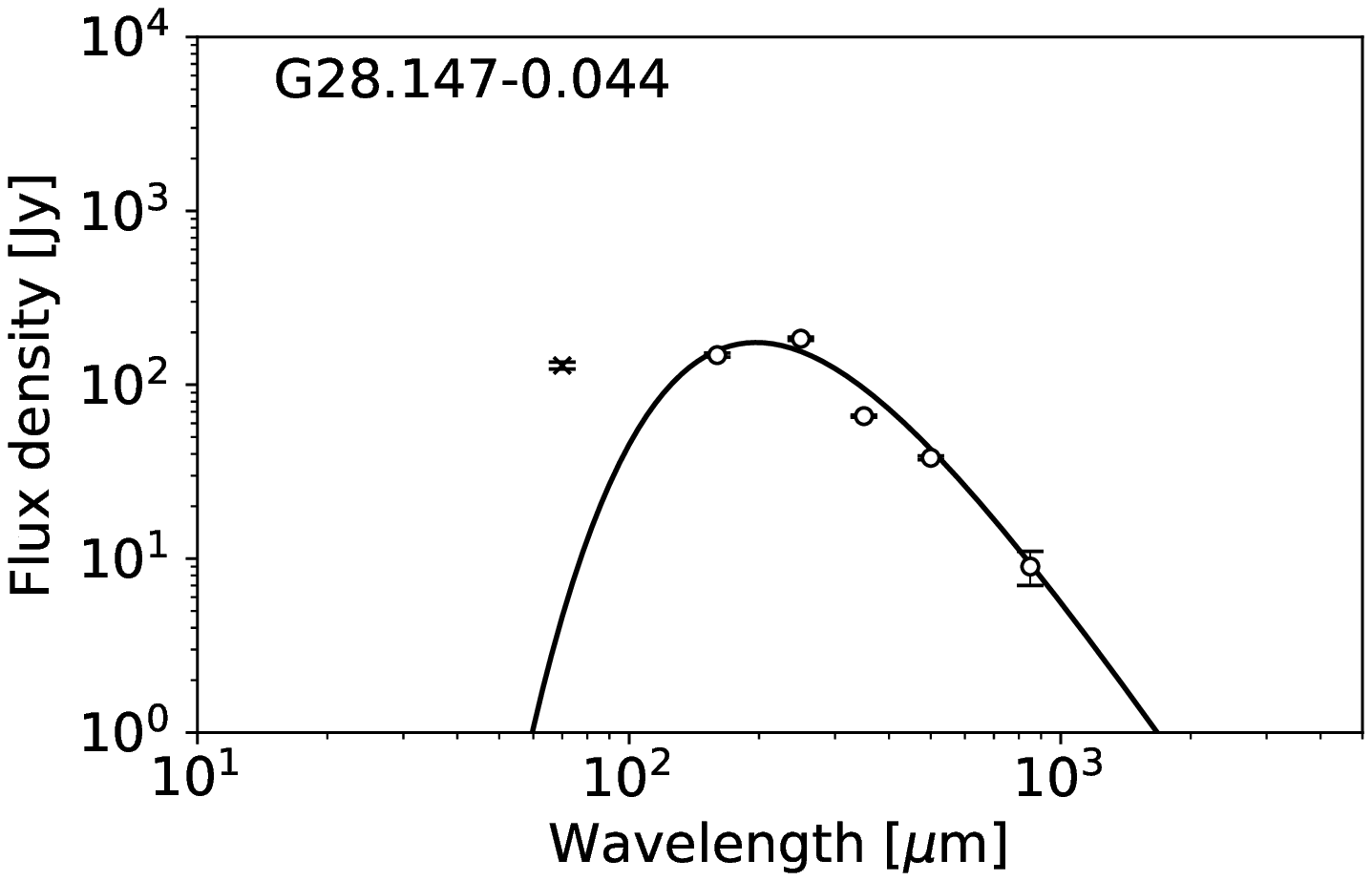}
    \centering
    \includegraphics[width=0.25\textwidth]{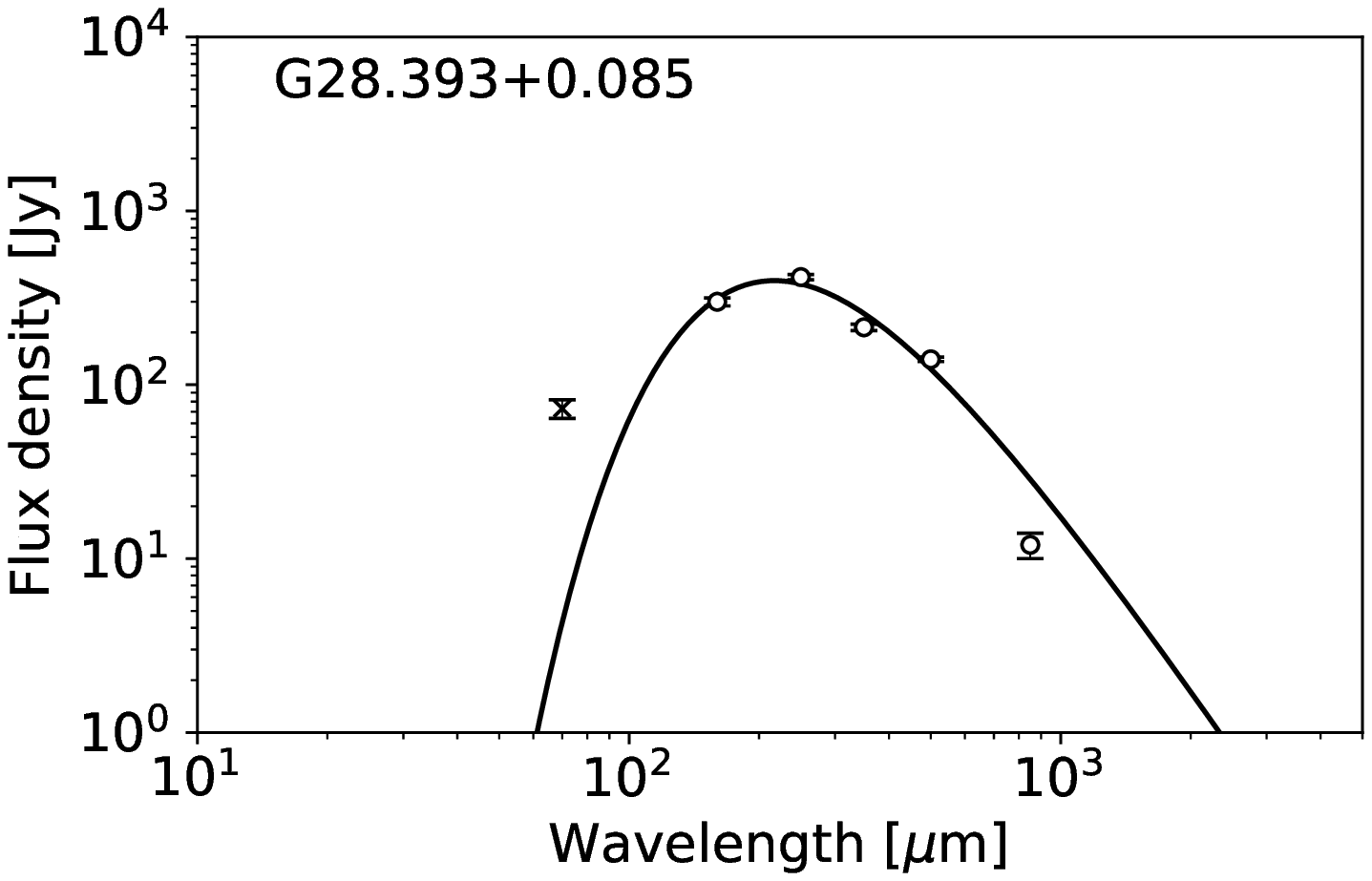}
    \centering
    \subfigure{
    \centering
    \includegraphics[width=0.25\textwidth]{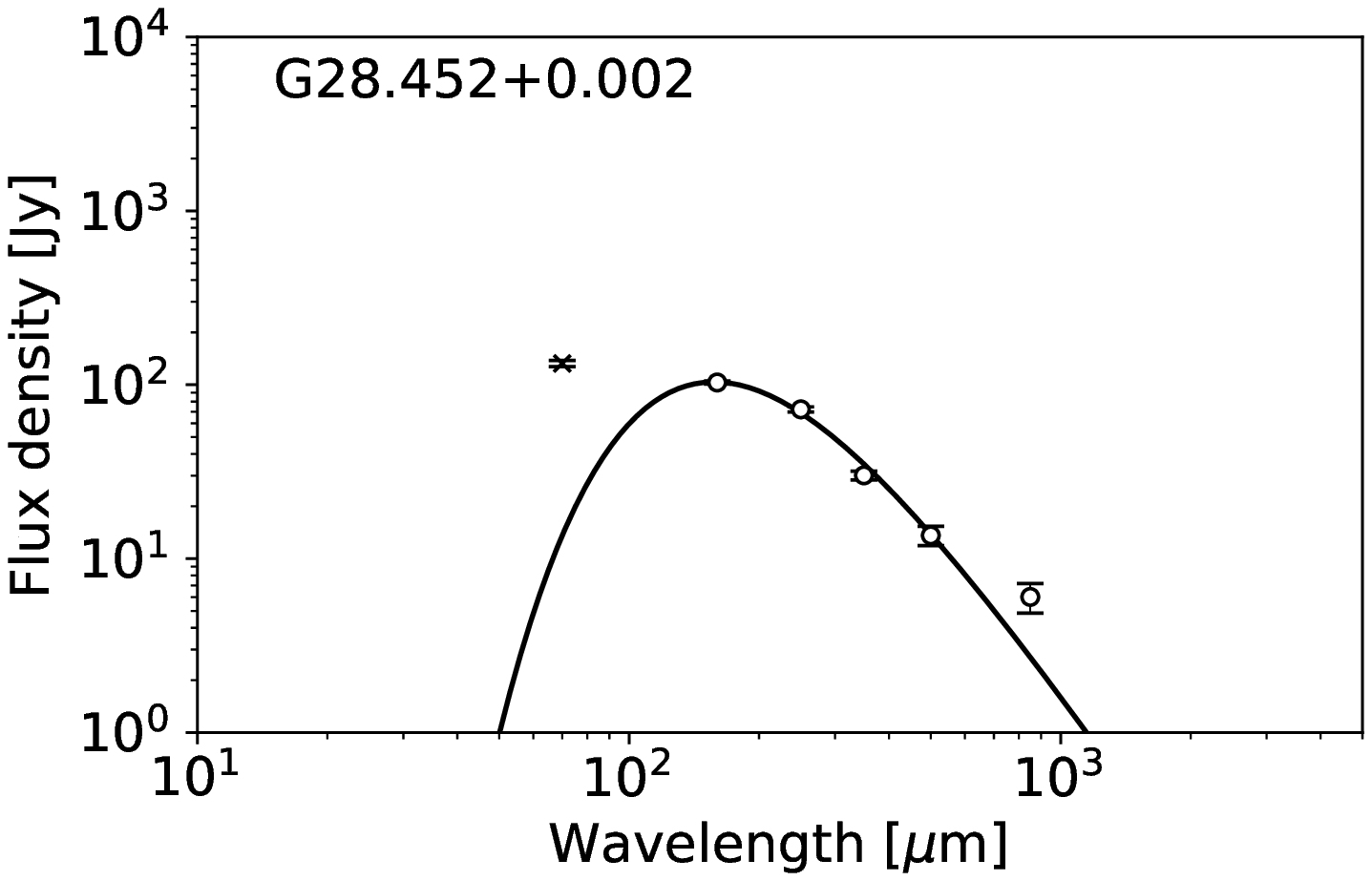}}
    \subfigure{
    \centering
    \includegraphics[width=0.25\textwidth]{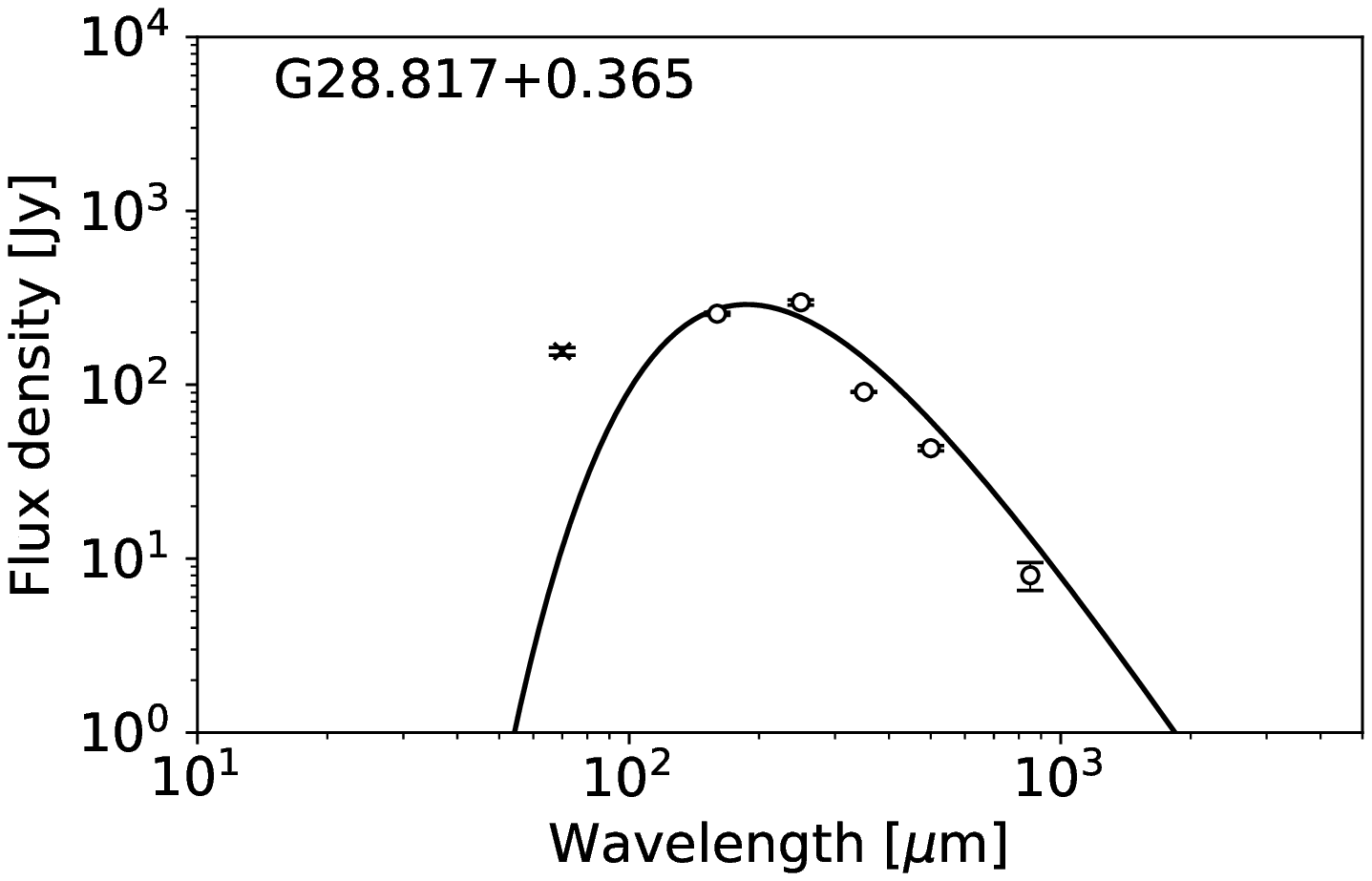}}
    \subfigure{
    \centering
    \includegraphics[width=0.25\textwidth]{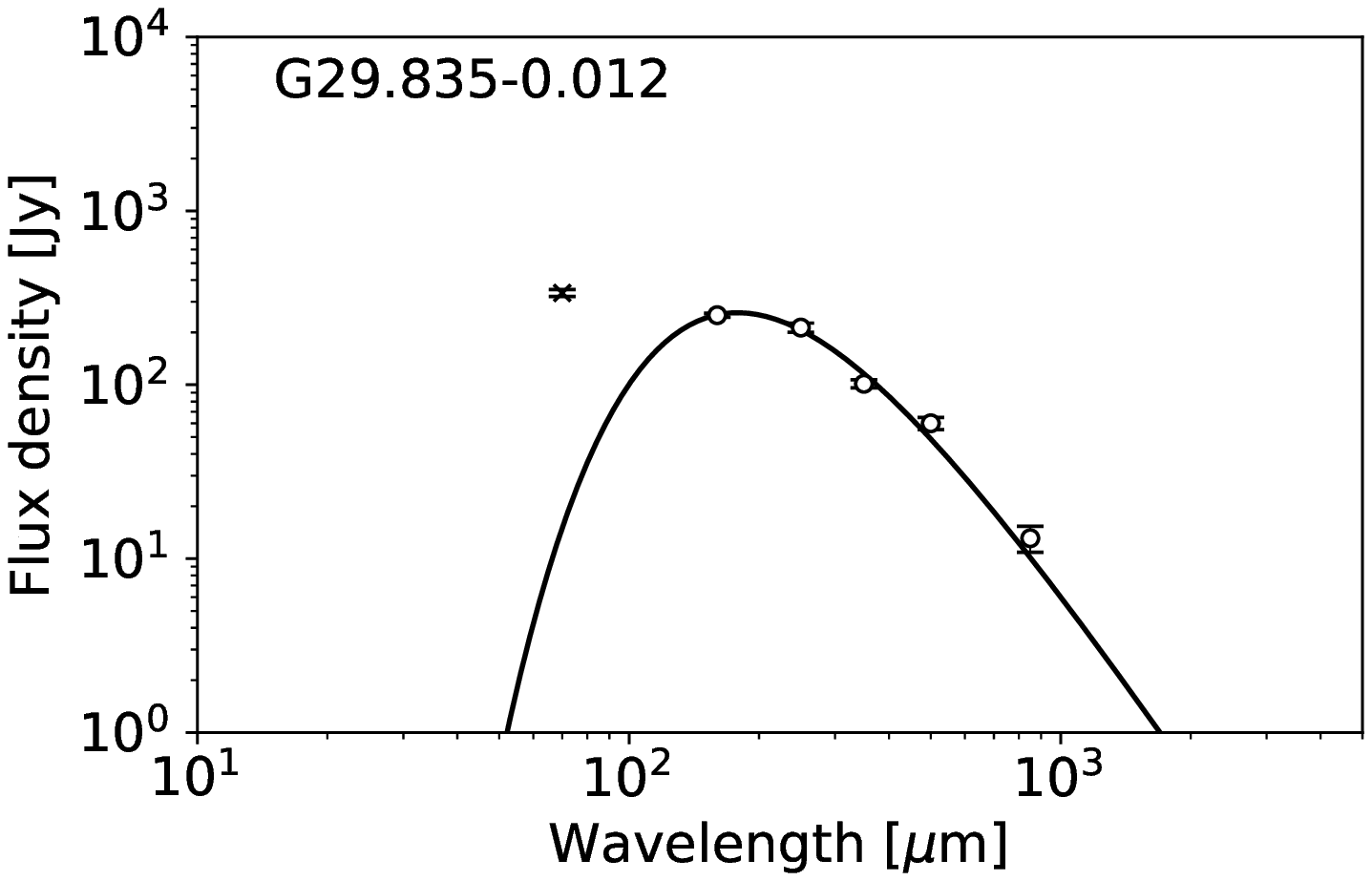}}
    \centering
    \subfigure{
    \centering
    \includegraphics[width=0.25\textwidth]{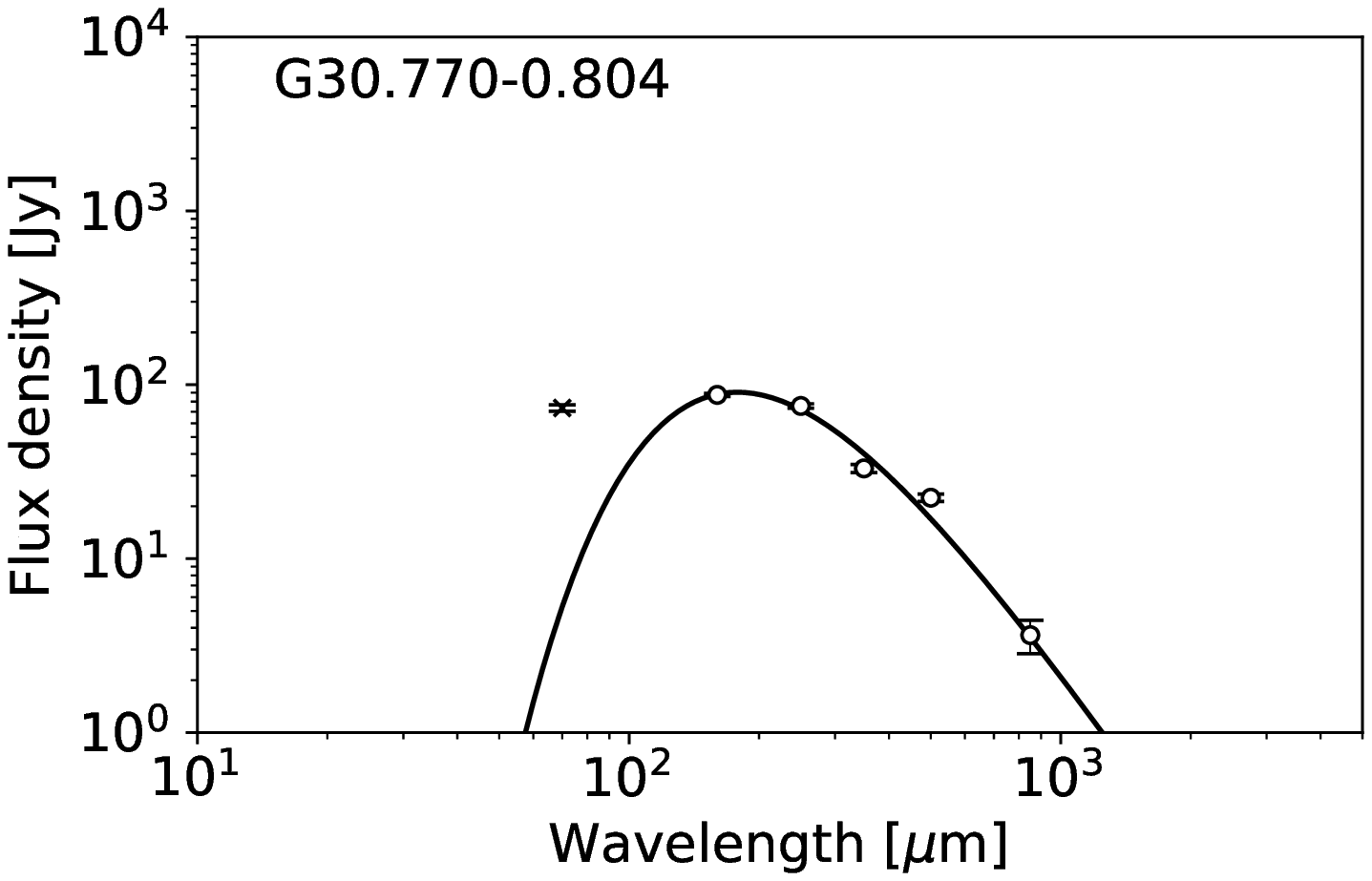}}
    \subfigure{
    \centering
    \includegraphics[width=0.25\textwidth]{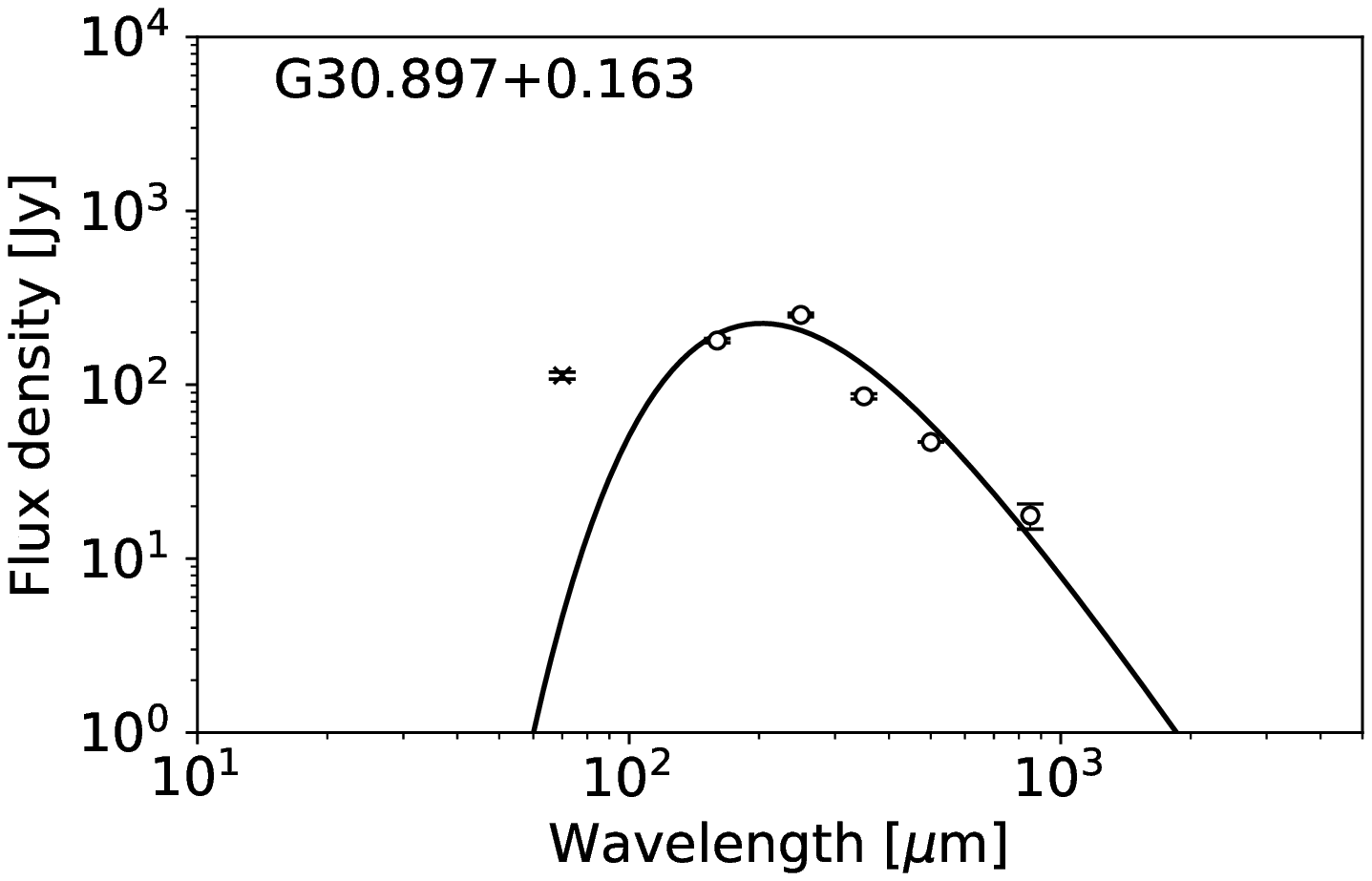}}
    \subfigure{
    \centering
    \includegraphics[width=0.25\textwidth]{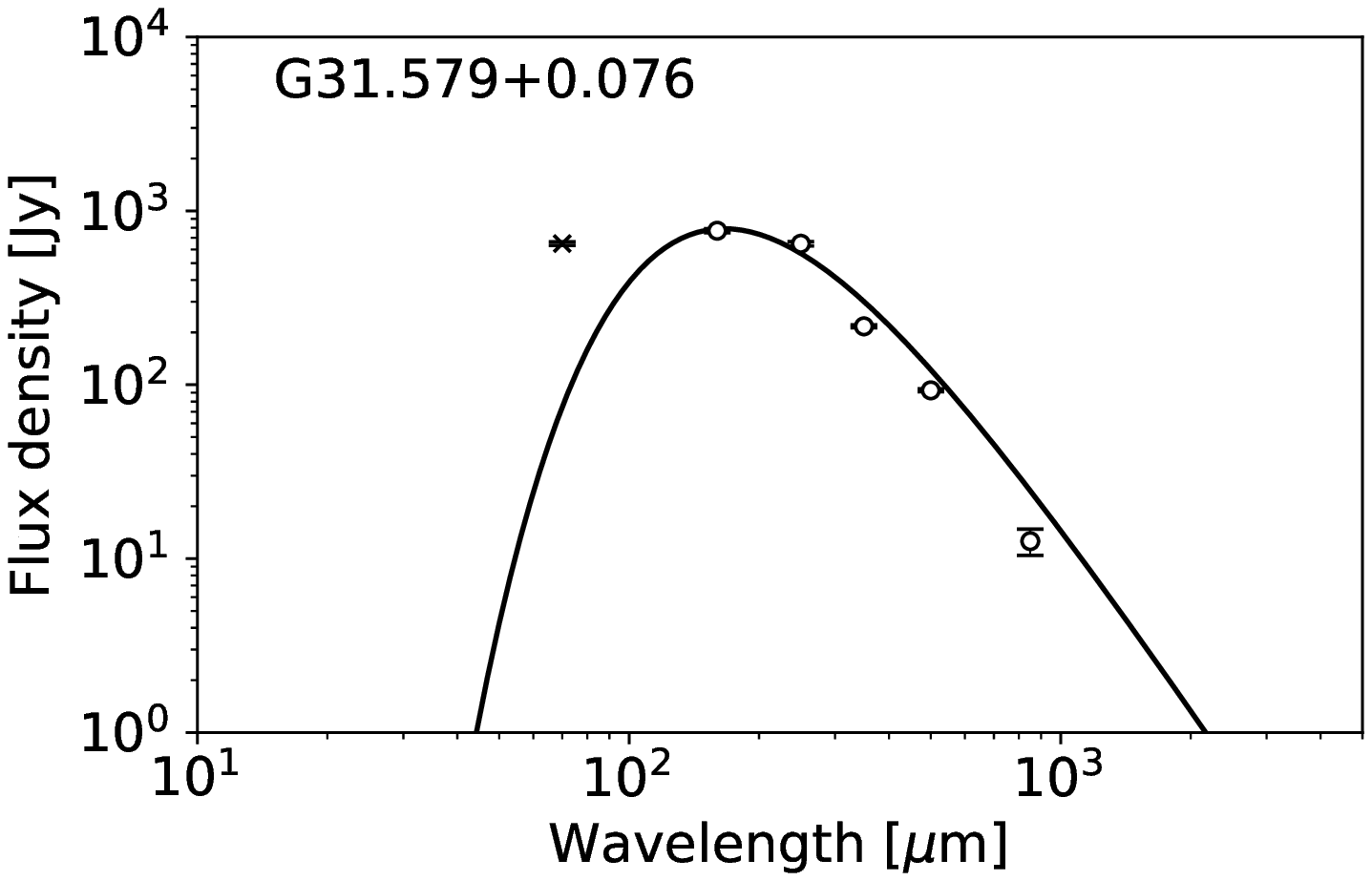}}
    \centering
    \subfigure{
    \centering
    \includegraphics[width=0.25\textwidth]{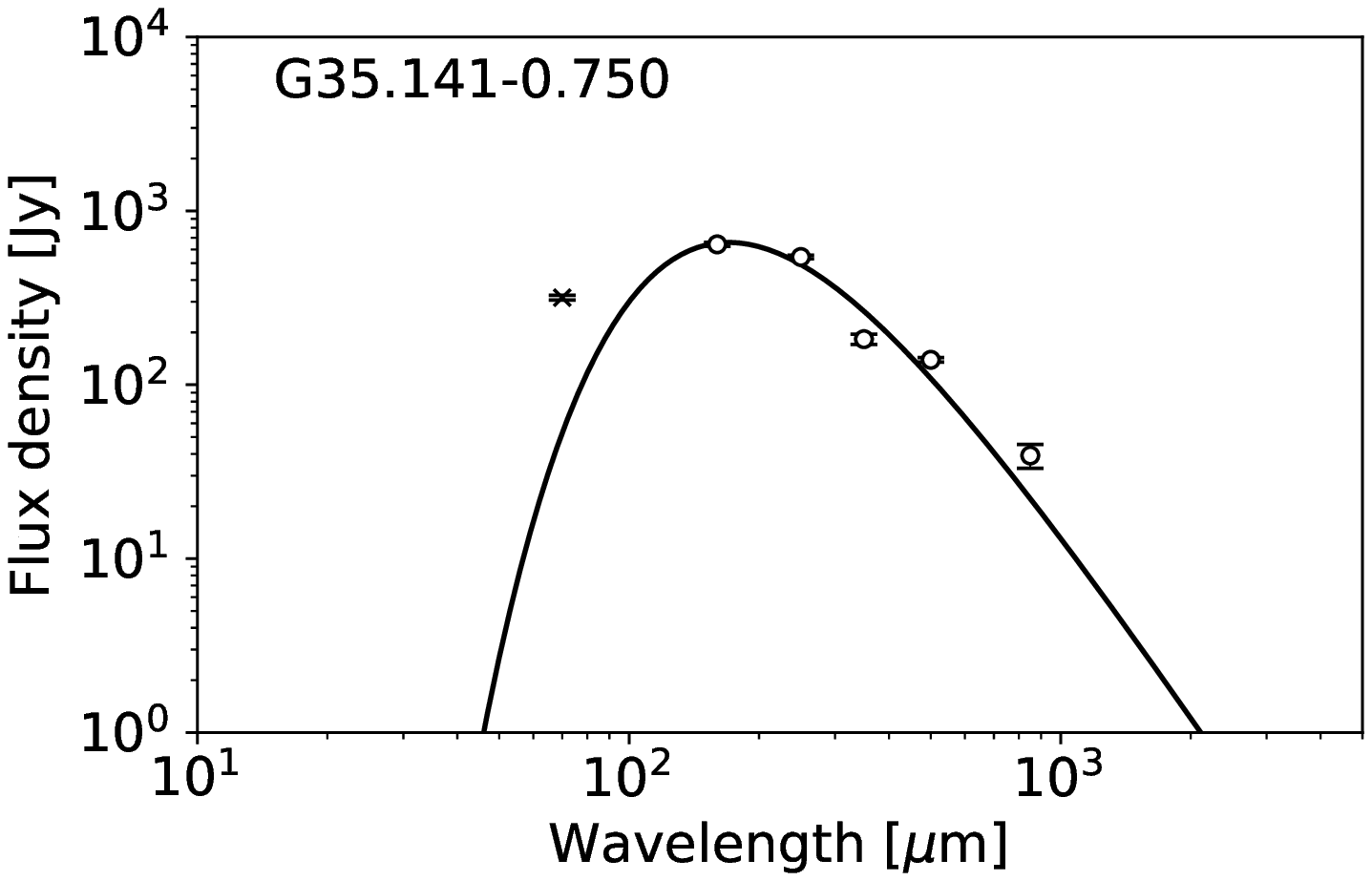}}
    \centering
    \subfigure{
    \centering
    \includegraphics[width=0.25\textwidth]{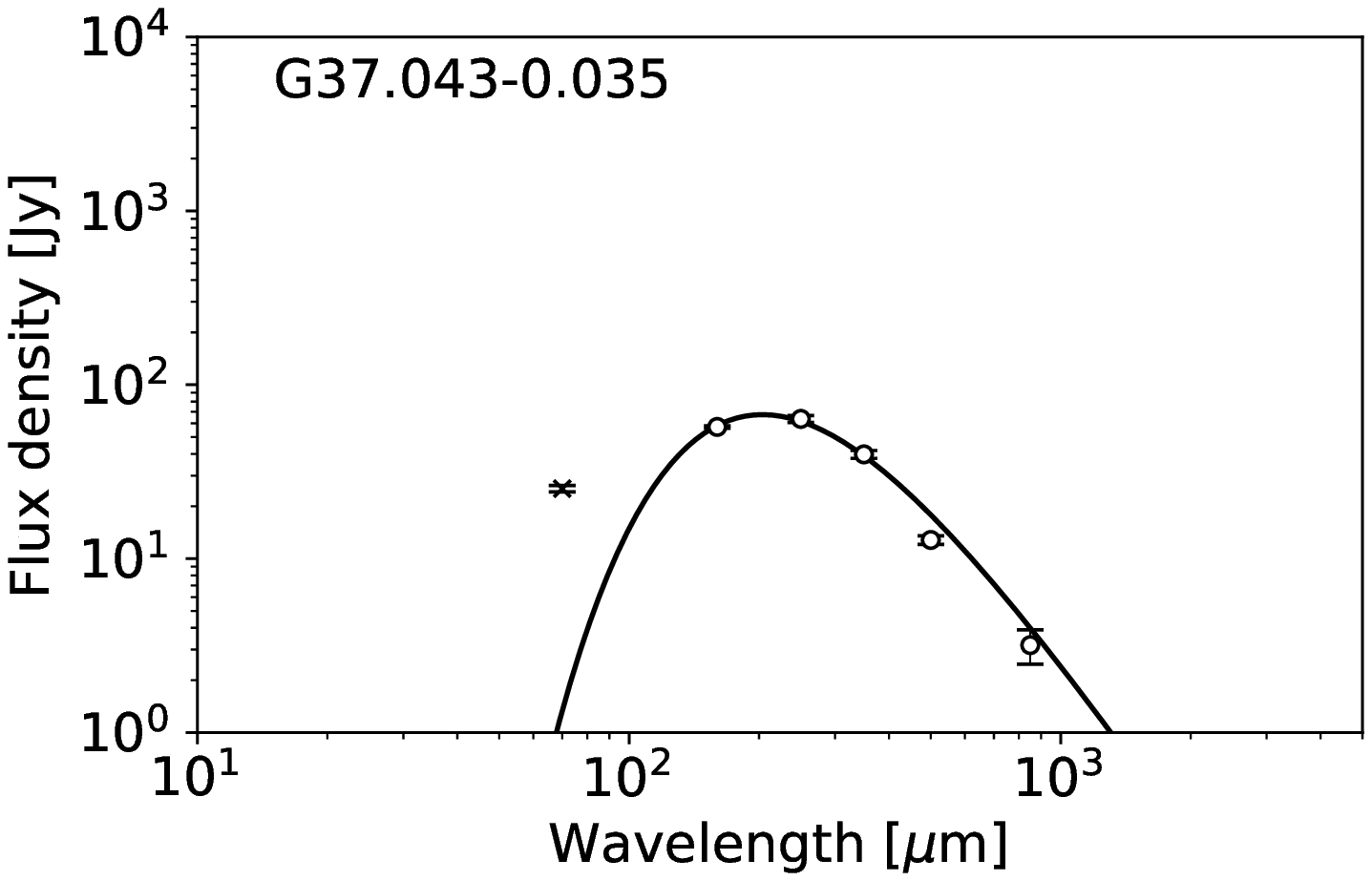}}
    \caption{FIR SED and its fit curve (the solid curve, from one single temperature grey-body model) of sources with cyanopolyynes detection. The cross indicates 70 $\mu$m emission, which is not involved in the SED fit.}
    \label{fig3}
\end{figure}

The Herschel infrared Galactic Plane Survey \citep[Hi-GAL,][]{2016A&A...591A.149M} compact-source catalogue was obtained by performing photometry of Herschel images (70, 160, 250, 350 and 500 $\mu$m). \cite{2014A&A...568A..41U} conducted a photometry of the Atacama Pathfinder Experiment \citep[APEX,][]{2006A&A...454L..13G} images (870 $\mu$m) with the source-extraction routine SExtractor\footnote{\href{https://sextractor.readthedocs.io/en/latest/License.html}{https://sextractor.readthedocs.io/en/latest/License.html}} and obtained the APEX Telescope Large Area Survey of the Galaxy \citep[ATLASGAL,][]{2009A&A...504..415S,2013A&A...549A..45C} compact-source catalogue. Using a search radius of 30" (roughly half the HPBW of the HC$_{3}$N), we cross-identified both catalogues with our HC$_{3}$N sources. Twenty-one sources were found in both catalogues and their IR flux data are presented in Table $\ref{tab5}$. In a number of previous works \citep[e.g.][]{2019MNRAS.488..495W,2010A&A...518L..97E}, a single-temperature grey-body model was used to fit the spectral energy distributions (SEDs) of these sources. The Levenberg-Marquardt algorithm provided in the Python package lmfit was applied \citep{2016ascl.soft06014N}. The grey-body model can be expressed as follows:
\begin{align}
S_{\rm \nu } = \frac{M\kappa_{\rm 0}}{D^{2}\gamma}(\frac{\nu }{\nu _{0}})^{\beta }B_{\rm \nu }(T_{\rm d}),
\end{align}
where \textit{M} is the total (dust + gas) mass, $\kappa_{0}$ = 1.85 cm$^{2}$ g$^{-1}$ is the dust opacity at 870 $\mu$m \citep{1994A&A...291..943O}, $B_{\rm \nu }$($T_{\rm d}$) is the Planck function at the dust temperature $T_{\rm d}$, $\beta$ is the dust emissivity index (the mean value of 1.75 was used for all dust models, \cite{1994A&A...291..943O}), and $\gamma$ is the gas-to-dust ratio, which depends on the galactocentric distance of the source \citep{2017A&A...606L..12G}: 
\begin{align}
\gamma = 10^{0.087\times D_{\rm GC} + 1.44}.
\label{eq8}
\end{align}
The derived values of $\gamma$ are listed in Table \ref{tab5}. Figure $\ref{fig3}$ shows the fitting results of the far-infrared (FIR) SEDs of our sources. Similar to the previous works, the Herschel 70 $\mu$m data were excluded in our grey-body fitting because of their origin from a warm dust component and a large fraction of this emission coming from very small grains \citep[e.g.][]{2019MNRAS.488..495W,2015ApJ...815..130G}. The dust temperature $T_{\rm d}$ and the total mass M were obtained from the SED fit, and are presented in Table $\ref{tab5}$. For comparison, we collected the results of $T_{\rm d}$ and M of our 21 sources (listed in Table \ref{tab5}), which were derived in ATLASGAL \citep{2018MNRAS.473.1059U}. There are eight sources with $M$ values consistent with those of \cite{2018MNRAS.473.1059U} within the error range, while all the others show a large difference in $M$ values between ours and theirs. The values of $T_{\rm d}$ for our sources are smaller than those of \cite{2018MNRAS.473.1059U}. The difference should be mainly caused by the different fitting models used, that is, a two-component (grey-body and black-body) model was used by these latter authors, while we used a single component model.

\subsection{Relative abundance of cyanopolyynes to H$_{2}$}
To obtain the relative abundance of cyanopolyynes with respect to H$_{2}$ for our sources, we first estimated $N_{\rm H_{2}}$ in these sources. Under the assumption of optically thin dust emission at 870 $\mu$m, the beam-averaged H$_{2}$ column density can be calculated using the following expression \cite{2009A&A...504..415S}:
\begin{align}
N\mathrm{_{H_{2}}} = \frac{F_{870}\gamma}{B_{870}(T_{\rm d})\Omega_{\rm app}\kappa_{0}\mu_{\rm H_{2}}m_{\rm H}},
\end{align}
where $F_{870}$ is the peak flux density (see Table \ref{tab5}) and $\Omega_{\rm app}$ is the beam solid angle. For a source located at D, $\Omega_{\rm app} = A/D^2$, where A = $R_{\rm eff}^{2}\pi$ is the the surface area of the source. The value of $R_{\rm eff}$ is presented in Table \ref{tab5}. Here, $m_{\rm H}$ is the mass of a hydrogen atom, and $\mu_{\rm H_{2}}$ = 2.8 is the mean molecular weight per H$_{2}$ molecule \citep{2008A&A...487..993K}. $B_{870}(T_{\rm d})$ is the intensity of the black-body at 870 $\mu$m at the dust temperature $T_{\rm d}$, which was determined from the SED fit (Sect. 3.2). We obtained the column densities of H$_{2}$ for 21 sources, including 6 Maser-only, 13 Maser-RRL, and 2 RRL-only sources (Table $\ref{tab5}$).

Using the results of the column density of cyanopolyynes (Table $\ref{tab4}$) and H$_{2}$ (Table $\ref{tab5}$), we derived the relative abundances of cyanopolyynes with respect to H$_{2}$ ($X(a) = N(a)/N_{\rm H_{2}}$ ---where $a$ refers to cyanopolyynes--- for 21 sources (Table $\ref{tab4}$). The mean relative abundance of HC$_{3}$N was found to be (1.22$\pm$0.52)$\times$10$^{-10}$ for Maser-only sources, (5.40$\pm$1.45)$\times$10$^{-10}$ for Maser-RRL sources, and (1.65$\pm$1.50)$\times$10$^{-10}$ for RRL-only sources. In Sect. 4.1, we discuss the differences in relative abundance between our three types of sources.

\section{Discussion}
\subsection{Evolution of HC$_{3}$N}
The definitions of the evolutionary stages of HMSFRs, including physical and chemical classifications, are still not clear \citep{2013ApJ...777..157H,2014A&A...562A...3M,2014A&A...563A..97G,2014MNRAS.443.1555U,2016Ap&SS.361..191Z,2017A&A...599A.139K,2019MNRAS.484.4444U,2022MNRAS.510.3389U}. Based on the physical properties of the clumps, the sequence of a HMSFR is usually divided into four evolutionary stages: quiescent, high-mass protostellar object (protostellar), massive young stellar object (YSO), and H{\sc ii} region \citep[see][and references therein]{2019MNRAS.484.4444U,2022MNRAS.510.3389U}. 

\cite{2019MNRAS.484.4444U} carried out a 3mm molecular-line survey towards a large sample consisting of 570 high-mass star-forming clumps and found a trend in the detection rate of HC$_{3}$N (J = 10-9), namely an increase from quiescent, to protostellar, to YSO, and then a slight decrease in H{\sc ii} region (including all H{\sc ii} region stages). \cite{2019MNRAS.489.4497Y} subdivided H{\sc ii} regions into ultra-compact (UC) H{\sc ii} regions and normal classical H{\sc ii} regions. By analysing the archival available data from the Hi-GAL and the Millimetre Astronomy Legacy Team Survey at 90 GHz (MALT90)\footnote{\href{http://atoa.atnf.csiro.au/MALT90}{http://atoa.atnf.csiro.au/MALT90}}, these latter authors found that the abundance of HC$_{3}$N could increase in UC H{\sc ii} regions, while it decreases or reaches a plateau in normal classical H{\sc ii} regions.

\begin{figure}
    \center
        \includegraphics[width=0.6\textwidth]{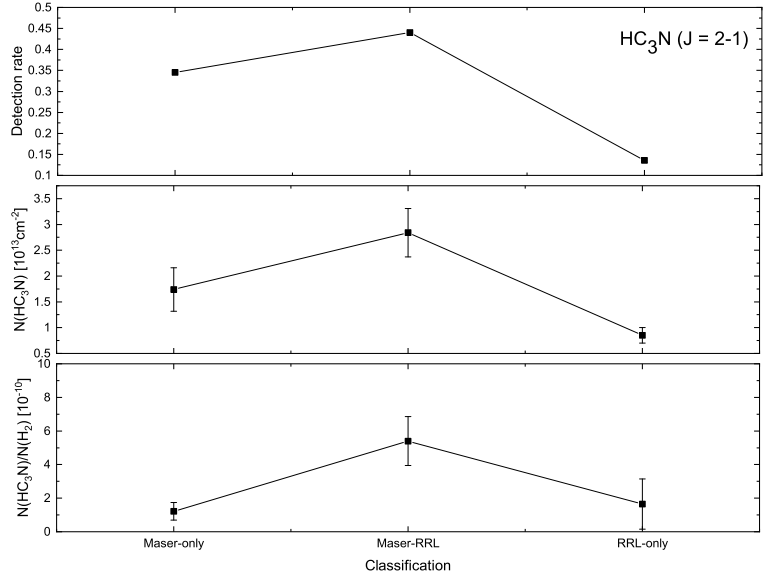}
    \caption{The detection rates, column densities, and relative abundances of HC$_{3}$N for different types of sources.}
    \label{fig4}
\end{figure}

The results of our analysis (Sect. 3) show that the detection rates of HC$_{3}$N are 34$\%$, 44$\%,$ and 14$\%$ for Maser-only, Maser-RRL, and RRL-only sources, respectively. The mean column density of HC$_{3}$N is found to be (1.75$\pm$0.42)$\times$10$^{13}$ cm$^{-2}$ for Maser-only sources, (2.84$\pm$0.47)$\times$10$^{13}$ cm$^{-2}$ for Maser-RRL sources, and (0.82$\pm$0.15)$\times$10$^{13}$ cm$^{-2}$ for RRL-only sources. The average values of the relative abundance of HC$_{3}$N are (1.22$\pm$0.52)$\times$10$^{-10}$, (5.40$\pm$1.45)$\times$10$^{-10}$, and (1.65$\pm$1.50)$\times$10$^{-10}$ for Maser-only, Maser-RRL, and RRL-only sources, respectively. Those results are plotted as a function of source classification in Fig. $\ref{fig4}$. All the results show the same trend, namely an increase from Maser-only to Maser-RRL sources and a decease from Maser-RRL to RRL-only ones.

The 6.7 GHz CH$_{3}$OH maser is excited according to the infrared radiative pumping mechanism in the disc when a formed protostar begins to warm its natal environment. This is normally considered to be related to the early stages of HMSFRs when there is significant mass accretion \citep[e.g.][]{1994A&A...291..569S,2003A&A...403.1095M,2006ApJ...638..241E,2008A&A...485..729X}. Based on the CH$_{3}$OH maser line and radio continuum mapping observations towards a sample of HMSFRs, \cite{1998MNRAS.301..640W} found that the 6.7 GHz CH$_{3}$OH maser is observable before the UC H{\sc ii} regions (YSOs) and are probably destroyed as the UC H{\sc ii} regions develop. Therefore, we assume our Maser-only sources in YSOs.

Based on the statistical investigations towards different types of masers, \cite{2007MNRAS.377..571E} and \cite{2010MNRAS.401.2219B} proposed an evolutionary sequence for masers in HMSFRs and suggested that 6.7 GHz CH$_{3}$OH maser can exist in UC H{\sc ii} regions, although most of them are located at earlier YSO stages. Considering this, as well as the fact that RRL appears after the formation of the H{\sc ii} region, we propose that our Maser-RRL sources are  found in the more evolved UC H{\sc ii} regions.

RRL are produced by scattering of free electrons off the ions, which come from the surrounding H{\sc i} gas ionised by high-energy ultraviolet (UV) photons (\textgreater 13.6 eV) from formed massive stars. RRL is a powerful tool to track H{\sc ii} regions \citep{1998MNRAS.301..640W,1996A&A...307..829B,2011ApJS..194...32A,2014ApJS..212....1A,2018ApJS..234...33A,2016era..book.....C,2018PASP..130h4101L,2022MNRAS.510.4998Z}. As RRLs are undetectable in young dense UC H{\sc ii} regions and at their previous stages due to the effect of beam dilution and high optical depth \citep[e.g.][]{2002ARA&A..40...27C,2020ApJS..248....3C}, our RRL-only sources should belong to more extended and evolved normal classical H{\sc ii} regions.

Based on these assumptions, our results concerning Maser-only sources (YSO, 34$\%$), which have a relatively high detection rate for HC$_{3}$N compared to Maser-RRL and RRL-only sources (H{\sc ii} region, $\sim$ 30$\%$), are identical to those of \cite{2019MNRAS.484.4444U}. Our results concerning Maser-RRL, namely with the largest abundance of HC$_{3}$N at this stage and a decrease from Maser-RRL (UC H{\sc ii} region) to RRL-only (normal classical H{\sc ii} region) sources, are consistent with those of \cite{2019MNRAS.489.4497Y}, that is a decrease from UC H{\sc ii} to normal classical H{\sc ii} regions. The UMIST database for astrochemistry suggested that HC$_{3}$N molecules could be destroyed by a reaction with a UV photon, which produces CN and CCH or HC$_{3}$N$^{+}$ and an electron \citep{2013A&A...550A..36M}. Therefore, the decrease in $N(\rm HC_{3}N)$ between UC H{\sc ii} regions and normal classical H{\sc ii} regions may reflect the possibility of a destruction of HC$_{3}$N by UV photons, which is consistent with the observations of \cite{2016ApJ...833..248Y}. The increase in the column density of HC$_{3}$N between YSOs and UC H{\sc ii} regions is likely correlated with the increase in dust temperature, which results in a more desorbed progenitor (e.g. C$_{2}$H$_{2}$, CH$_{4}$) of HC$_{3}$N from the dust grain and a further increase in the amount of HC$_{3}$N by gas phase reactions \citep{2019MNRAS.489.4497Y,2019ApJ...872..154T}.

However, this evolutionary picture needs to be confirmed by comparing the physical parameters (such as luminosity, mass, and $T_{\rm dust}$) of these three types of samples, as in \cite{2008A&A...481..345M}. This issue will be discussed in a forthcoming paper (Chen et al. In prep.; private communication).

\subsection{HC$_{3}$N: another tracer of shock?}

Based on its spectra with large line-width and wing emissions \citep{2004A&A...416..631B,2015MNRAS.451.2507Y,2018ApJ...854..133T,2019MNRAS.489.4497Y,2021PASJ...73..467F}, and some theoretical modelling works on L1157-B1 \citep{2011ApJ...740L...3V,2013MNRAS.436..179B,2018MNRAS.475.5501M}, HC$_{3}$N was proposed to be a  species originating from outflow shock within active star-forming regions. Here, by analysing the relation between HC$_{3}$N and other shock tracers, we study whether or not HC$_{3}$N is a good tracer of shock.

Towards our HC$_{3}$N sample, we collected the H$_{2}$CO line intensity data from the TMRT C band survey (private communication, Dr. X. Chen). Using these data, we further calculated the column density of H$_{2}$CO of our sample, as for cyanopolyynes in sample Sect. 3.1. Our column density results for H$_{2}$CO are presented in Table \ref{tab6}. The derived column densities range from 0.45 to 61.49 $\times$10$^{13}$ cm$^{-2}$, the mean value being equal to (7.78$\pm$3.25)$\times$ 10$^{13}$ cm$^{-2}$. Our results are consistent with values calculated from previous observations towards one ATLASGAL HMSFR sample, which range from 0.48 to 61 $\times$ 10$^{13}$ cm$^{-2}$, the mean value being (7.8$\pm$1.3)$\times$ 10$^{13}$ cm$^{-2}$, and the predictions of the RADEX non-LTE model \citep{2018A&A...611A...6T}. 

\begin{figure}
    \includegraphics[width=0.5\textwidth]{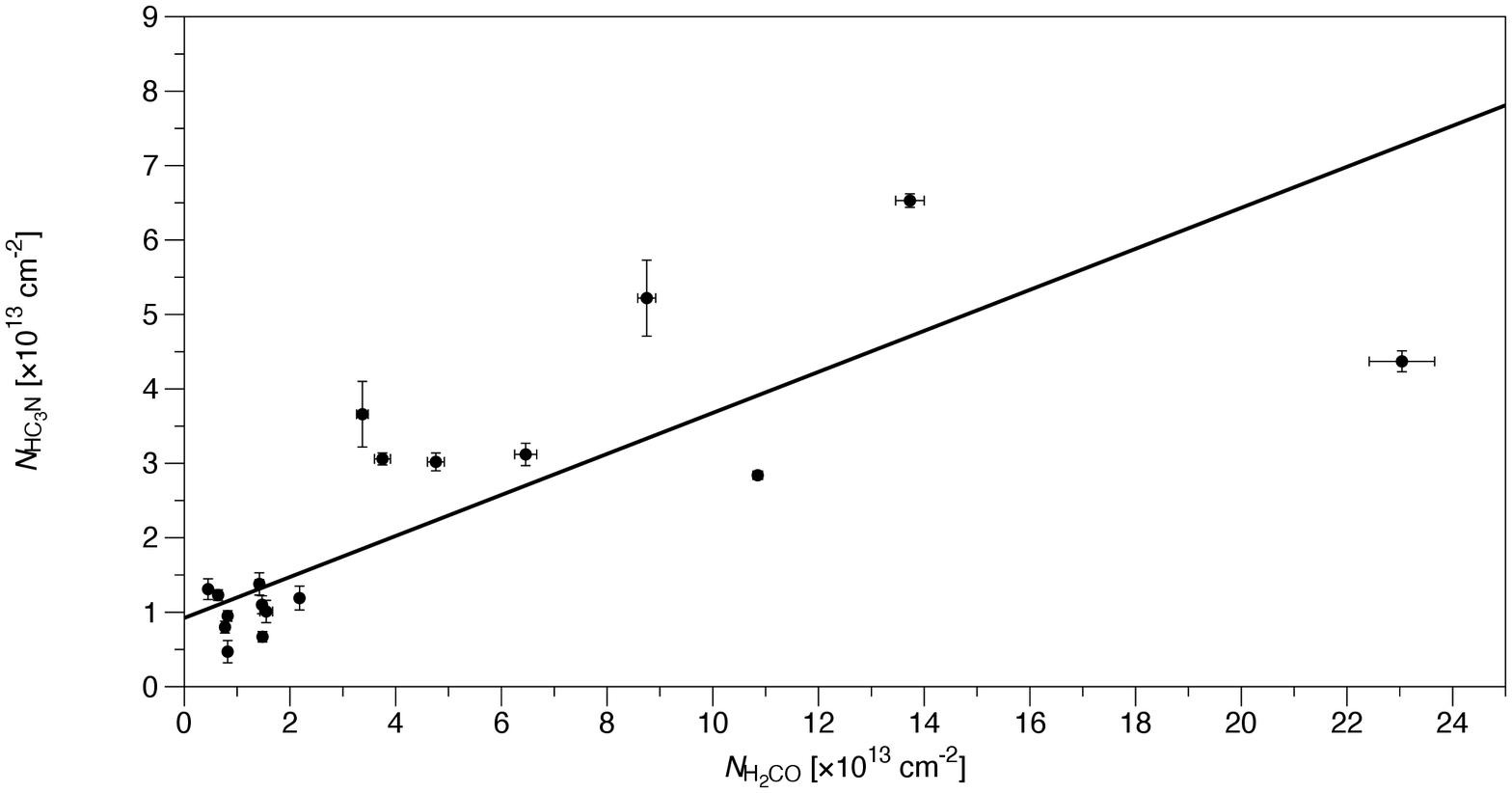}
    \includegraphics[width=0.5\textwidth]{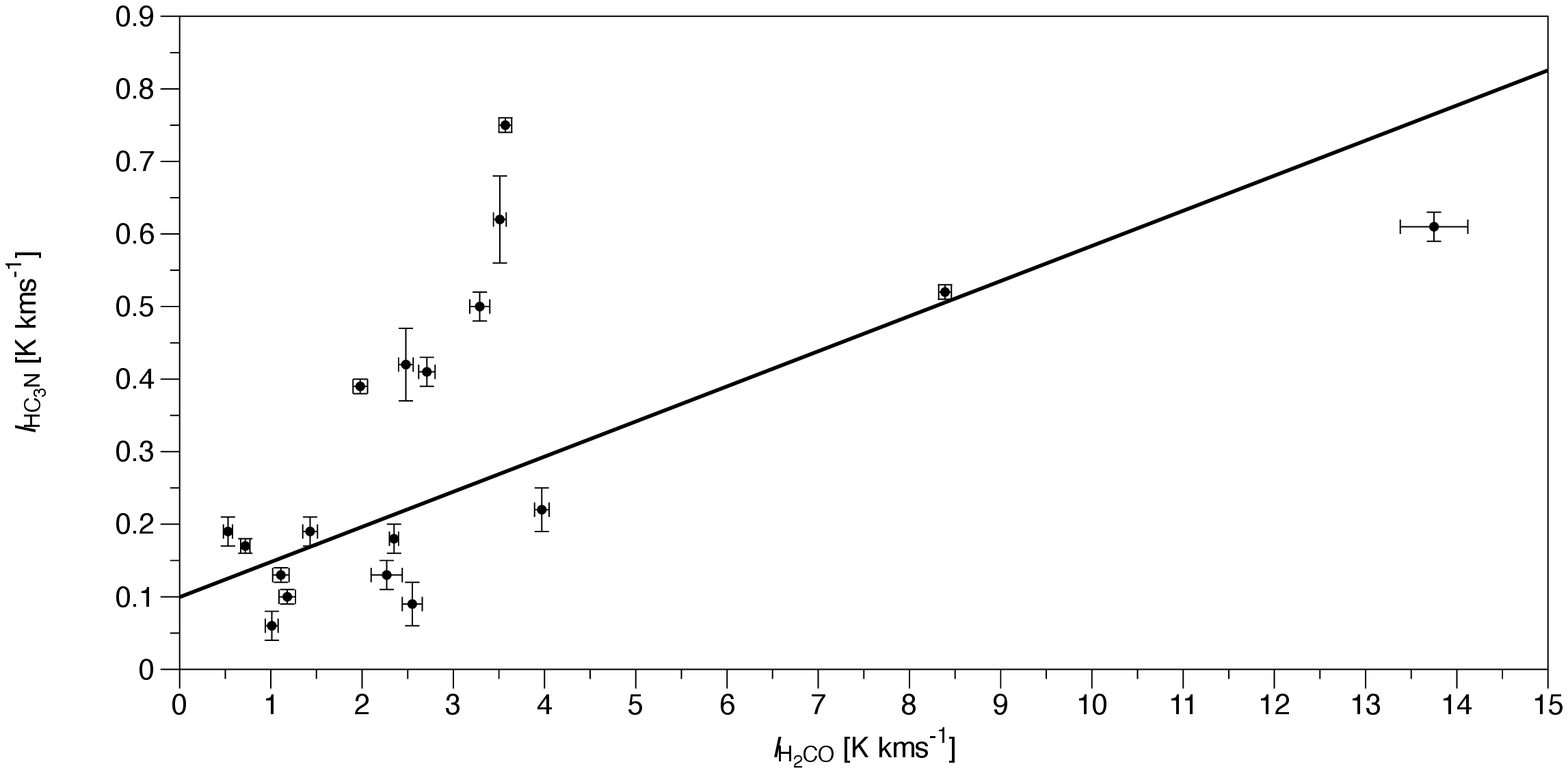}
    \includegraphics[width=0.5\textwidth]{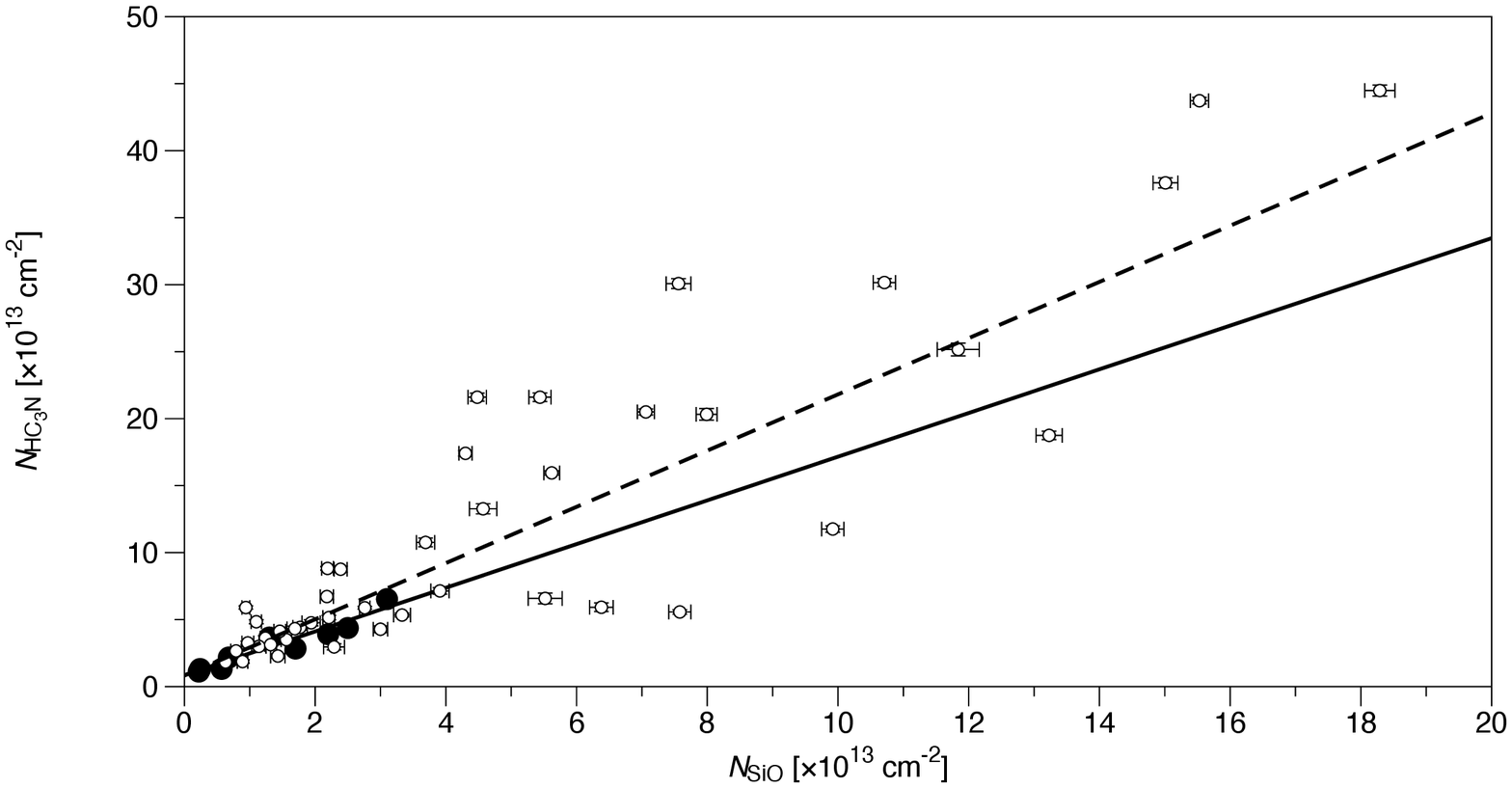}
    \includegraphics[width=0.5\textwidth]{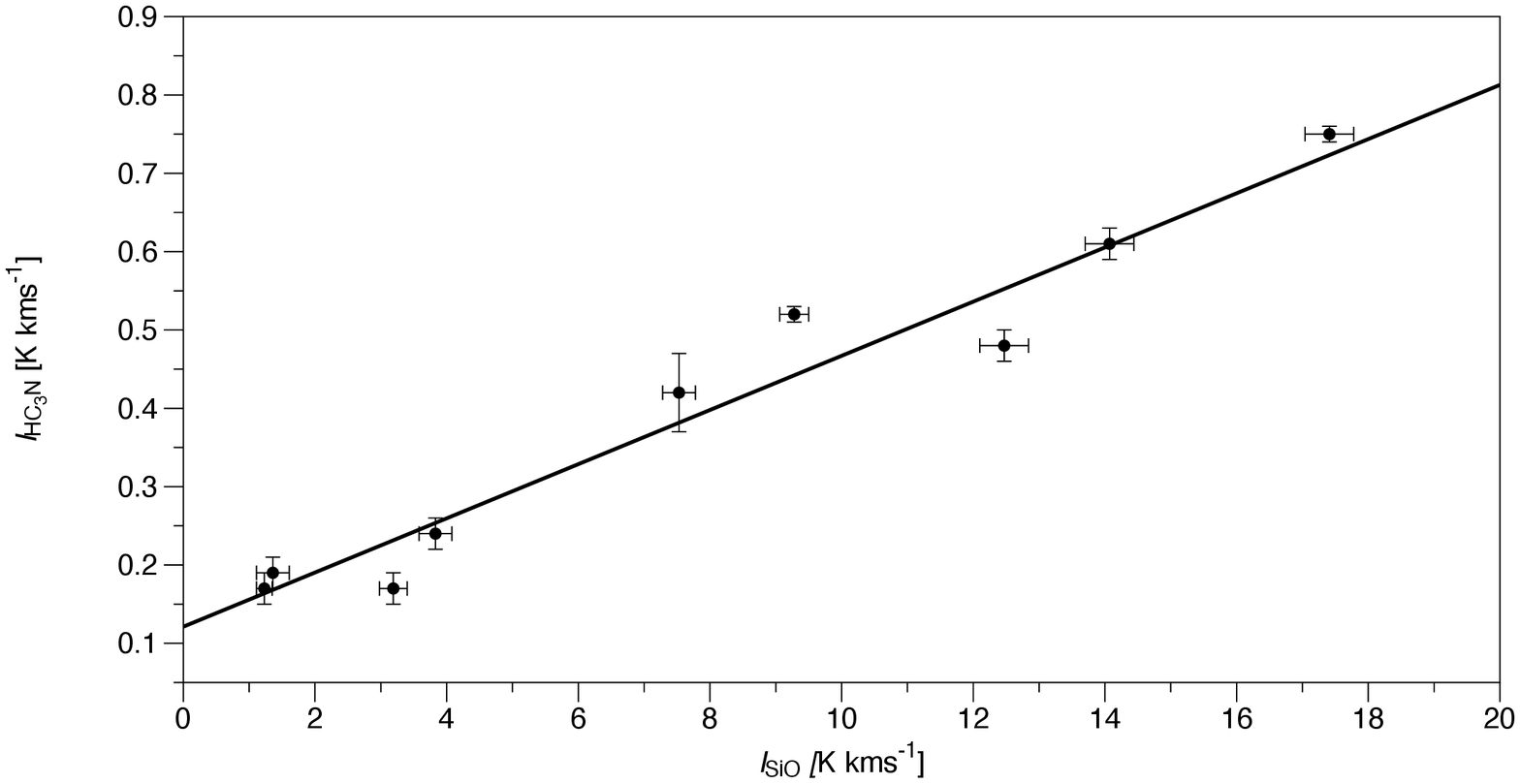}
    \caption{Column density (left panels) and the integrated line intensity (right panels) of HC$_{3}$N are plotted against those of shock-tracing molecules (H$_{2}$CO and SiO). The circles and empty circles indicate our sources and sources reported in \cite{2021ApJS..253....2H}, respectively. The solid lines and the dashed line show the weighted linear fitting line for our data and theirs.}
    \label{fig5}
\end{figure}

Figure \ref{fig5} (upper left panel) presents the column density of H$_{2}$CO versus that of HC$_{3}$N for our sample. A correlation between these characteristics can be seen, albeit with large scatter; that is, $N_{\rm HC_{3}N}$ tends to be stronger in the sources with larger $N_{\rm H_{2}CO}$. A weighted least-squared linear fit gives $N_{\rm HC_{3}N} = (0.28\pm0.06) \times N_{\rm H_{2}CO} + (0.92\pm0.28)$, with a correlation coefficient of 0.77, and the errors representing 1$\sigma$ standard deviations. A similar correlation can also be found between the line intensities (with a Pearson correlation of 0.57, upper right panel), which reflects non-significant Malmquist bias on the correlation of their column densities.   

For another shock tracer, SiO, we collected the data for our sample, including the integrated line intensity and the corresponding column density \citep{2016A&A...586A.149C}. In Fig. \ref{fig5}, we plotted the column density (bottom left panel) and the integrated line intensity (bottom right panel) of HC$_{3}$N versus those for SiO for our sample. Both panels show a significant correlation between SiO and HC$_{3}$N, which reflects no significant selection bias. For the column density, a weighted least-squares linear fit (solid line) gives $N_{\rm HC_{3}N} = (1.63\pm0.22)\times N_{\rm SiO} + (0.85\pm0.26)$, with a larger correlation coefficient of 0.94 than that found for H$_{2}$CO and HC$_{3}$N.

Similar strong correlations between the parameters for HC$_{3}$N and SiO were also found before based on the results for a sample of 43 southern HMSFRs. These parameters included the line width, the integrated line intensity, the column density, and the relative abundance \citep{2021ApJS..253....2H}. For comparison, the column density results (empty circles) with the weighted least-squares linear fitting line (the dashed line) are also shown in Fig. \ref{fig5} (bottom left panel). The strong correlation between HC$_{3}$N and SiO is consistent, although their column densities cover larger ranges and their fit line has a larger slope. Both results in the literature and ours suggest a strong correlation between HC$_{3}$N and SiO in terms of integrated line intensity and column density.

Taking H$_{2}$CO and SiO molecules as good tracers of shock, positive correlations between them and HC$_{3}$N support the idea of a shock origin of HC$_{3}$N. Under the impact of shocks, C$_{2}$H$_{2}$, being the precursor molecule of HC$_{3}$N, is released from dust grains to the gas phase and reacts with CN to form HC$_{3}$N. This is in agreement with the findings  of \cite{2019ApJ...881...57T}  from chemical models that the sublimation of CH$_{4}$ and C$_{2}$H$_{2}$ from dust grains enhances key reactions of the formation of cyanopolyynes. Therefore, our results indirectly suggest that the neutral--neutral reaction between C$_{2}$H$_{2}$ and CN is the dominant formation pathway of HC$_{3}$N. In addition, shock models indicate that HC$_{3}$N abundance would be strongly enhanced as a consequence of the passage of the shock \citep{2018MNRAS.475.5501M}. The presence of a C-type shock with a pre-shock density \textgreater 10$^{4}$ cm$^{-3}$ and a velocity of $\sim$ 40 km s$^{-1}$ (to reach the maximum shock temperature of 4000 K) can produce the observed high abundance of HC$_{3}$N \citep{2013MNRAS.436..179B}. Meanwhile, silicon monoxide (SiO) mainly originates from jets probing accretion processes \citep[e.g.][]{1999A&A...343..585C,2014A&A...570A...1D,2018MNRAS.474.3760C}. Under the action of high-velocity (20-50 km s$^{-1}$) shocks, Si atoms and Si-bearing molecules from the dust grains are evaporated to the gas phase and are subsequently oxidised to SiO \citep{1992A&A...254..315M,2008A&A...482..809G,2016ApJ...822...85L}. However, H$_{2}$CO is formed on the surface of dust grains by consecutive hydrogenation of CO, which can be released to the gas phase by shocks with relatively low velocity ($\sim$ 15 km s$^{-1}$) \citep{1984ApJS...54...81M,2010A&A...518L.112C,2010A&A...522A..91T}. Therefore, SiO and HC$_{3}$N may trace similar high-velocity shock regions, while H$_{2}$CO traces relatively low-velocity shock ones. This is in agreement with our results, which show a stronger correlation between SiO and HC$_{3}$N than that seen between H$_{2}$CO and HC$_{3}$N.

To further confirm the dominant formation pathway of HC$_{3}$N, we can investigate the correlation between C$_{2}$H$_{2}$ and HC$_{3}$N towards sources with shocks and carry out observations of HC$_{3}$N and its three isotopologues towards a large sample. Moreover, observations of the spatial distribution of different shock tracers are needed to find the similarities and differences in terms of their physical and chemical properties.

\subsection{Unsaturated iCOMs in HMSFRs with a formed protostar}
As mentioned in Sect. 1, unsaturated iCOMs were previously thought to be destroyed in 'hot core' chemistry. However, some observational and theoretical studies seem to indicate that HC$_{5}$N and even HC$_{7}$N can be present in the protostellar or hot molecule cores \citep[HMCs, which are the sources around YSOs with a rich molecular line spectrum;][]{2000prpl.conf..299K,2005IAUS..227...59C,2009ApJ...691..823H,2011ApJ...741..120R,2021MNRAS.505.2801L}. Unlike LMSFRs, the formation pathway of unsaturated iCOMs in HMSFRs with a formed protostar is not yet clear.

Our survey successfully detected HC$_{5}$N J = 6-5 transition ($E_{\rm u}/k_{B}$ $\sim$ 2.68 K) in 11 sources (10 of them are new) among 123 targeted HMSFRs (Sect. 3). Combined with previous observations \citep{2014MNRAS.443.2252G,2018ApJ...854..133T}, HC$_{5}$N was detected in 60 HMSFRs with a formed protostar. Furthermore, our first detection of the HC$_{7}$N J = 15-14 transition ($E_{\rm u}/k_{B}$ $\sim$5.68 K) in G24.790+0.084 leads to three HC$_{7}$N sources in three HMSFRs with a formed protostar (Orion KL in Feng et al. 2015 and G28.28-0.36 in Taniguchi et al. 2018a). Our detections enlarge the HC$_{5}$N and HC$_{7}$N samples and confirm that unsaturated iCOMs can exist in YSOs and UC H{\sc ii} regions, which is consistent with the predictions of the chemical model of \cite{2009MNRAS.394..221C}. However, our observations may be contaminated by the low-temperature gas without star-forming activity because of the large beam size ($\sim$53") and low J transition with low level energy of molecular transitions ($\textless$ 10 K). Therefore, follow-up observations of high J molecular transitions with high spatial resolution would help to confirm the origin of HC$_{5}$N and HC$_{7}$N. More modelling works on iCOM chemistry are also needed to improve our understanding of unsaturated iCOMs.

It is also interesting that we find a significant difference in the column density ratio HC$_{3}$N/HC$_{5}$N between our subsamples (Table \ref{tab4}). The mean value of the ratio of 10.45$\pm$2.89 for Maser-RRL sources exceeds that of Maser-only sources of 4.11$\pm$0.46 by a factor of about three. However, the ratio of $N(\rm HC_{3}N)$/$N(\rm HC_{5}N)$ seems to decrease with time as obtained from the simulation result \citep{2019ApJ...881...57T}. Further investigations on large samples and more targeted chemical simulations are required to check whether the $N(\rm HC_{3}N)$/$N(\rm HC_{5}N)$ ratio can be used as a chemical clock, given that the relative abundances of different molecules changes with time \citep{2004A&A...422..159W}.

\section{Summary}
We present a cyanopolyyne (HC$_{\rm 2n+1}$N) line survey in the Ku band (12-18 GHz) towards a large sample of 123 HMSFRs using the TMRT. The sample was divided into 29 sources with detection of the 6.7 GHz CH$_{3}$OH maser alone (Maser-only), 44 sources with detection of RRL alone (RRL-only), and 50 sources with both maser and RRL (Maser-RRL). Our main results can be summarised as follows:   \begin{enumerate}
      \item Among 123 targets, HC$_{3}$N is detected in 38 sources (including 34 new ones), which include 10 Maser-only, 22 Maser-RRL, and 6 RRL-only sources. The detection rates are 34$\%$, 44$\%,$ and 14$\%$ for Maser-only, Maser-RRL, and RRL-only sources, respectively.
      \item The mean column densities of HC$_{3}$N are (1.75$\pm$0.42)$\times$10$^{13}$ cm$^{-2}$, (2.84$\pm$0.47)$\times$10$^{13}$ cm$^{-2}$, and (0.82$\pm$0.15)$\times$10$^{13}$ cm$^{-2}$ for Maser-only, Maser-RRL, and RRL-only sources, respectively. Based on the FIR data obtained from Hi-GAL and ATLASGAL, we constructed the FIR SED for our sample and obtained a number of physical parameters by fitting the SED, including the dust temperature, the H$_{2}$ column density, and the relative abundance of cyanopolyynes. The mean relative abundances of HC$_{3}$N are (1.22$\pm$0.52)$\times$10$^{-10}$, (5.40$\pm$1.45)$\times$10$^{-10}$, and (1.65$\pm$1.50)$\times$10$^{-10}$ for Maser-only, Maser-RRL, and RRL-only sources, respectively.
      \item The detection rate, the mean column density, and the mean relative abundance of HC$_{3}$N for three types of our sources show the same trend, that is, an increase from Maser-only to Maser-RRL sources and a decrease from Maser-RRL to RRL-only ones. This trend is consistent with the proposed evolutionary trend of HC$_{3}$N \citep[e.g.][]{2019MNRAS.484.4444U,2019MNRAS.489.4497Y} under the assumption that our Maser-only, Maser-RRL, and RRL-only sources correspond to YSOs, UC H{\sc ii} regions, and normal classical H{\sc ii} regions, respectively.
      \item Our statistical analysis of the integrated line intensity and the column density of HC$_{3}$N and shock-tracing molecules (SiO, H$_{2}$CO) enabled us to find positive correlations between them. Such findings suggest that HC$_{3}$N may be another tracer of shock. This supports the results of chemical models put forward by \cite{2019ApJ...881...57T}, which suggest that the sublimation of CH$_{4}$ and C$_{2}$H$_{2}$ from dust grains enhances key reactions of the formation of cyanopolyynes. We indirectly show that the neutral--neutral reaction between C$_{2}$H$_{2}$ and CN is the dominant formation pathway of HC$_{3}$N.
      \item We detect HC$_{5}$N in 11 HMSFRs with a formed protostar, including 2 Maser-only and 9 Maser-RRL sources. Among these 11 sources, 10 sources are new detections (except G24.790+0.084). We detect HC$_{7}$N in G24.790+0.084, which results in three HC$_{7}$N sources in HMSFRs with a formed protostar. Our detections enlarge the samples of HC$_{5}$N and HC$_{7}$N and support the idea that unsaturated iCOMs can exist in YSOs and UC H{\sc ii} regions.  
   \end{enumerate}

\begin{acknowledgements}
        This work is supported by the Natural Science Foundation of China (No. 12041302, 11590782). We thank the operators and staff at the TMRT stations for their assistance during our observations. We also thank Dr. J.Z. Wang, Dr. X.D. Tang and Dr. X. Chen for their nice comments and suggestions. Y. T. Y. is a member of the International Max Planck Research School (IMPRS) for Astronomy and Astrophysics at the Universities of Bonn and Cologne. Y. T. Y. would like to thank the China Scholarship Council (CSC) for support. J. J. Q. thanks for support from the NSFC (No. 12003080), the China Postdoctoral Science Foundation funded project (No. 2019M653144), the Guangdong Basic and Applied Basic Research Foundation (No. 2019A1515110588), and the Fundamental Research Funds for the Central Universities, Sun Yat-sen University (No. 2021qntd28). The authors would like to express their gratitude to EditSprings (\href{https://www.editsprings.cn}{https://www.editsprings.cn}) for the expert linguistic services provided. 
\end{acknowledgements}

%
%


\clearpage
\onecolumn

\begin{table}
    \begin{center}
    \caption{Summary of target lines.}
\setlength{\tabcolsep}{5mm}{
        \begin{tabular}{cccccc} 
       \hline\hline
       Species  &  Transition & Frequency  &  $S\mu ^{2}$  & $E_{\rm u}/k$ & HPBW \\
         & & (MHz) & (Debye$^{2}$) & (K) & (") \\
        (1) & (2) & (3) & (4) & (5) & (6) \\
        \hline
       HC$_{3}$N &  J = 2-1, F = 2-2 & 18194.92 & 2.32 & 1.3 & 52 \\
        & J = 2-1, F = 1-0 & 18195.14 & 3.09 & 1.3 & 52 \\
        & J = 2-1, F = 2-1 & 18196.22 & 6.96 & 1.3 & 52 \\
        & J = 2-1, F = 3-2 & 18196.31 & 13.00 & 1.3 & 52\\
        & J = 2-1, F = 1-2 & 18197.08 & 0.18 & 1.3 & 52\\
        & J = 2-1, F = 1-1 & 18198.38 & 2.32 & 1.3 & 52\\
        HC$_{5}$N & J = 6-5 & 15975.93 & 337.45 & 2.68 & 60 \\
        HC$_{7}$N & J = 15-14 & 16919.98 & 1045.44 & 6.50 & 56\\
        \hline 
        \end{tabular}}
        \tablefoot{(1): Molecular species; (2): Transition; (3): Rest frequency; (4-5): The product of the line strength and the square of the electric dipole moment and the upper level energy in K, which are taken from CDMS; (6): Half power beam width.}
    \label{tab1}
    \end{center}
\end{table}

\begin{landscape}
\begin{table}
        \centering
        \caption{Far infrared properties of our HC$_{3}$N sources.}
    \setlength{\tabcolsep}{0.5mm}
    \begin{threeparttable}
        \begin{tabular}{ccccccccccccccc} 
        \hline\hline
{Classifications} & Source Name &       F$_{70}$        &       F$_{160}$         &       F$_{250}$       &       F$_{350}$       &       F$_{500}$         &       F$_{870}$       &       $R_{\rm eff}$ & $\gamma$ &      $T_{\rm dust}$  &       M & $T_{\rm dust}^{a}$ & M$^{a}$ &      N(H$_{2}$)      \\
&       &       (Jy)    &       (Jy)    &       (Jy)    &       (Jy)    &       (Jy)    &       (Jy)    &       (")     &  &      (K)     &       (M$\odot$) & (K)        &       (M$\odot$)      &       (10$^{23}$ cm$^{-2}$)      \\
(1)     &       (2)     &       (3)     &       (4)     &       (5)     &       (6)     &       (7)     &       (8)     &       (9)     &       (10)    &       (11) &  (12) & (13) & (14) & (15)    \\
\hline
Maser-only & G20.234+0.085      &       47$\pm$1        &       176$\pm$4       &       157$\pm$5       &       66$\pm$1        &       35$\pm$1        &       5$\pm$1 &       27      & 65 &    17.26$\pm$0.76  &        593$\pm$115 & 21.4     &   687  &  1.84$\pm$0.05       \\
 & G23.484+0.097        &       144$\pm$4       &       257$\pm$6       &       282$\pm$10      &       110$\pm$4       &       58$\pm$4        &       12$\pm$2        &       41      & 63 &    16.09$\pm$1.04  &        1571$\pm$462   & 21.4 & 2148  &         0.71$\pm$0.02   \\
 & G25.649+1.050        &       1140$\pm$48     &       1053$\pm$21     &       752$\pm$4       &       413$\pm$10      &       177$\pm$6       &       39$\pm$6        &       55      & 75 &    18.38$\pm$0.16  &        4122$\pm$155   & 23.4 & 743  &  1.06$\pm$0.03  \\
 & G28.817+0.365        &       156$\pm$8       &       256$\pm$5       &       298$\pm$10      &       91$\pm$1        &       43$\pm$1        &       8$\pm$1 &       29      & 71 &    16.35$\pm$1.63  &        1381$\pm$624   & 23.0 & 807&    1.47$\pm$0.03  \\
 & G30.770-0.804        &       73$\pm$3        &       87$\pm$2        &       76$\pm$2        &       33$\pm$2        &       22$\pm$1        &       4$\pm$1 &       24      & 79 &    17.20$\pm$0.75  &        259$\pm$50     & 19.8 & 399  & 2.77$\pm$0.09   \\
 & G37.043-0.035        &       25$\pm$1        &       57$\pm$1        &       64$\pm$3        &       40$\pm$2        &       13$\pm$1        &       3$\pm$1 &       5       & 79 &    15.03$\pm$0.41  &  548$\pm$69   & 18.0 & 695 &   67.87$\pm$1.58 \\
Maser-RRL & G23.271-0.256       &       70$\pm$2        &       104$\pm$3       &       235$\pm$12      &       132$\pm$5       &       81$\pm$4        &       25$\pm$4        &       62      & 56 &    12.40$\pm$0.80  &        4922$\pm$1520  & 19.3 & 3020  &  0.40$\pm$0.01 \\
 & G23.389+0.185        &       270$\pm$13      &       173$\pm$3       &       181$\pm$4       &       83$\pm$1        &       43$\pm$1        &       9$\pm$2 &       29      & 64 &    15.94$\pm$0.66  &        1055$\pm$201   & 20.2 & 2612  &         1.28$\pm$0.03   \\
 & G23.436-0.184        &       147$\pm$4       &       446$\pm$20      &       630$\pm$28      &       253$\pm$9       &       128$\pm$4       &       40$\pm$6        &       51      & 57 &    14.77$\pm$1.16  &        5997$\pm$2200  & 24.6 & 5702  & 0.36$\pm$0.01  \\
 & G23.965-0.110        &       207$\pm$9       &       336$\pm$7       &       377$\pm$9       &       152$\pm$4       &       83$\pm$2        &       20$\pm$3        &       49      & 62 &    15.86$\pm$0.97  &  2330$\pm$648 &  20.1 & 4169  & 0.43$\pm$0.01 \\
 & G24.328+0.144        &       523$\pm$20      &       810$\pm$16      &       623$\pm$1       &       253$\pm$6       &       116$\pm$2       &       17$\pm$3        &       43      & 57 &    18.68$\pm$0.75  &        4076$\pm$695   & 23.8 & 4416  &         0.29$\pm$0.01   \\
 & G24.790+0.084        &       2000$\pm$300    &       3000$\pm$600    &       1000$\pm$200    &       600$\pm$100     &       200$\pm$100     &       57$\pm$9        &       42      & 55 &    32.10$\pm$3.89  &       1639$\pm$571    & 26.5 & 7638  & 0.53$\pm$0.01  \\
 & G25.709+0.044        &       352$\pm$17      &       431$\pm$9       &       473$\pm$18      &       140$\pm$8       &       64$\pm$1        &       9$\pm$2 &       28      & 59 &    16.77$\pm$1.69  &        2844$\pm$128   & 29.0 &        2041  & 1.02$\pm$0.02   \\
 & G28.147-0.004        &       129$\pm$6       &       148$\pm$4       &       184$\pm$4       &       66$\pm$1        &       38$\pm$1        &       9$\pm$2 &       33      & 60 &    15.55$\pm$1.27  &        1773$\pm$665   & 22.2 & 1596  & 0.41$\pm$0.02  \\
 &  G28.393+0.085       &       73$\pm$9        &       300$\pm$16      &       416$\pm$14      &       214$\pm$9       &       140$\pm$4       &       12$\pm$2        &       41      & 72  &   14.11$\pm$0.66  &        3590$\pm$798   & 18.5 & 2189  & 0.31$\pm$0.02  \\
 & G29.835-0.012        &       337$\pm$16      &       251$\pm$7       &       213$\pm$13      &       101$\pm$5       &       60$\pm$5        &       13$\pm$2        &       37      & 62 &    17.17$\pm$0.53  &        2491$\pm$337   & 29.6 & 1119  &         0.89$\pm$0.02   \\
 & G30.897+0.163        &       113$\pm$5       &       179$\pm$5       &       252$\pm$6       &       86$\pm$3        &       47$\pm$1        &       18$\pm$3        &       43      & 64 &    15.09$\pm$1.46  &        3474$\pm$1560  & 21.0 & 578 &   0.70$\pm$0.01  \\
 & G31.579+0.076        &       648$\pm$16      &       769$\pm$20      &       647$\pm$18      &       217$\pm$4       &       93$\pm$2        &       13$\pm$2        &       26      & 68 &    18.44$\pm$1.32  &        2993$\pm$915   & 27.0 & 1216  & 1.43$\pm$0.03  \\
 & G35.141-0.750        &       317$\pm$10      &       642$\pm$17      &       543$\pm$13      &       183$\pm$12      &       139$\pm$4       &       39$\pm$6        &       40      & 100 &   18.00$\pm$1.26  &        677$\pm$204    & 19.4 & 1086  &         1.07$\pm$0.02   \\
RRL-only & G24.546-0.245        &       64$\pm$3        &       114$\pm$6       &       183$\pm$12      &       105$\pm$6       &       64$\pm$3        &       17$\pm$3        &       55      & 59 &    13.27$\pm$0.53  &        2908$\pm$546   & 21.1 & 2904 & 0.44$\pm$0.02   \\
 & G28.452+0.002        &       133$\pm$5       &       103$\pm$2       &       72$\pm$2        &       30$\pm$2        &       14$\pm$2        &       6$\pm$1 &       29      & 139 &   19.47$\pm$0.70  &       1400$\pm$204 & 21.6 & 7047 & 3.74$\pm$0.17      \\
        \hline 
        \end{tabular}
\tablefoot{(1): Classifications; (2): Source name; (3-7): Flux density at each  band from  the Herschel telescope; (8-9): Flux density and effective radius of sources at 870 $\mu$m from APEX telescope, which are given by \cite{2014A&A...568A..41U}; (10): The gas-to-dust ratio;} (11): Dust temperature; (12): Total (gas + dust) core mass; (13-14): Dust temperature and total mass, from \cite{2018MNRAS.473.1059U}; (15): Column density of H$_{2}$.
      \end{threeparttable}
      \label{tab5}
\end{table}
\end{landscape}

\onecolumn
\begin{appendix}
\section{Spectra of HC$_{3}$N, HC$_{5}$N, and HC$_{7}$N}
\begin{figure}[htp]
    \centering
    \subfigure{
    \centering
    \includegraphics[width=0.26\textwidth]{G20-234.eps}}
    \subfigure{
    \centering
    \includegraphics[width=0.26\textwidth]{G23-271.eps}}
    \centering
    \subfigure{
    \centering
    \includegraphics[width=0.26\textwidth]{G23-389.eps}}
    \subfigure{
    \centering
    \includegraphics[width=0.26\textwidth]{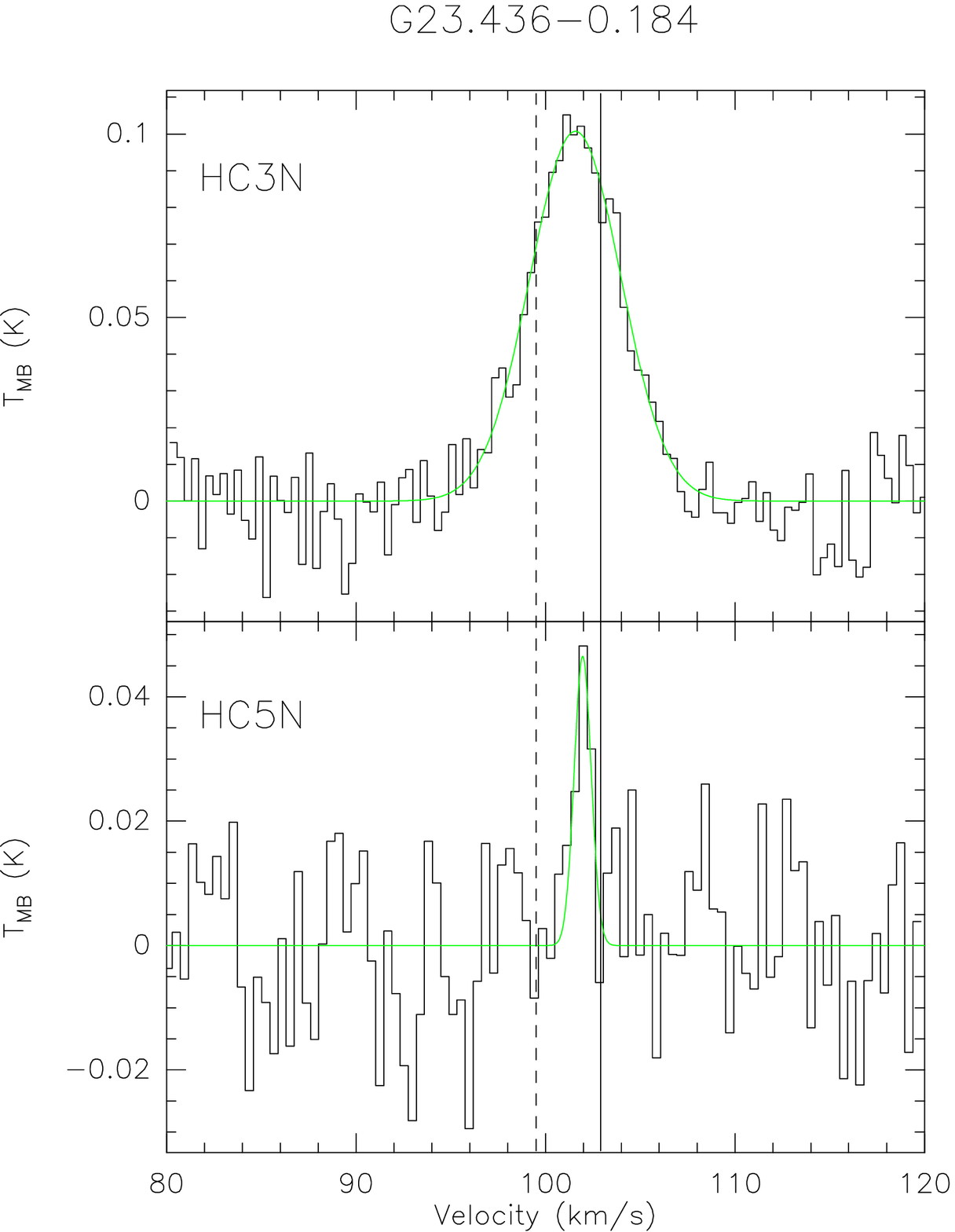}}
    \centering
    \subfigure{
    \centering
    \includegraphics[width=0.26\textwidth]{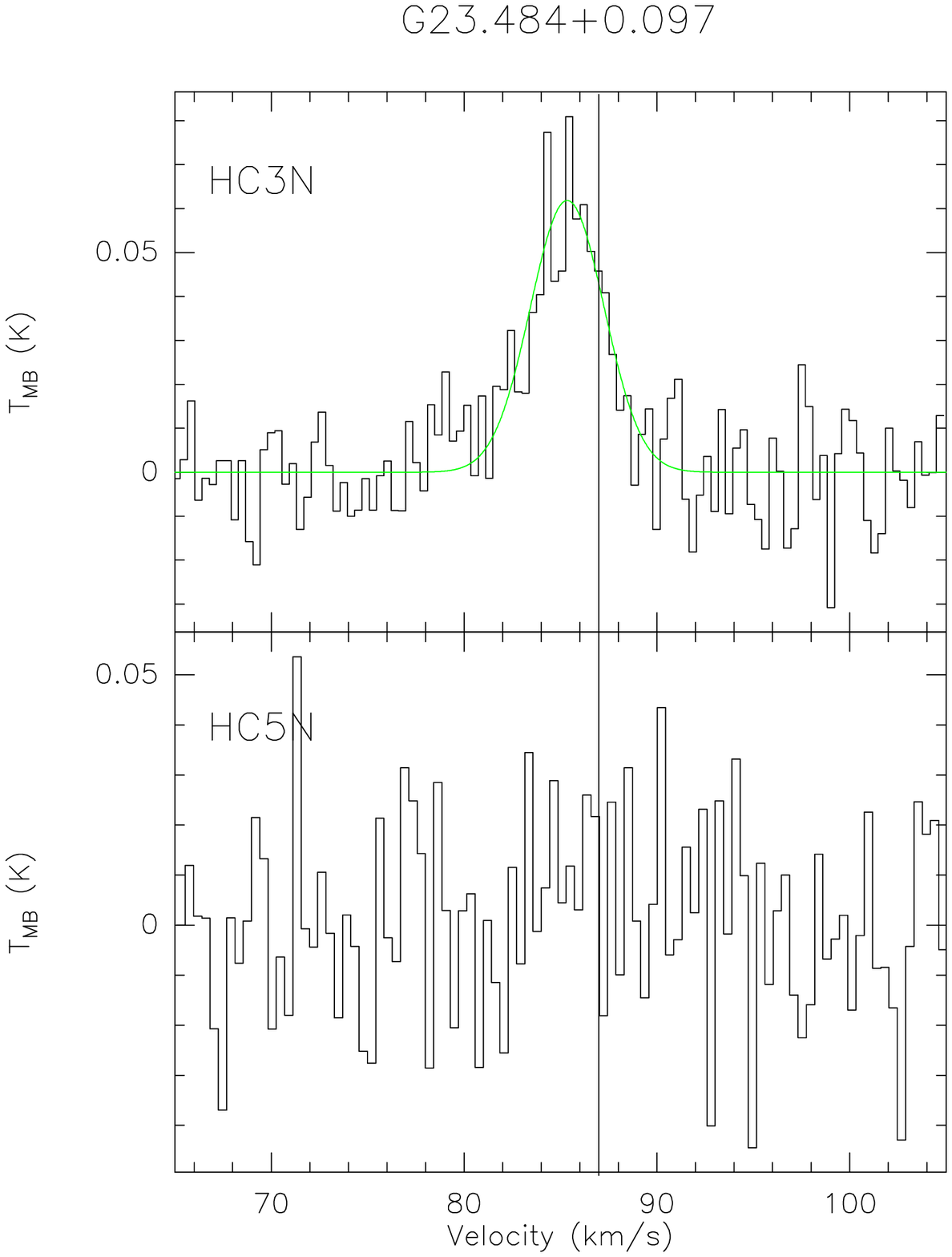}}
    \subfigure{
    \centering
    \includegraphics[width=0.26\textwidth]{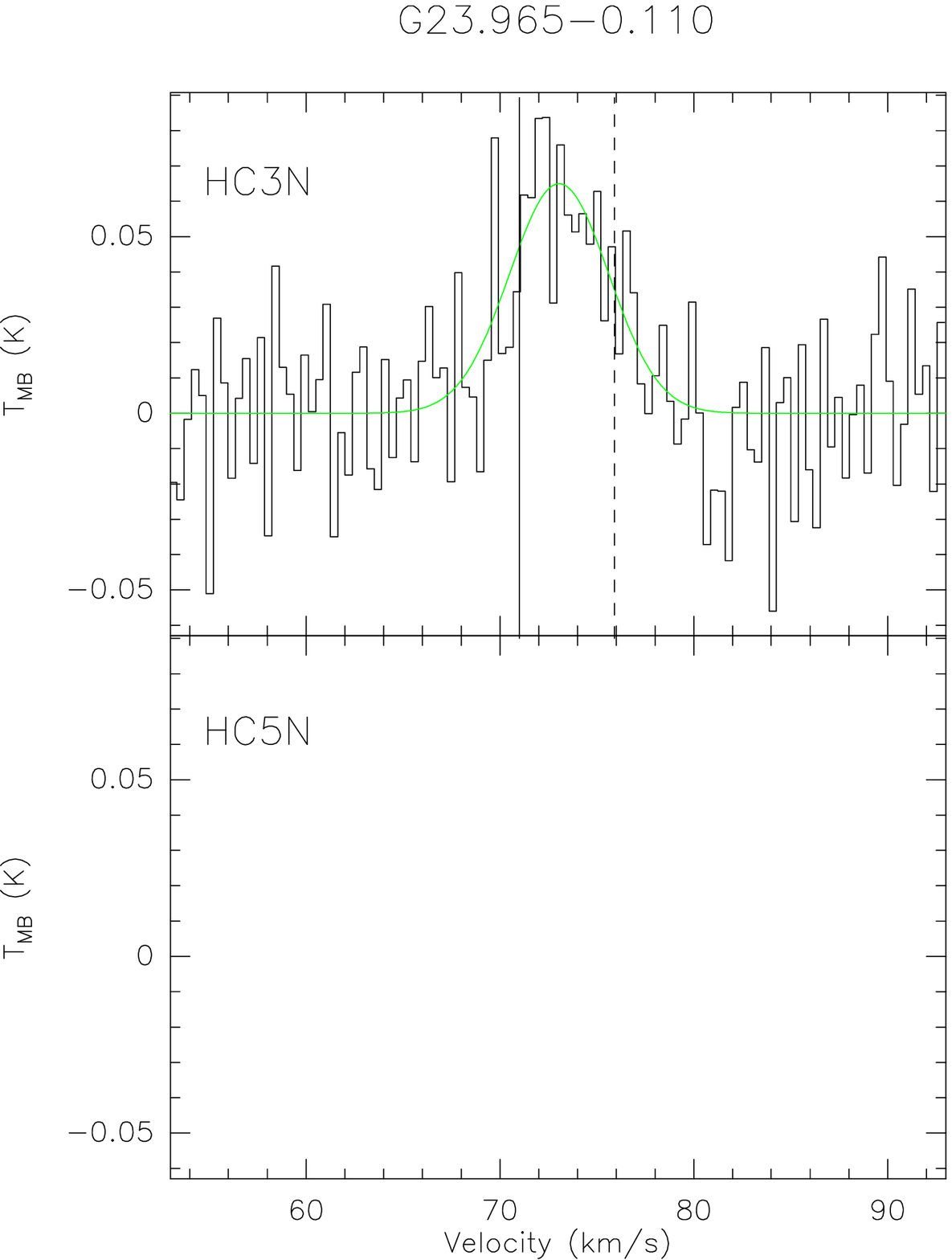}}
    \centering
    \subfigure{
    \centering
    \includegraphics[width=0.26\textwidth]{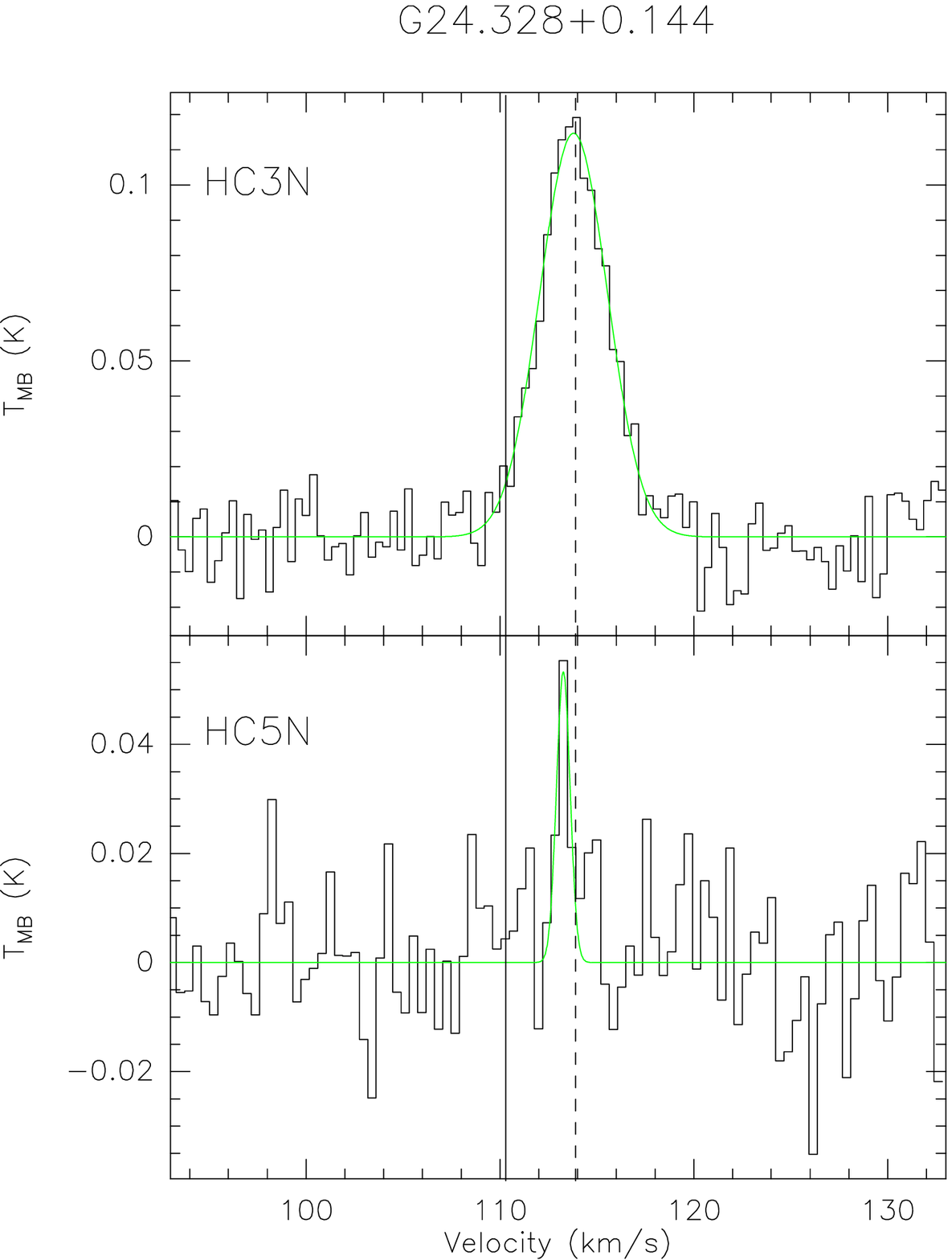}}
    \subfigure{
    \centering
    \includegraphics[width=0.26\textwidth]{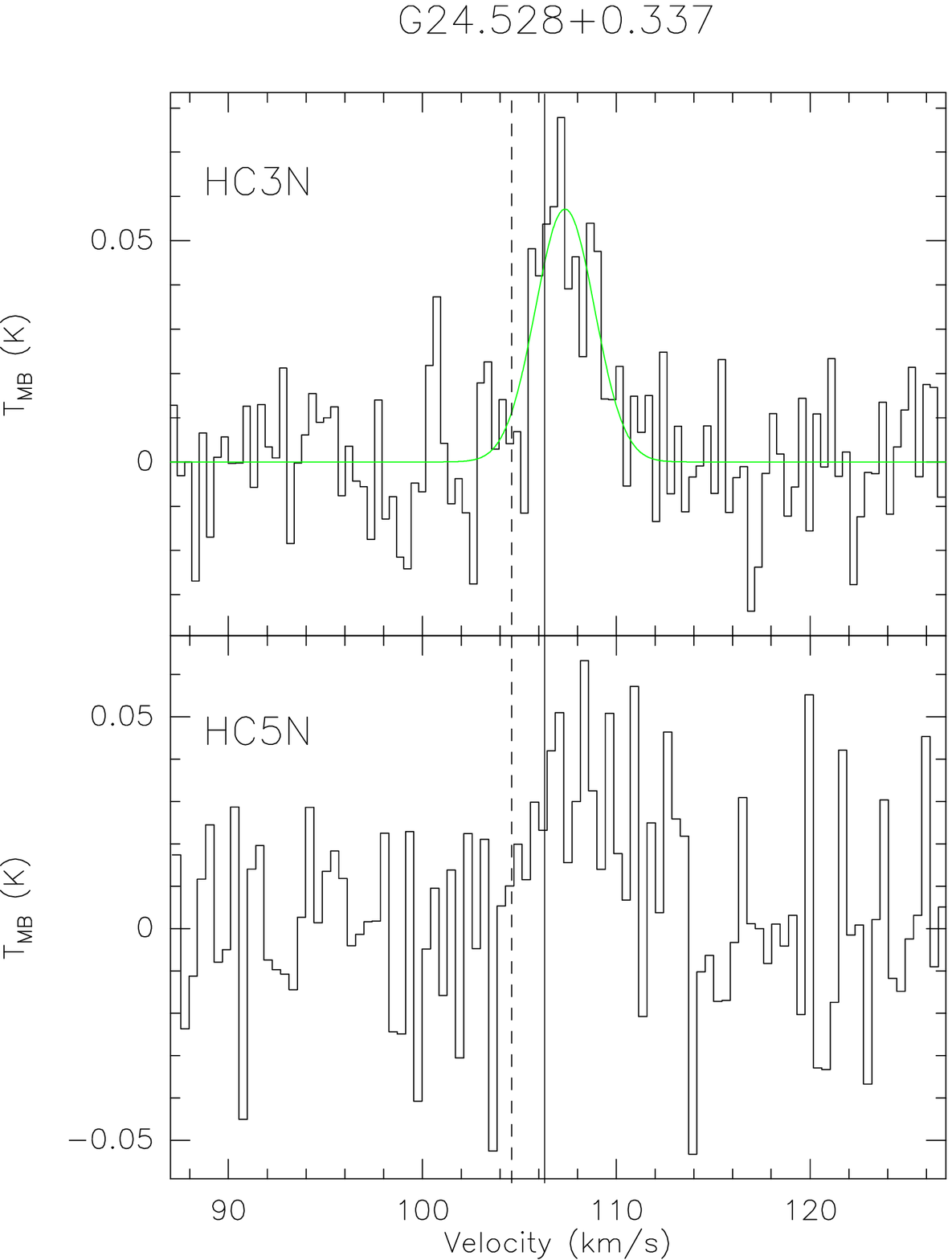}}
    \centering
    \subfigure{
    \centering
    \includegraphics[width=0.26\textwidth]{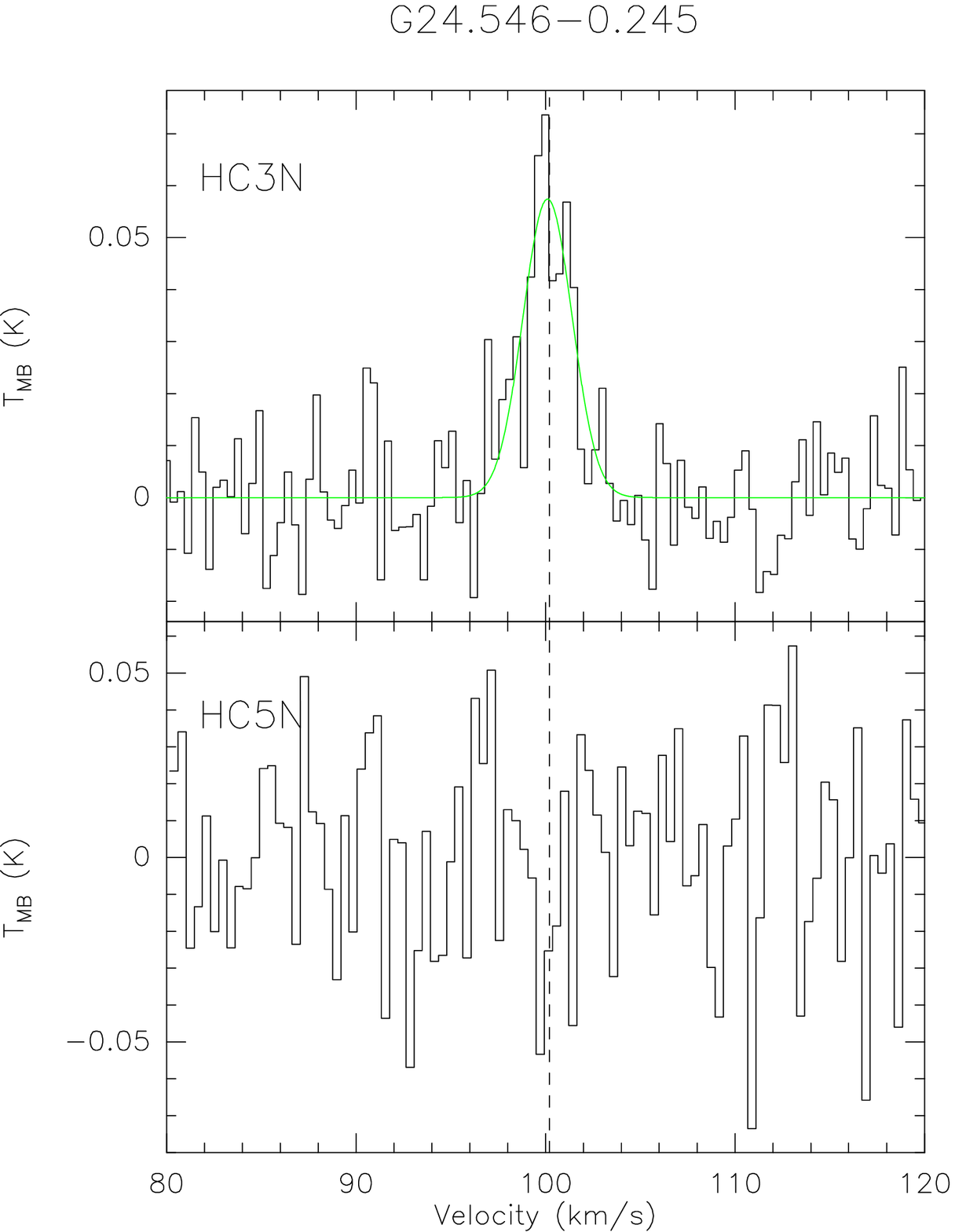}}
    \caption{All spectra for HC$_{3}$N, HC$_{5}$N, and HC$_{7}$N with TMRT telescope. Solid and dashed lines show the LSR velocity values of the 6.7 GHz CH$_{3}$OH maser \citep{2017ApJ...846..160Y,2019ApJS..241...18Y}, and of RRL \citep{2020ApJS..248....3C}, respectively. Blank panels represent  non-detections.}
\end{figure}

\addtocounter{figure}{-1}
\begin{figure}
    \subfigure{
    \centering
    \includegraphics[width=0.26\textwidth]{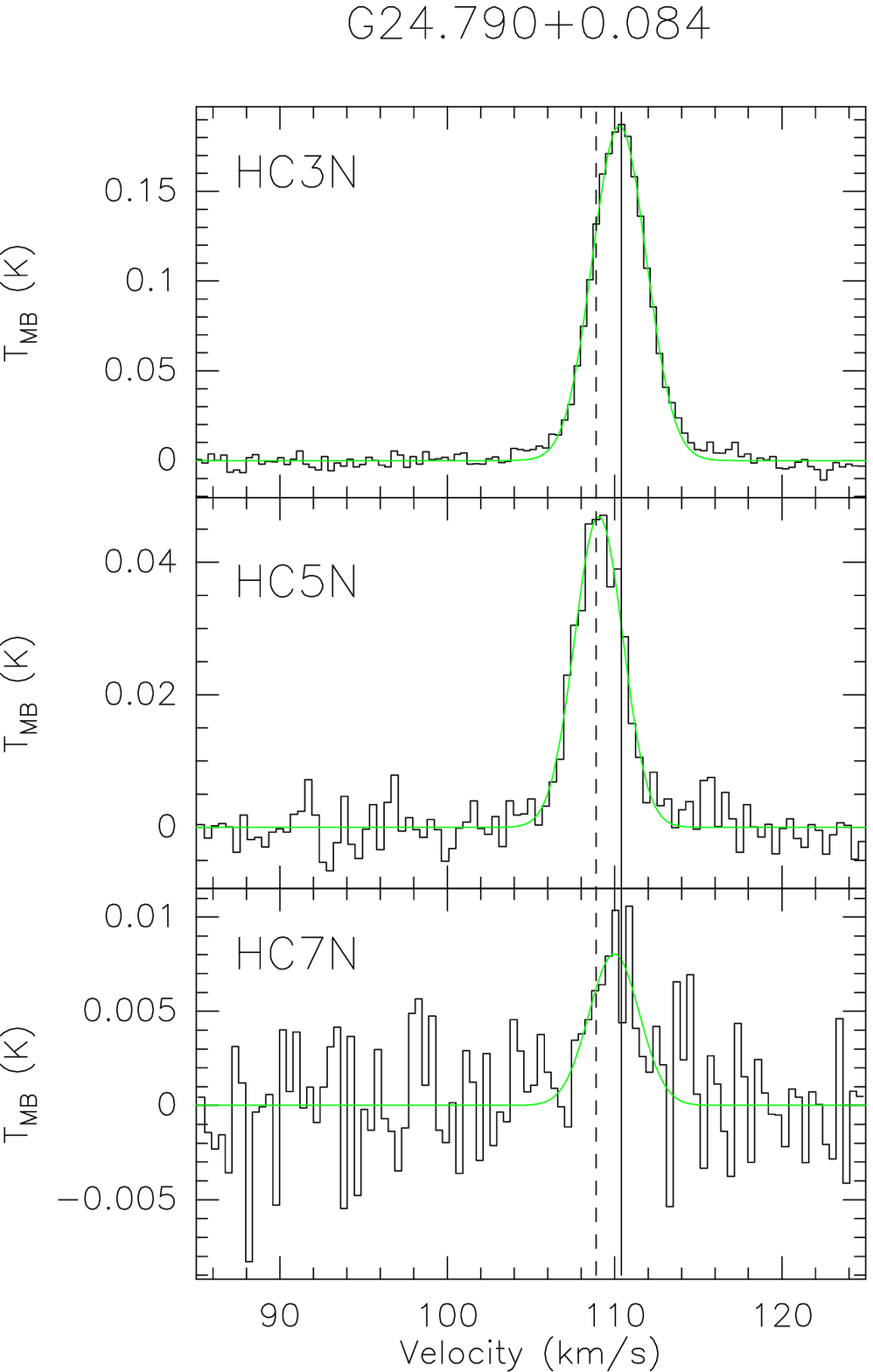}}
    \centering
    \subfigure{
    \centering
    \includegraphics[width=0.26\textwidth]{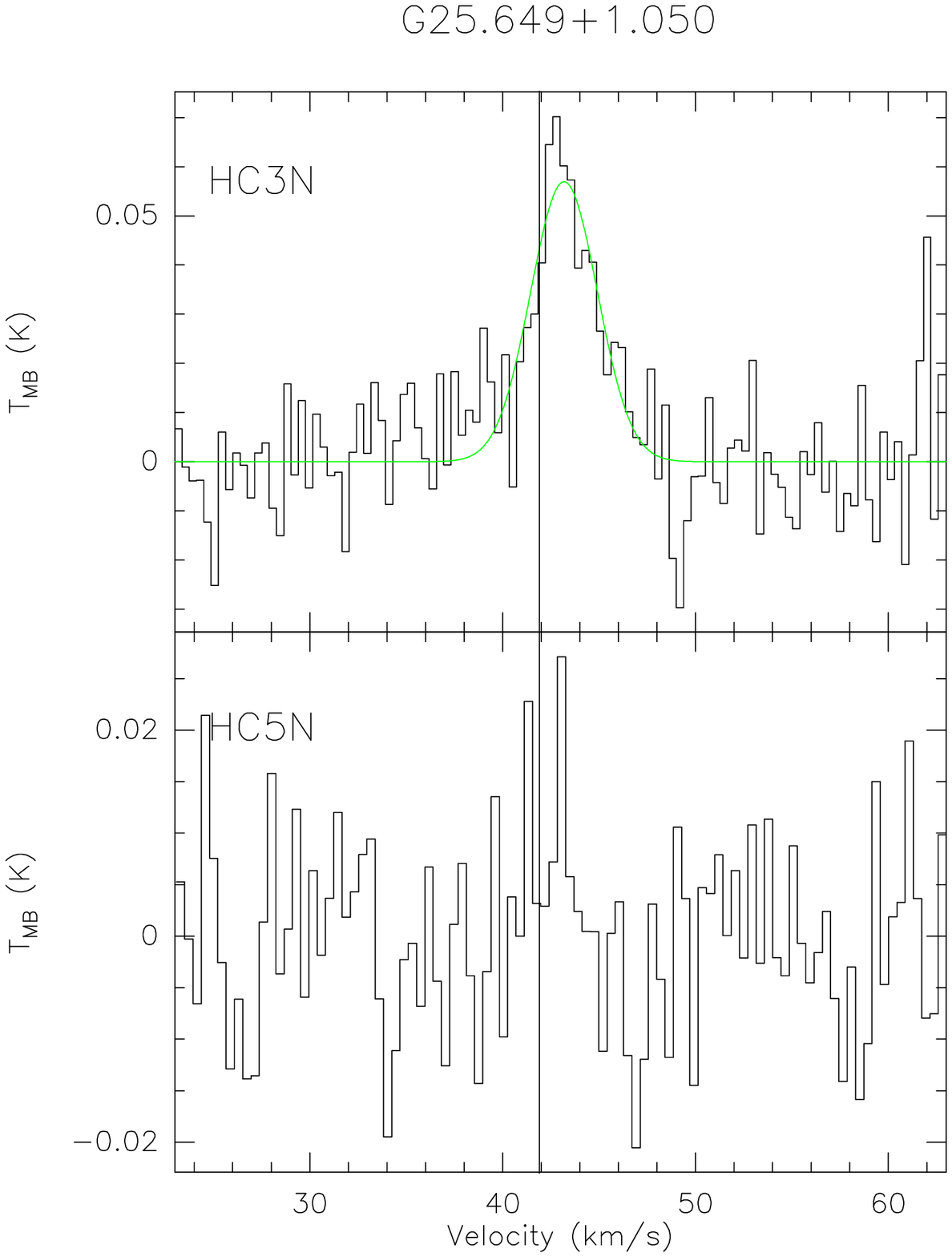}}
    \subfigure{
    \centering
    \includegraphics[width=0.26\textwidth]{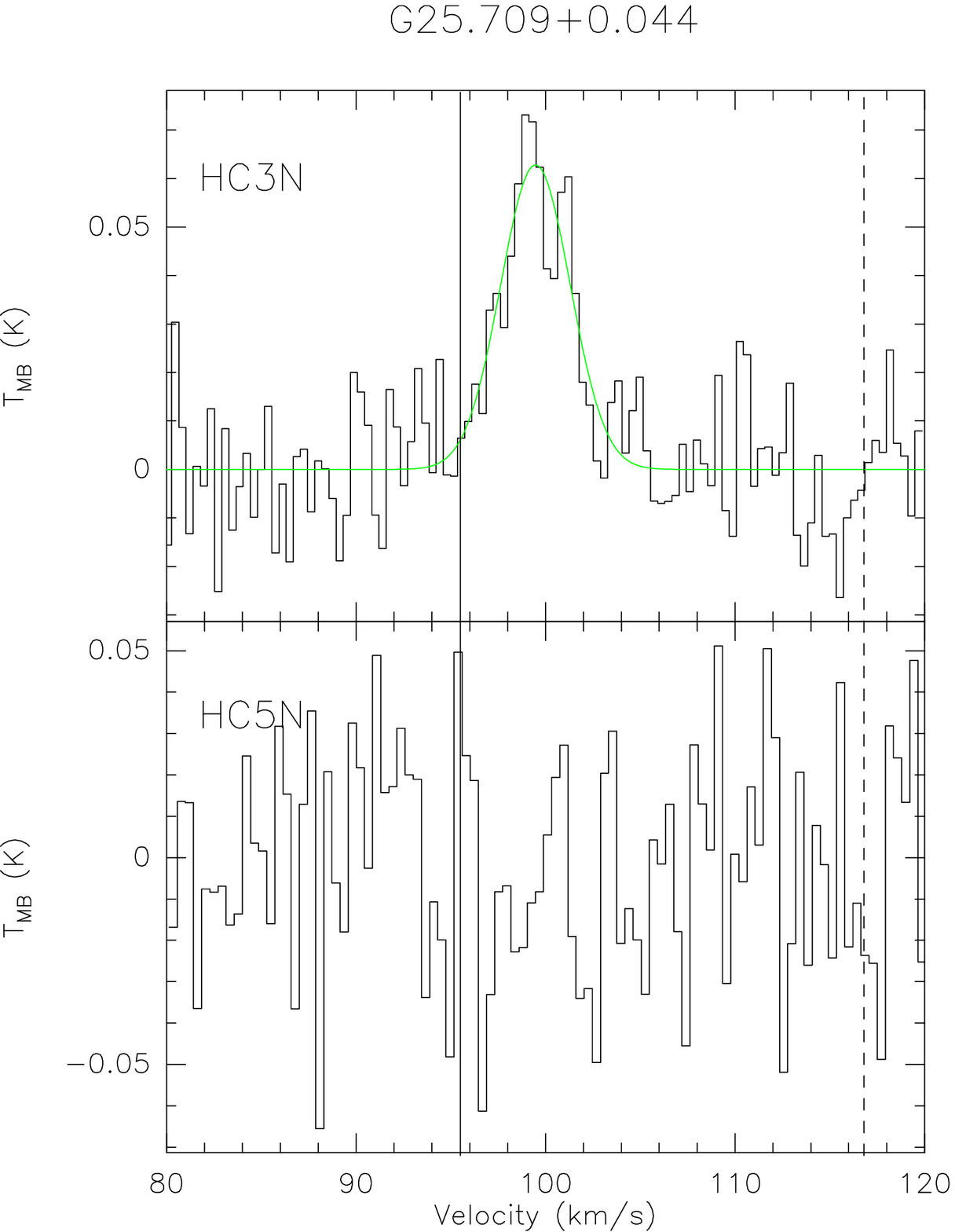}}
    \centering
    \subfigure{
    \centering
    \includegraphics[width=0.26\textwidth]{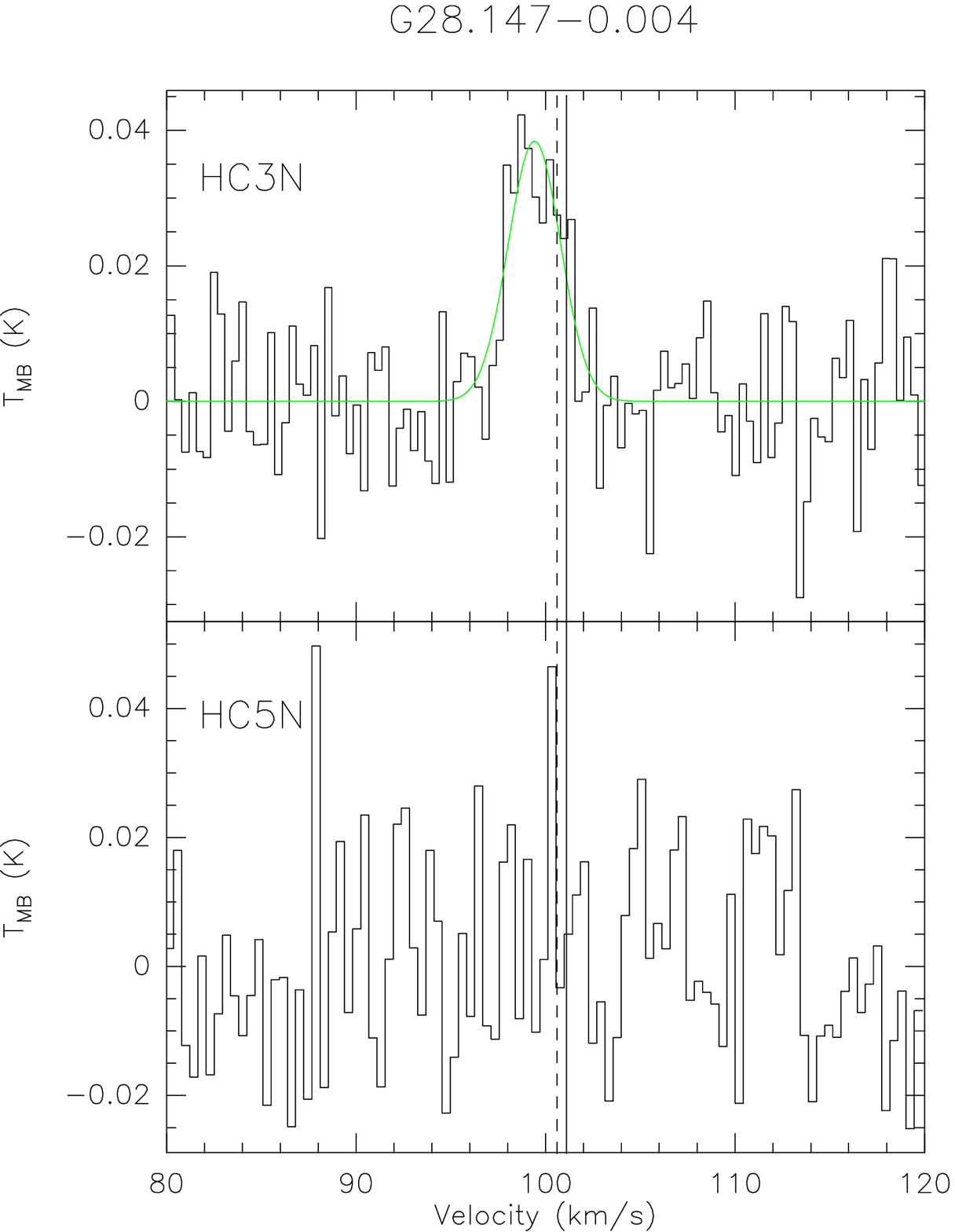}}
    \subfigure{
    \centering
    \includegraphics[width=0.26\textwidth]{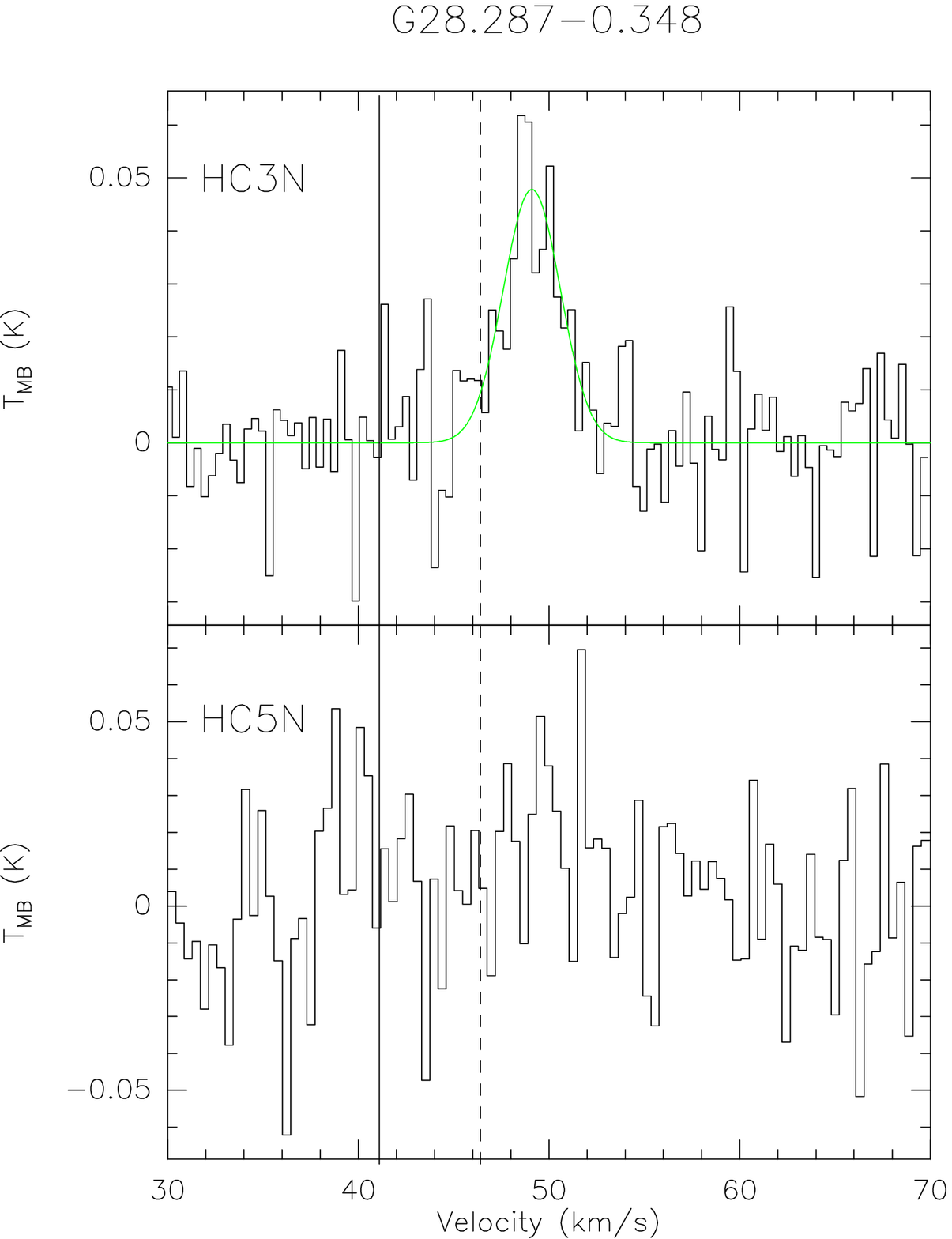}}
    \centering
    \subfigure{
    \centering
    \includegraphics[width=0.26\textwidth]{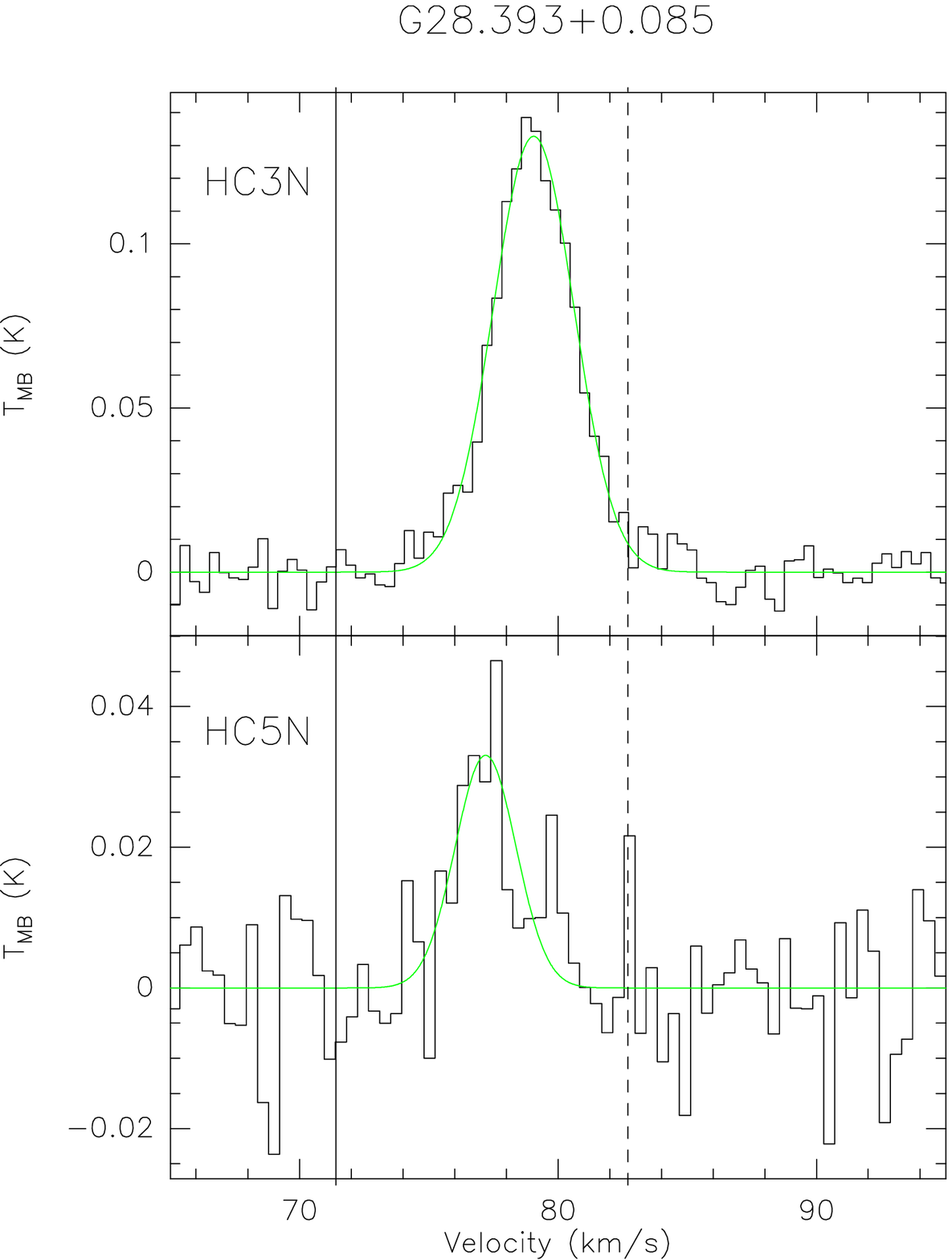}}
    \subfigure{
    \centering
    \includegraphics[width=0.26\textwidth]{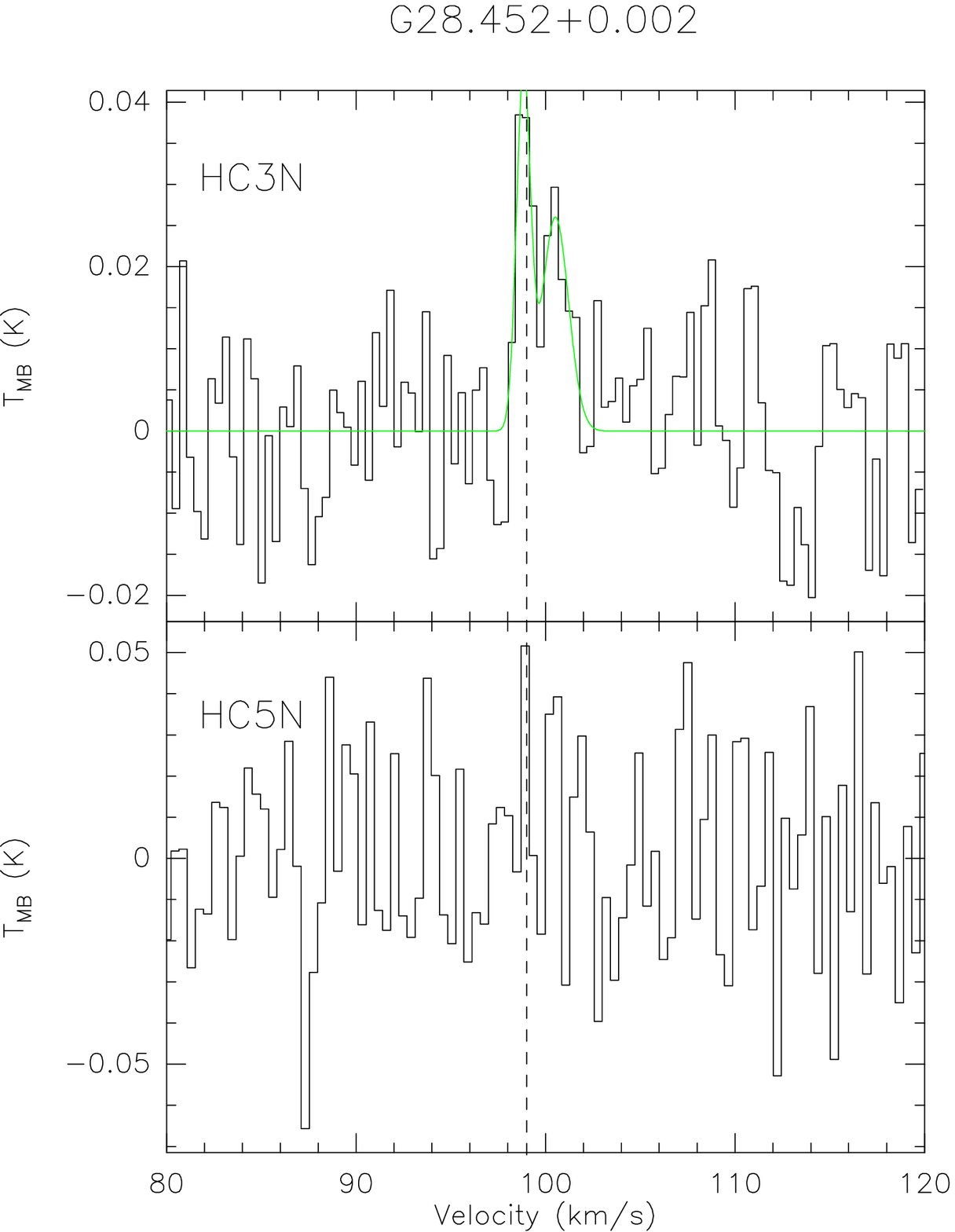}}
    \centering
    \subfigure{
    \centering
    \includegraphics[width=0.26\textwidth]{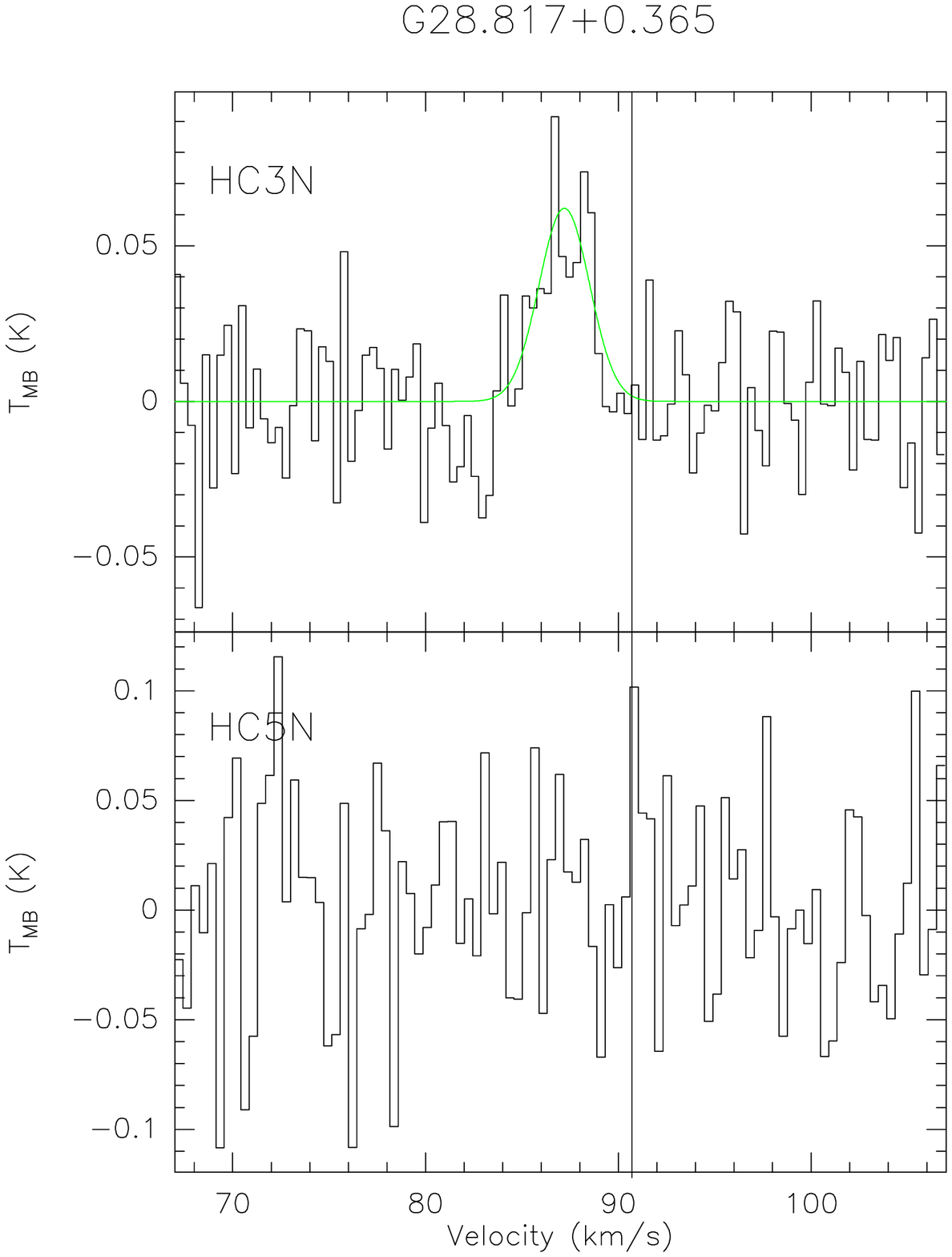}}
    \subfigure{
    \centering
    \includegraphics[width=0.26\textwidth]{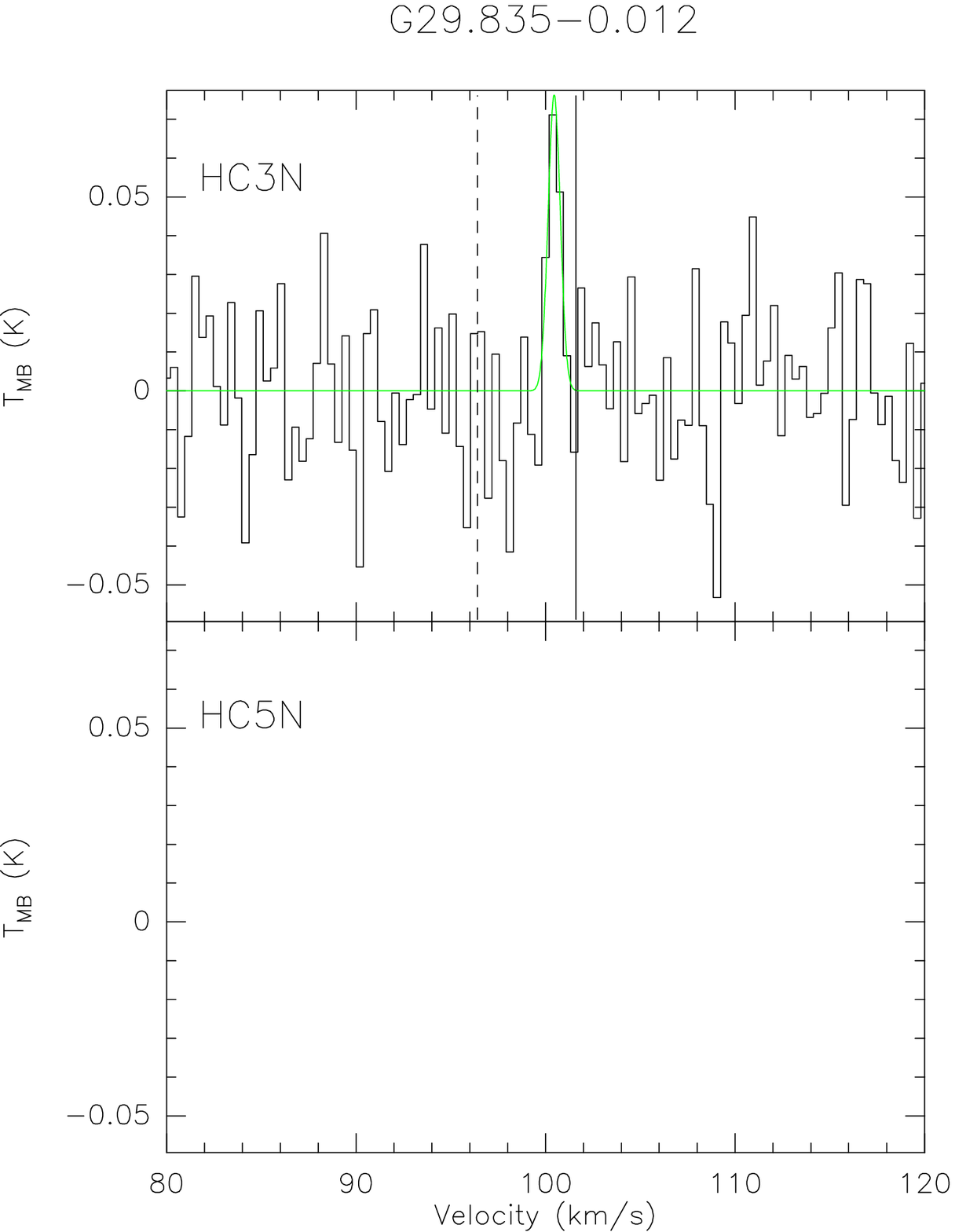}}
    \caption{Continued.}
\end{figure}

\addtocounter{figure}{-1}
\begin{figure}
    \centering
    \subfigure{
    \centering
    \includegraphics[width=0.26\textwidth]{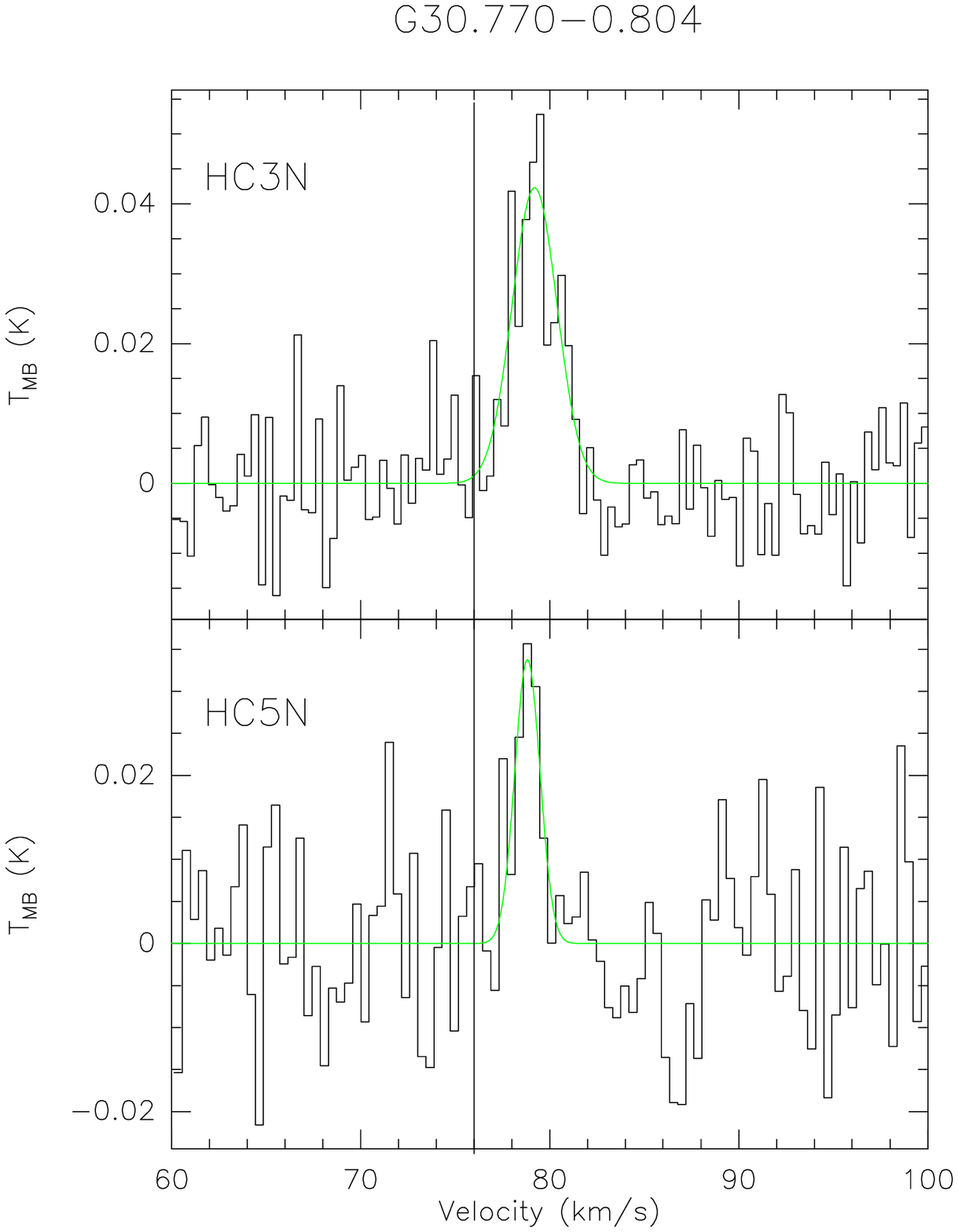}}
    \subfigure{
    \centering
    \includegraphics[width=0.26\textwidth]{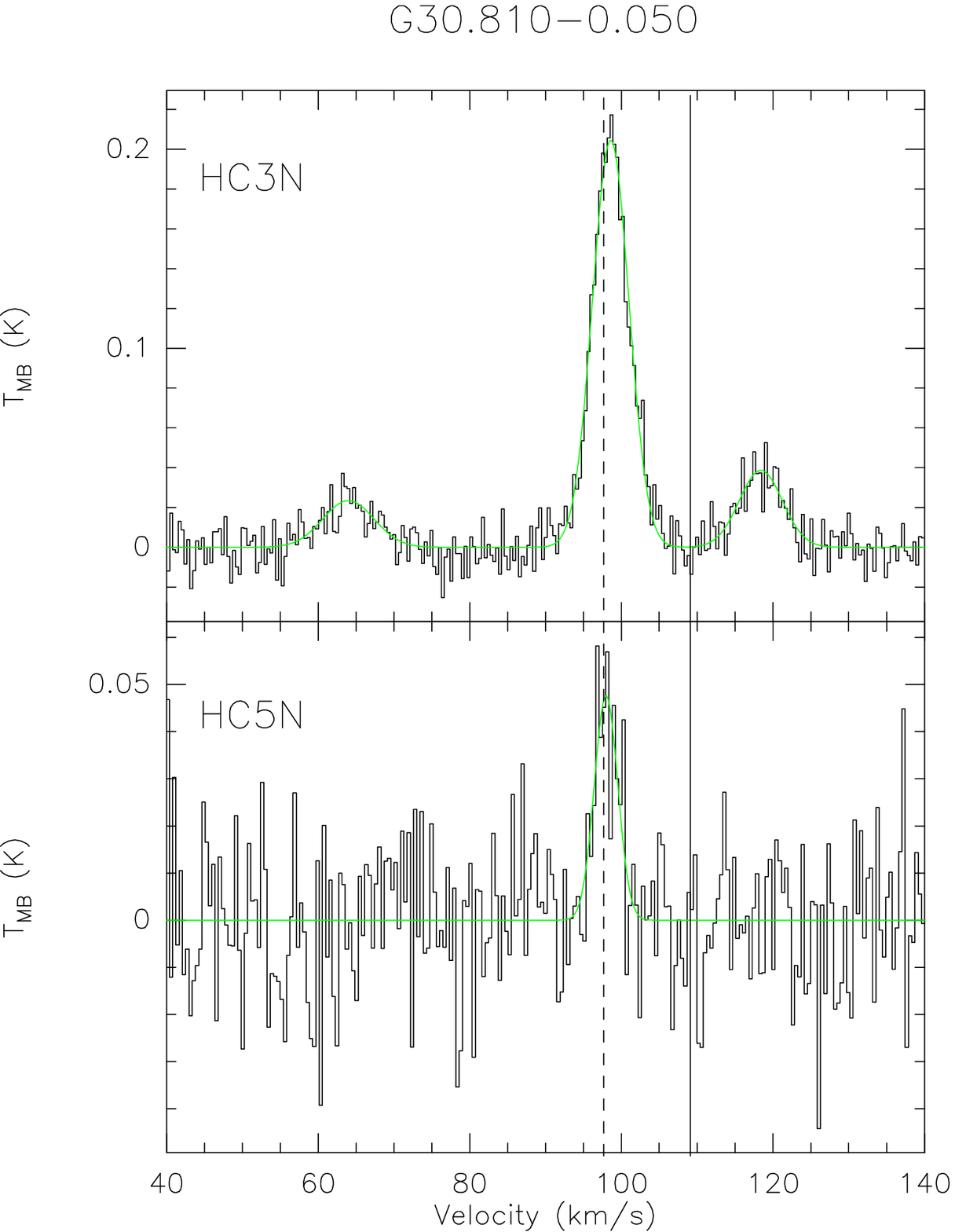}}
    \centering
    \subfigure{
    \centering
    \includegraphics[width=0.26\textwidth]{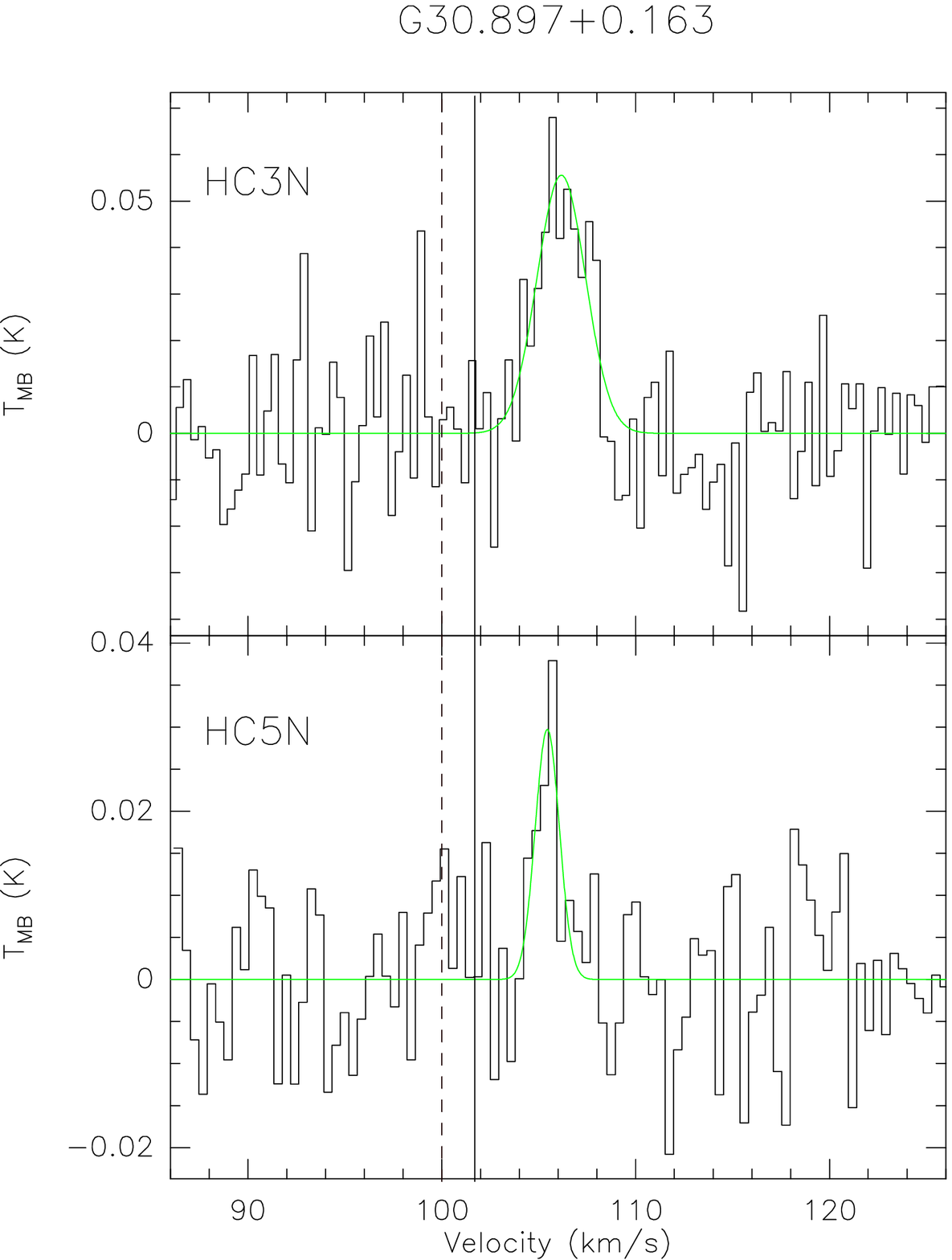}}
    \subfigure{
    \centering
    \includegraphics[width=0.26\textwidth]{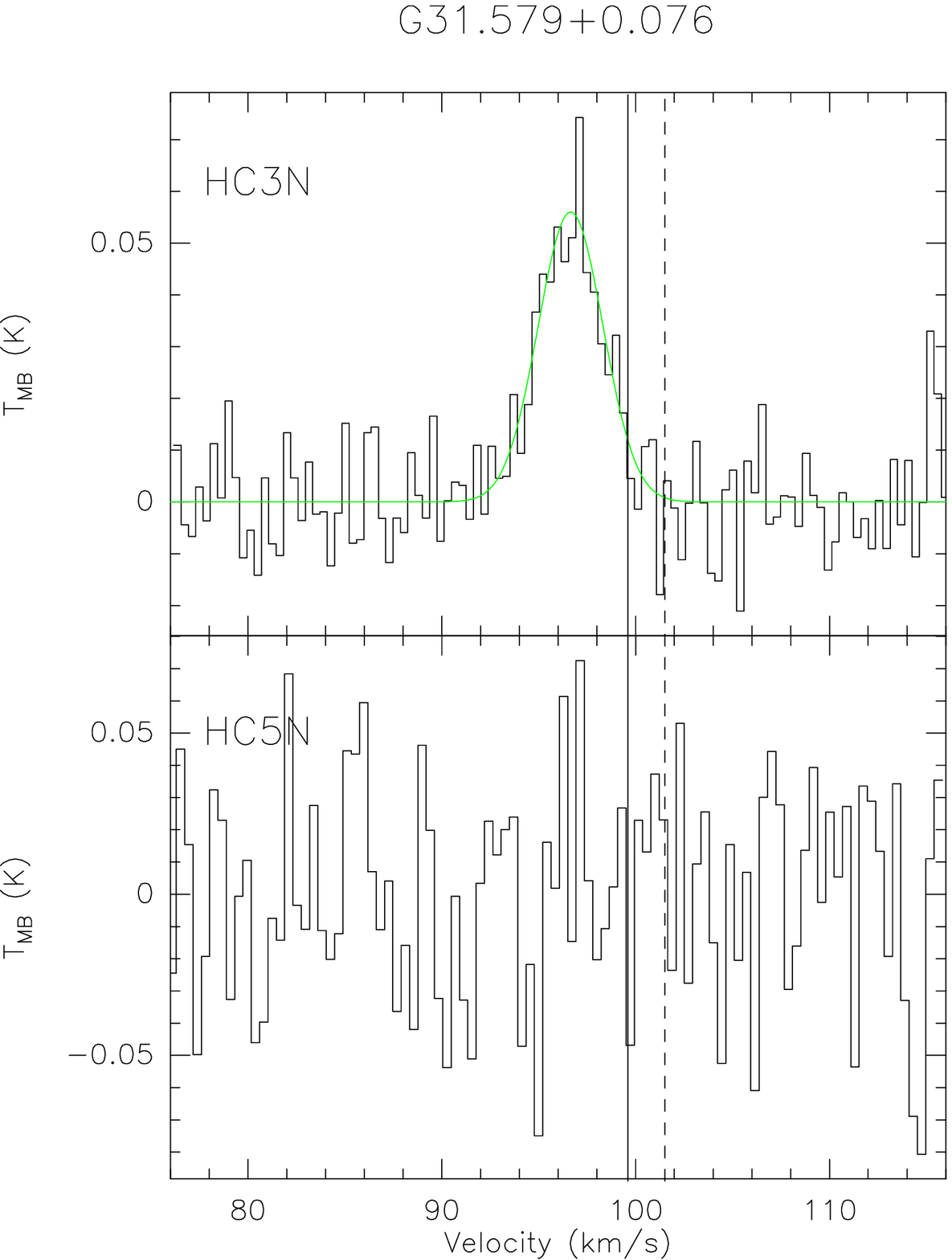}}
    \centering
    \subfigure{
    \centering
    \includegraphics[width=0.26\textwidth]{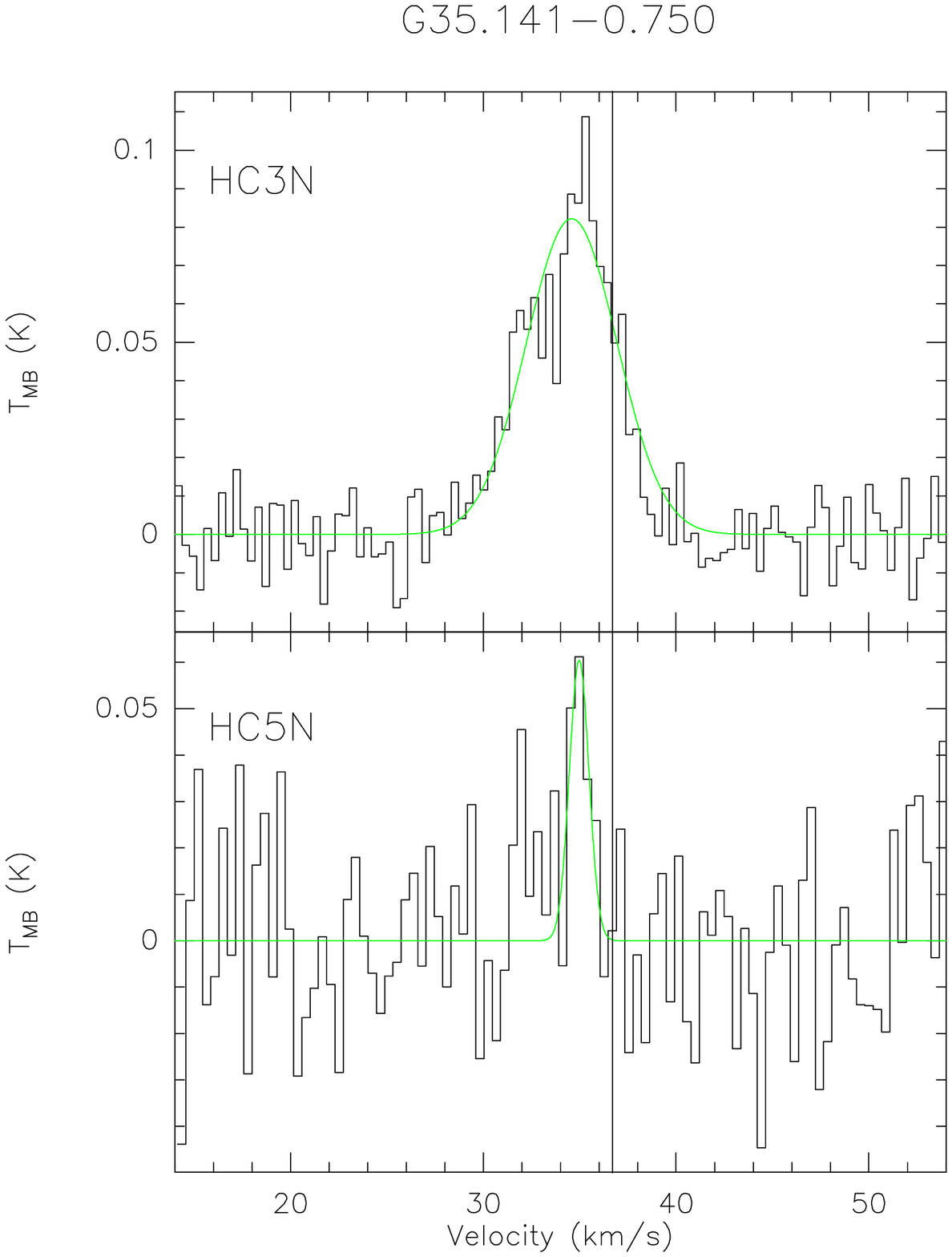}}
    \subfigure{
    \centering
    \includegraphics[width=0.26\textwidth]{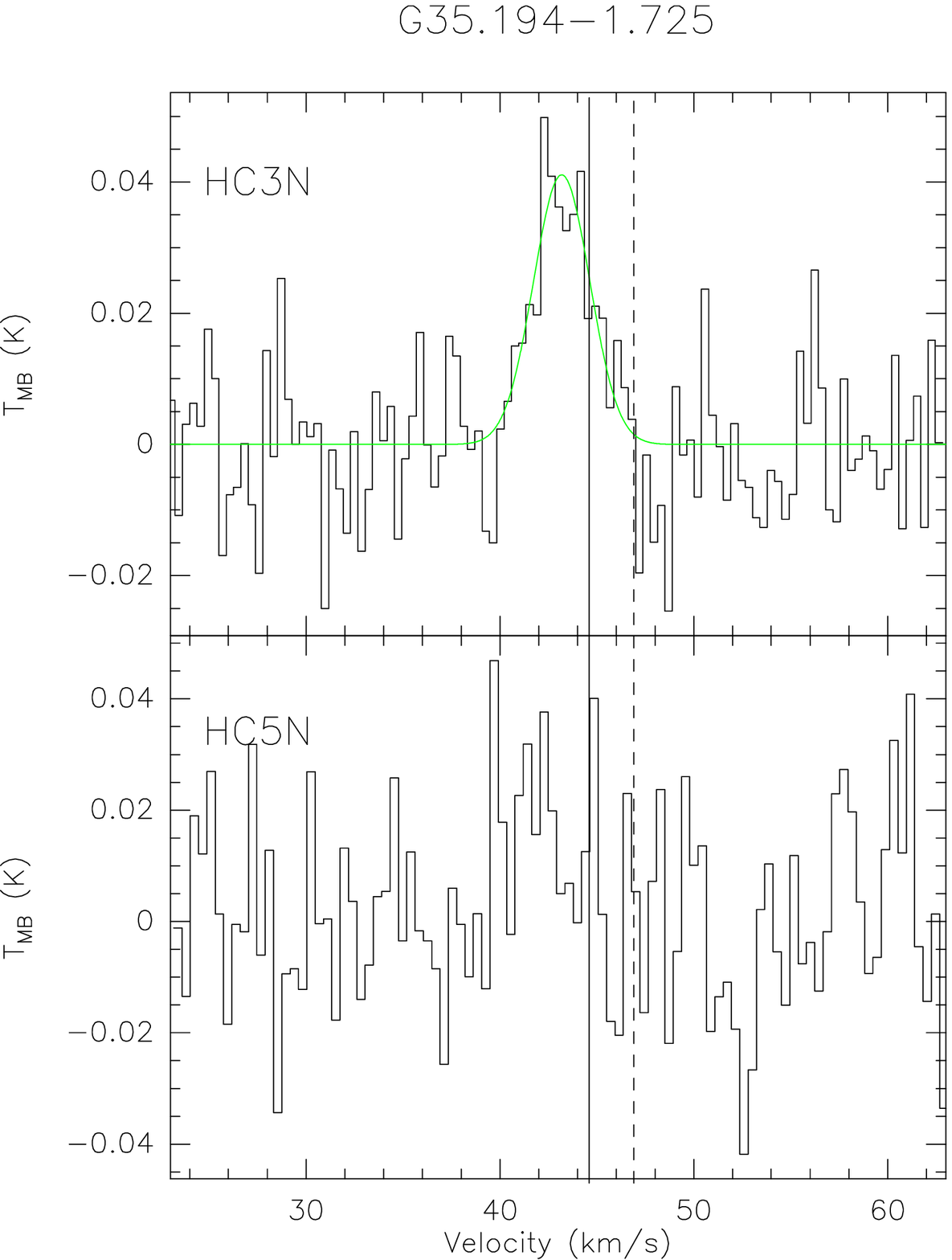}}
    \centering
    \subfigure{
    \centering
    \includegraphics[width=0.26\textwidth]{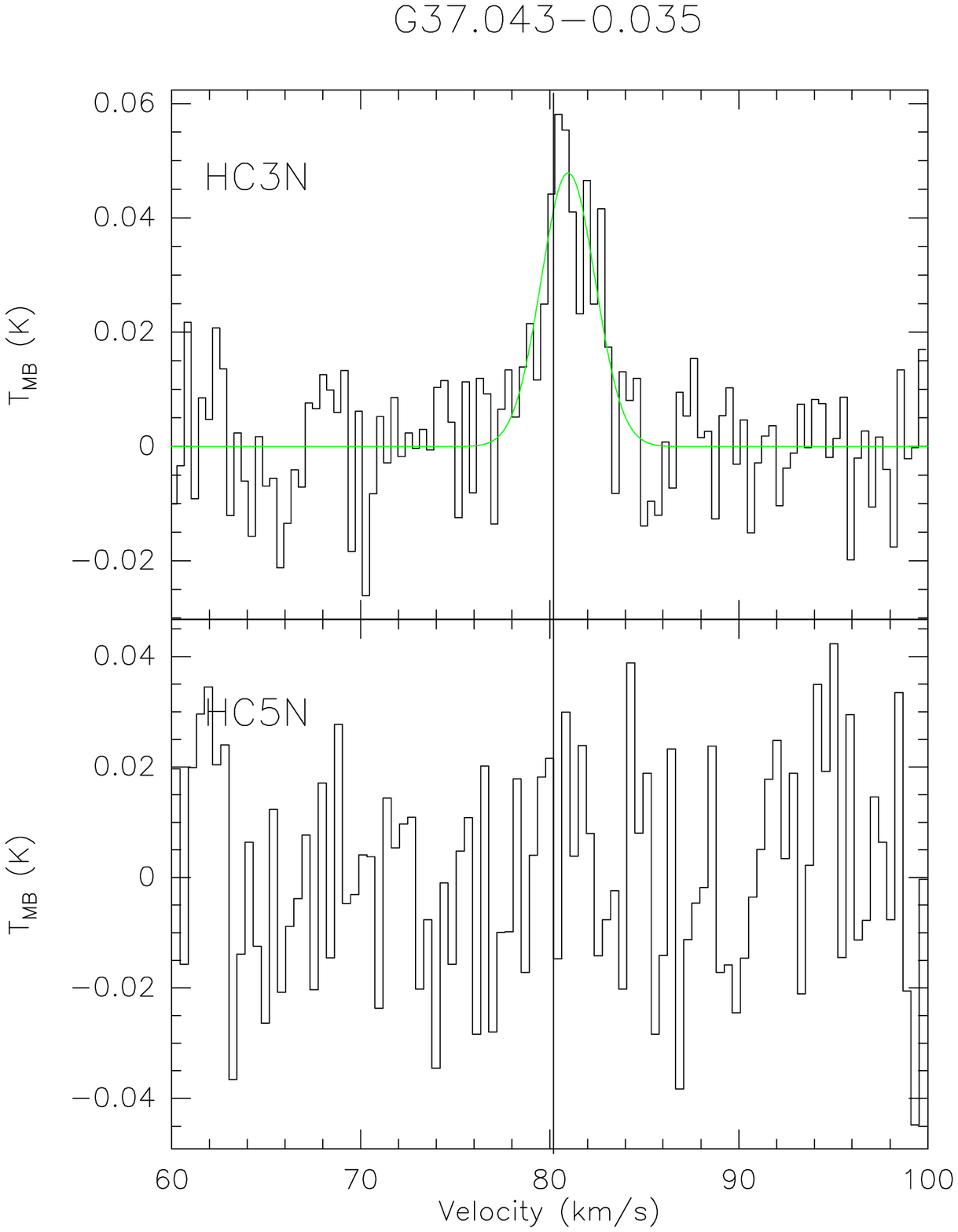}}
    \subfigure{
    \centering
    \includegraphics[width=0.26\textwidth]{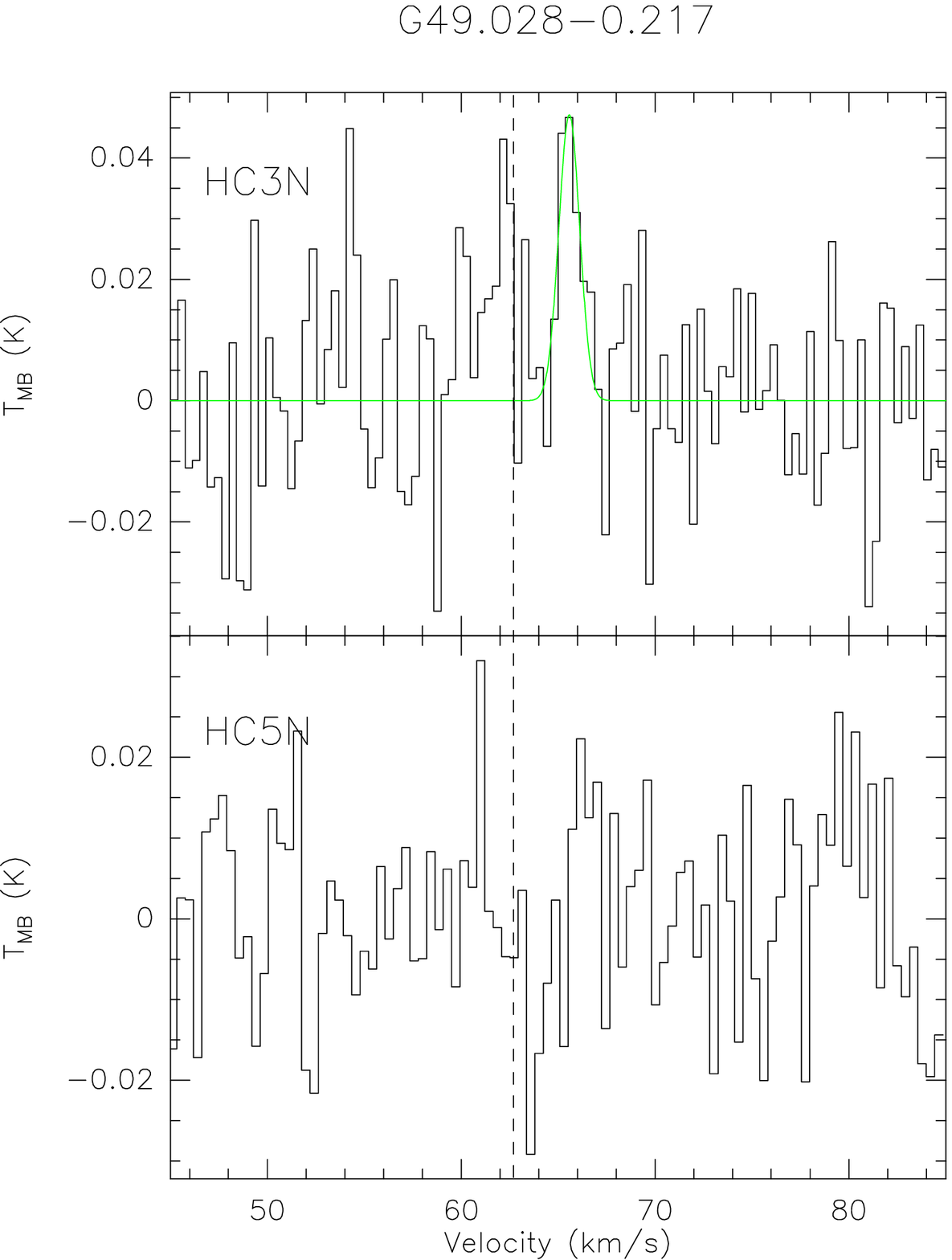}}
    \centering
    \subfigure{
    \centering
    \includegraphics[width=0.26\textwidth]{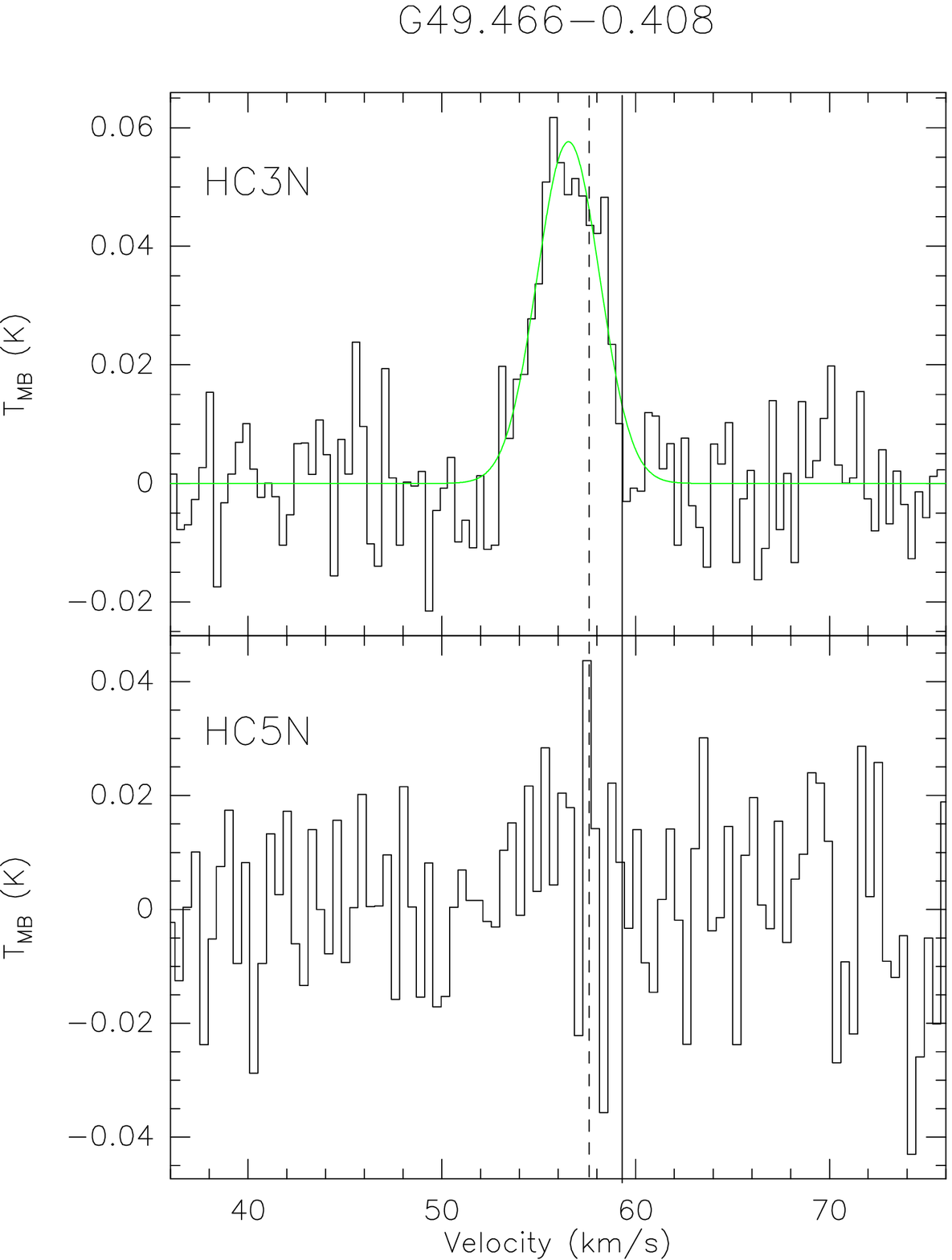}}
    \caption{Continued.}
\end{figure}

\addtocounter{figure}{-1}
\begin{figure}
    \subfigure{
    \centering
    \includegraphics[width=0.26\textwidth]{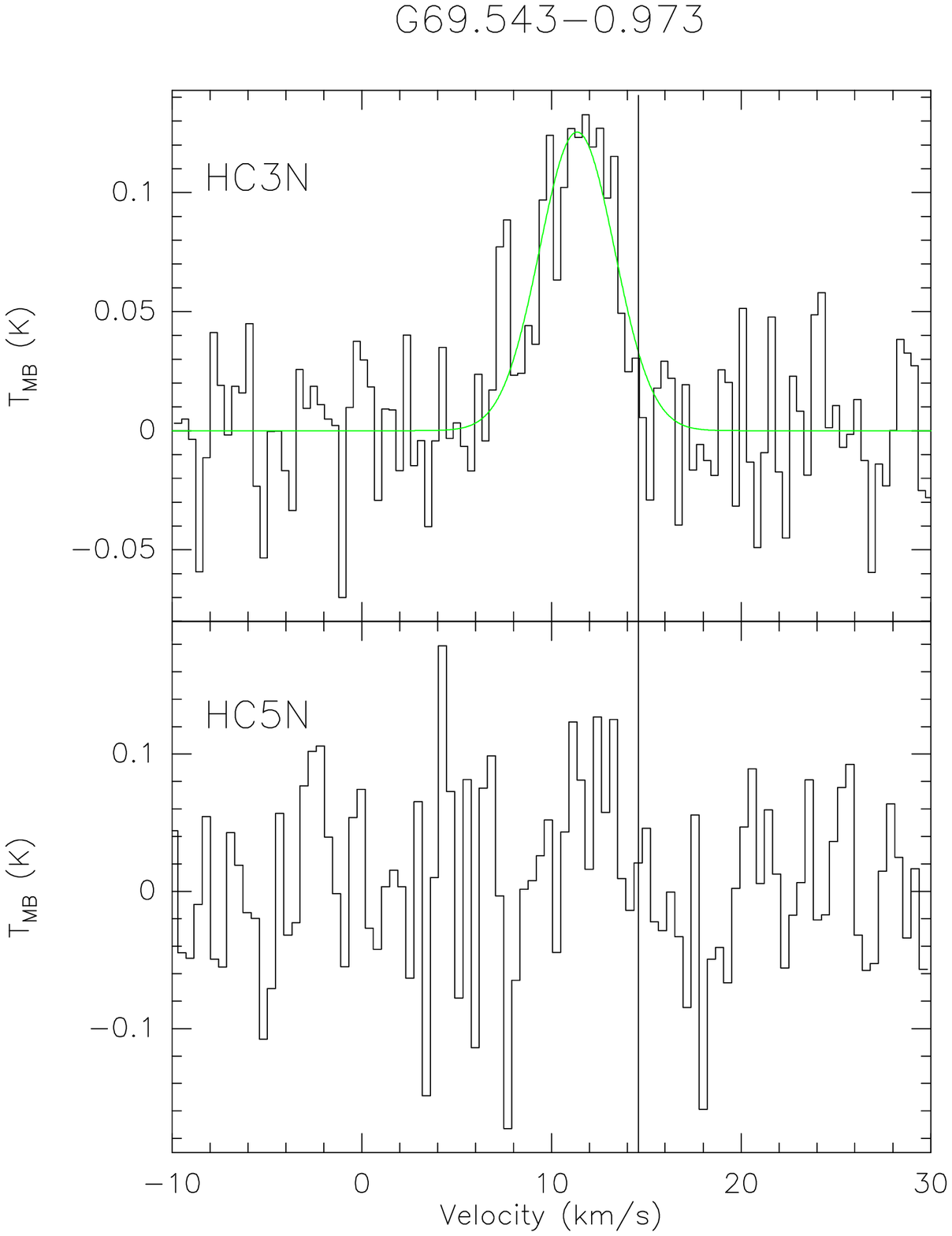}}
    \centering
    \subfigure{
    \centering
    \includegraphics[width=0.26\textwidth]{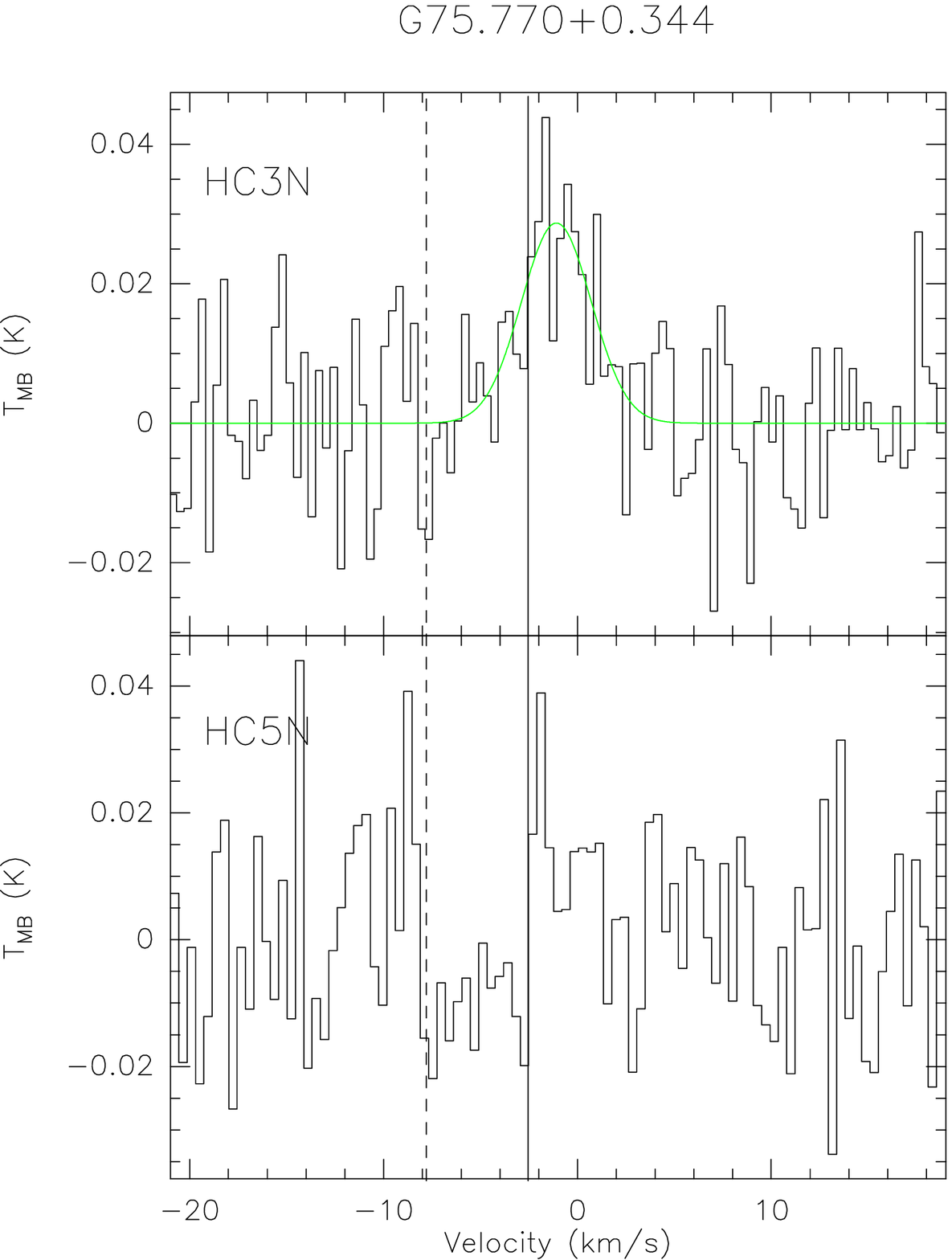}}
    \subfigure{
    \centering
    \includegraphics[width=0.26\textwidth]{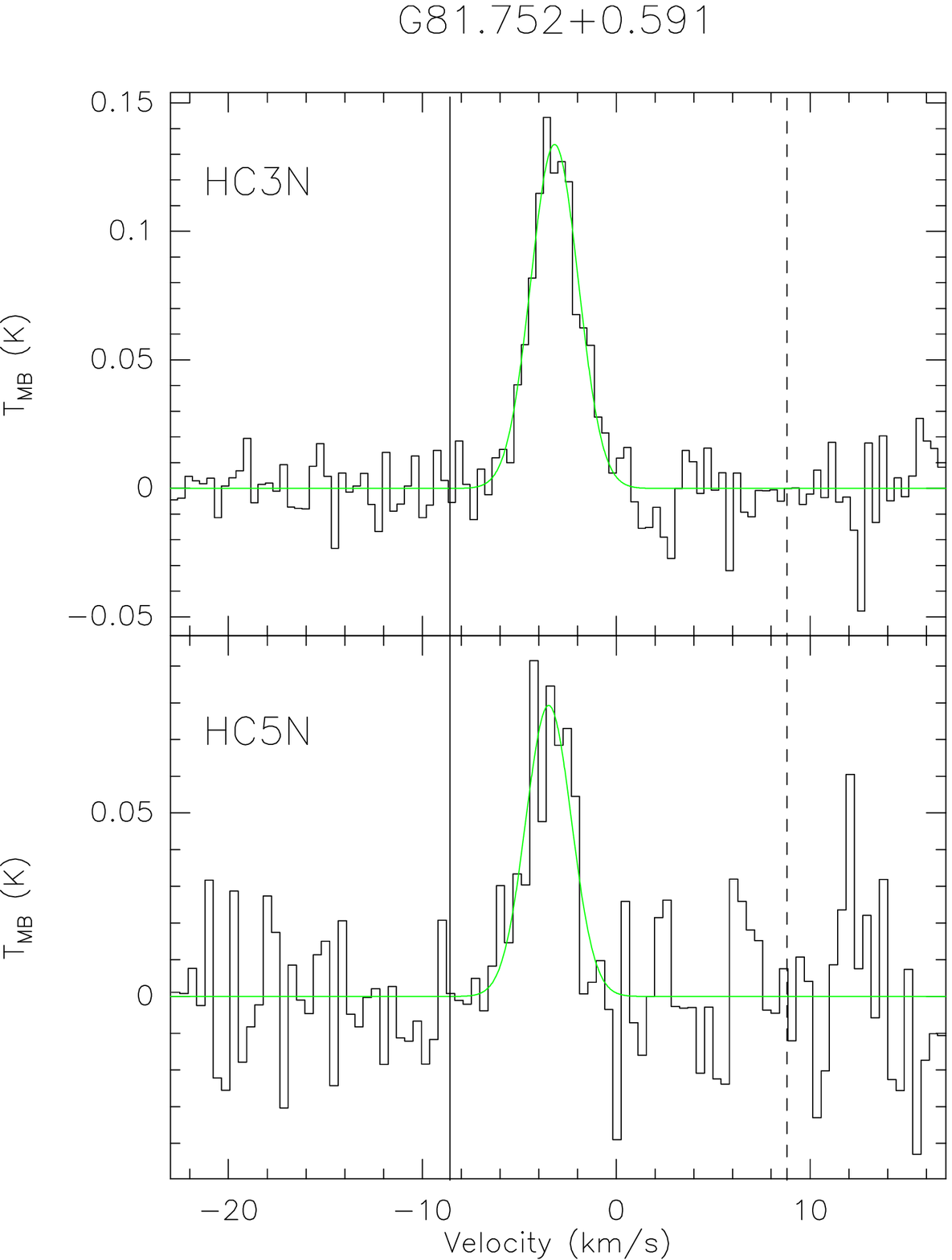}}
    \centering
    \subfigure{
    \centering
    \includegraphics[width=0.26\textwidth]{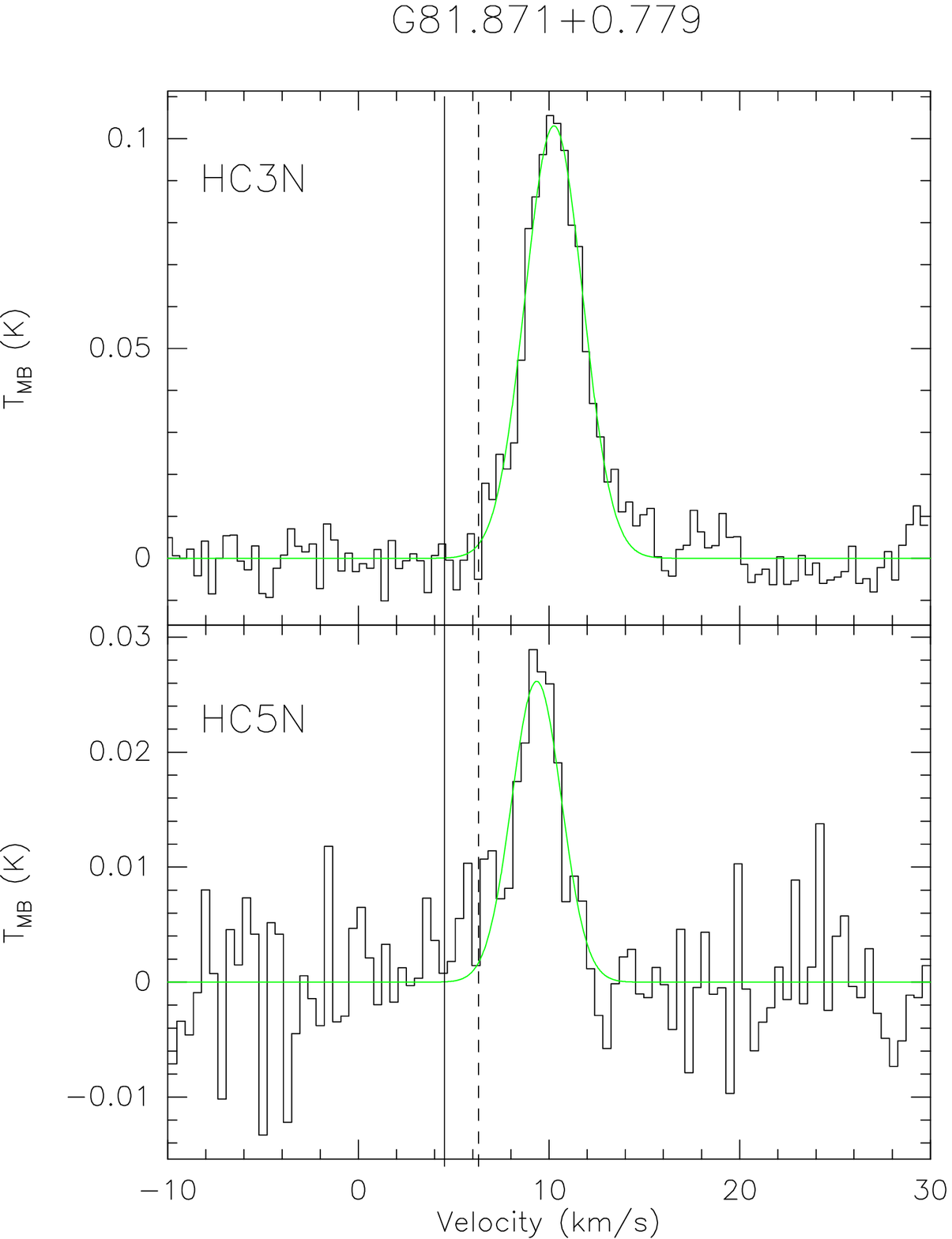}}
    \subfigure{
    \centering
    \includegraphics[width=0.26\textwidth]{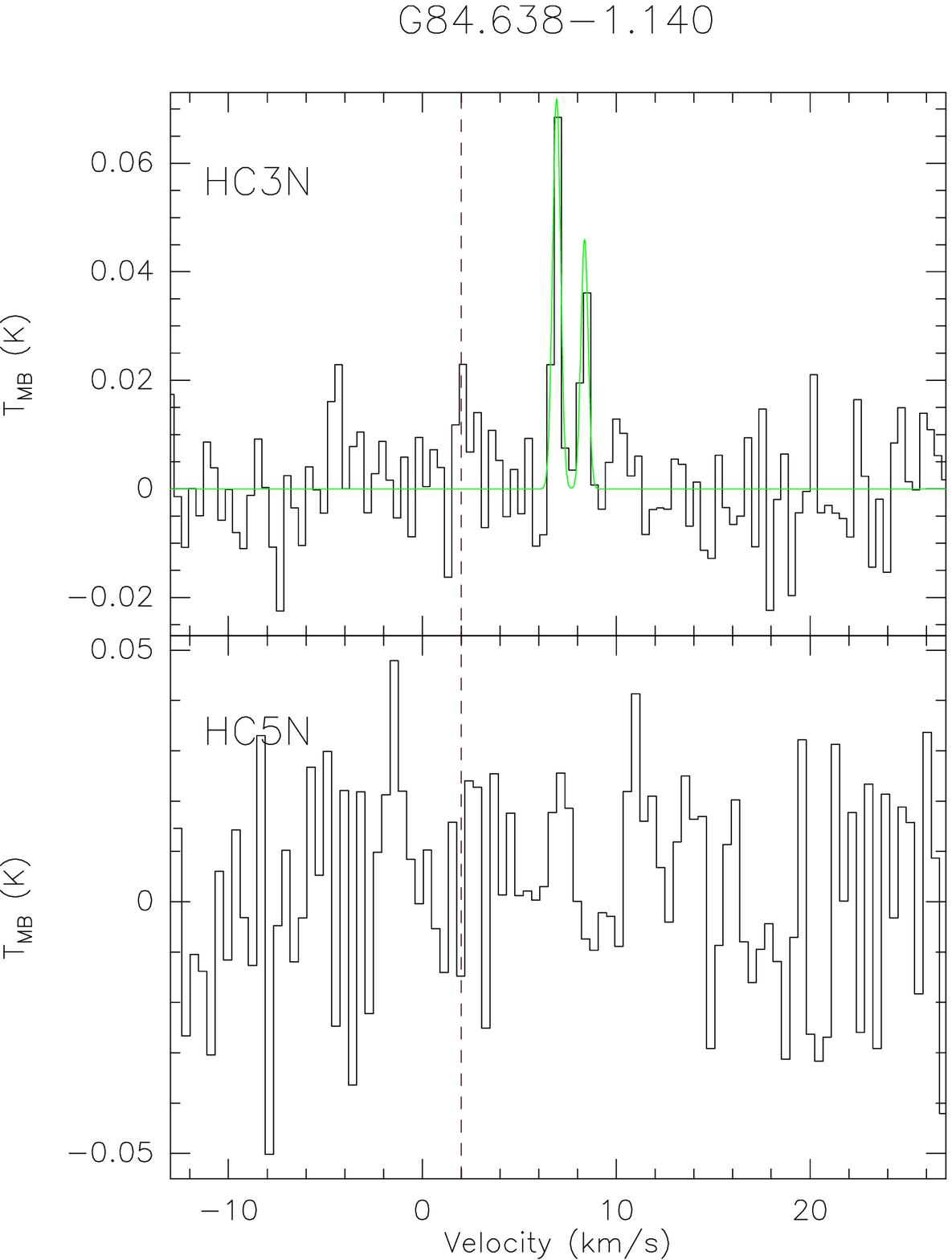}}
    \centering
    \subfigure{
    \centering
    \includegraphics[width=0.26\textwidth]{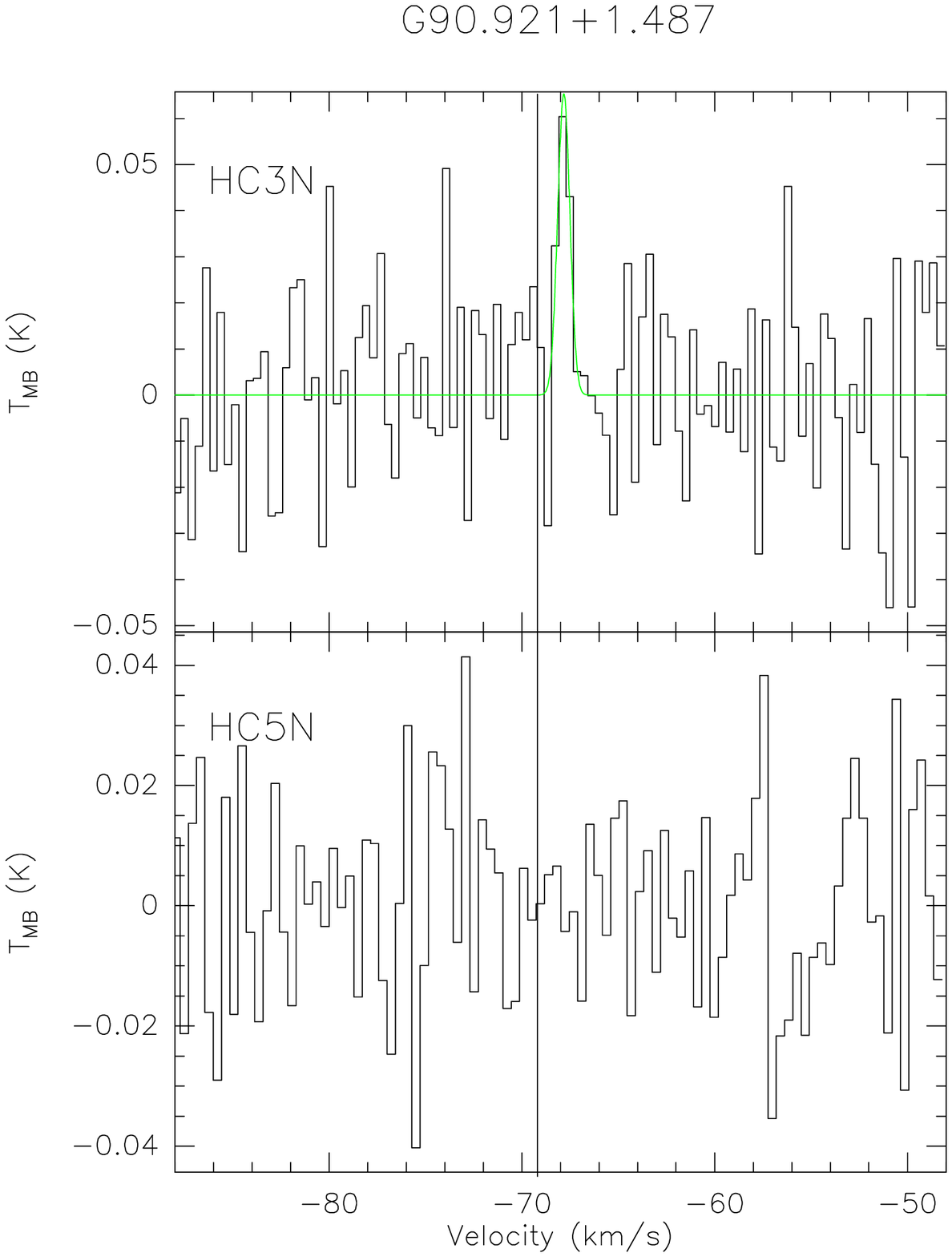}}
    \subfigure{
    \centering
    \includegraphics[width=0.26\textwidth]{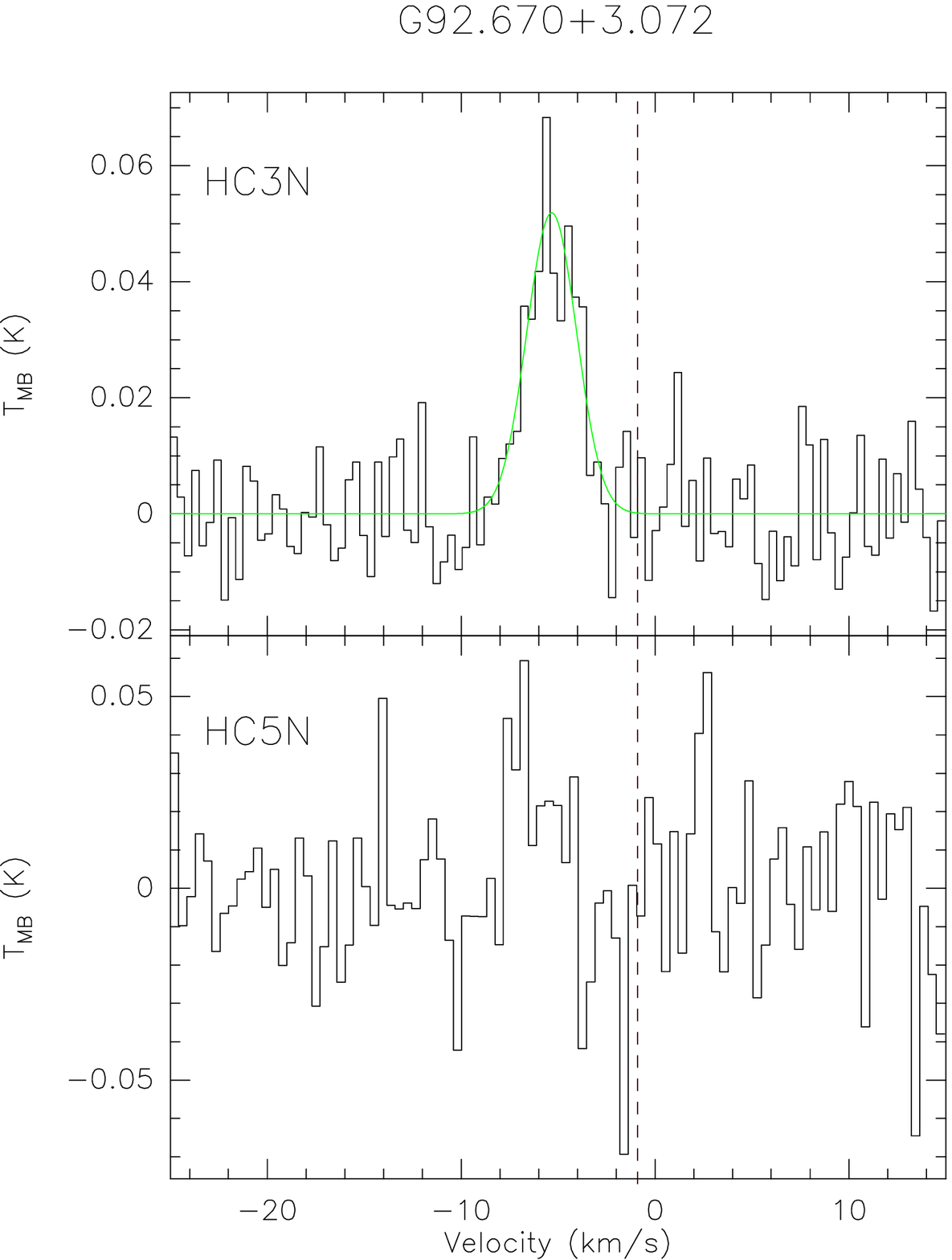}}
    \centering
    \subfigure{
    \centering
    \includegraphics[width=0.26\textwidth]{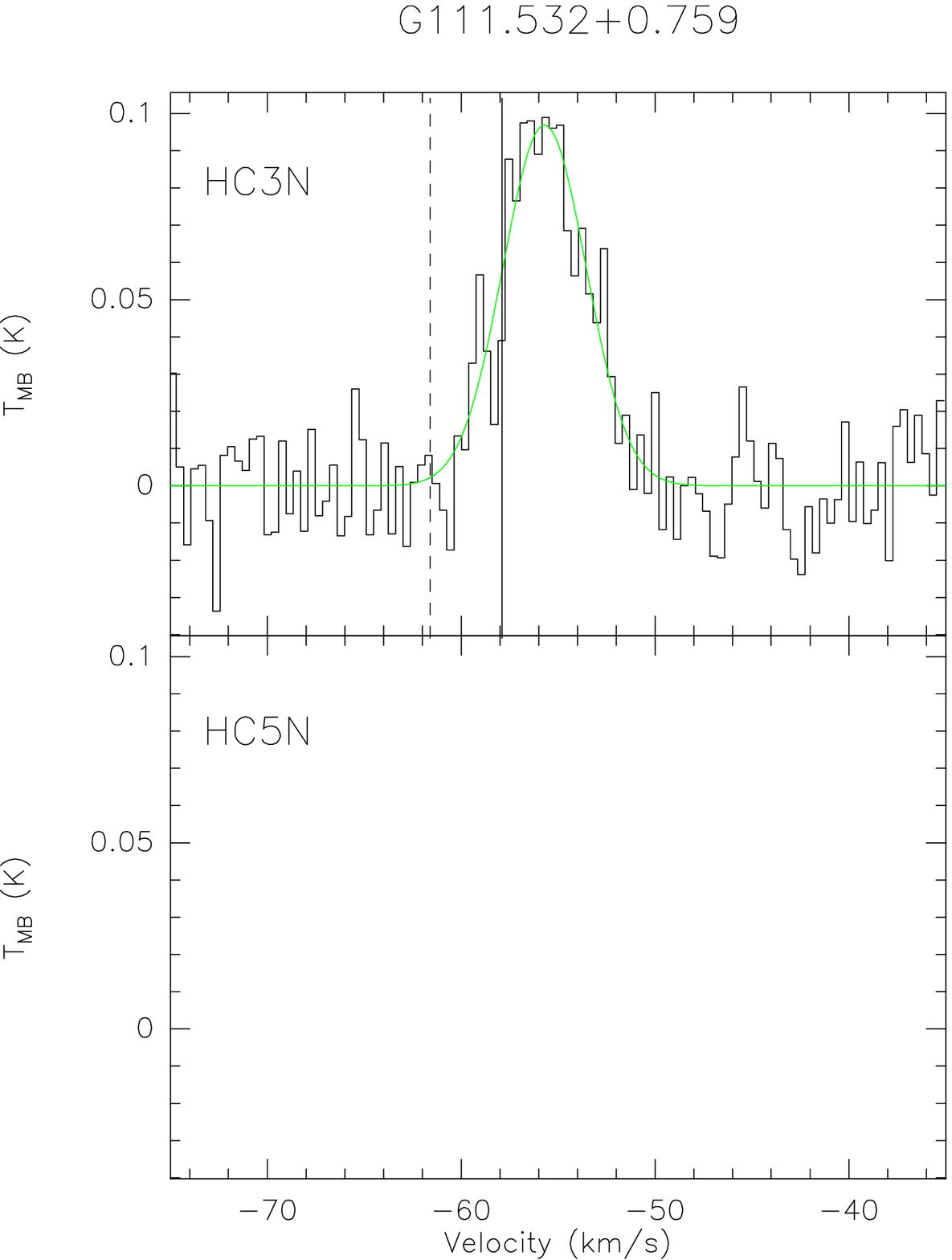}}
    \subfigure{
    \centering
    \includegraphics[width=0.26\textwidth]{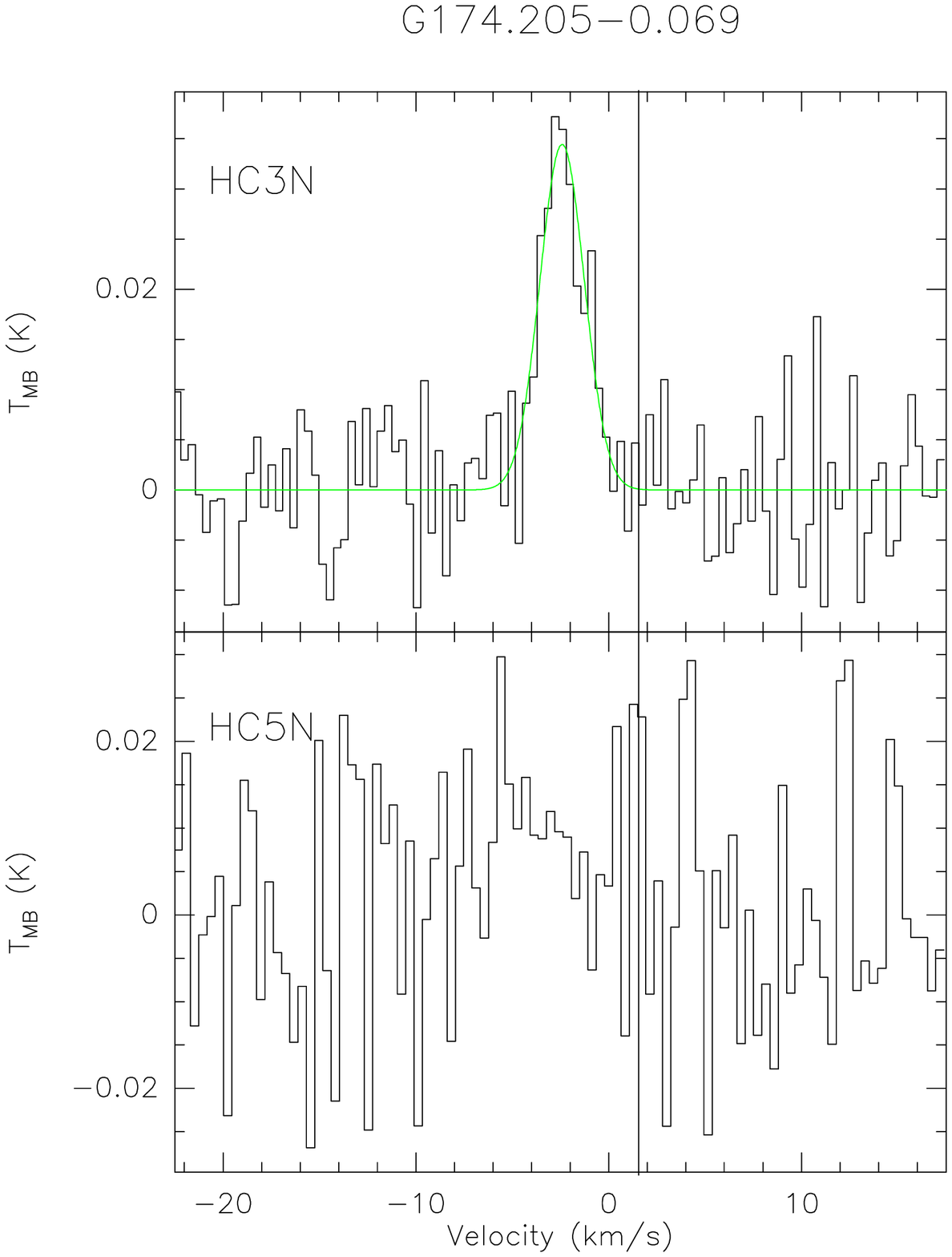}}
    \quad
    \caption{Continued.}
\end{figure}

\addtocounter{figure}{-1}
\begin{figure}
    \subfigure{
    \centering
    \includegraphics[width=0.26\textwidth]{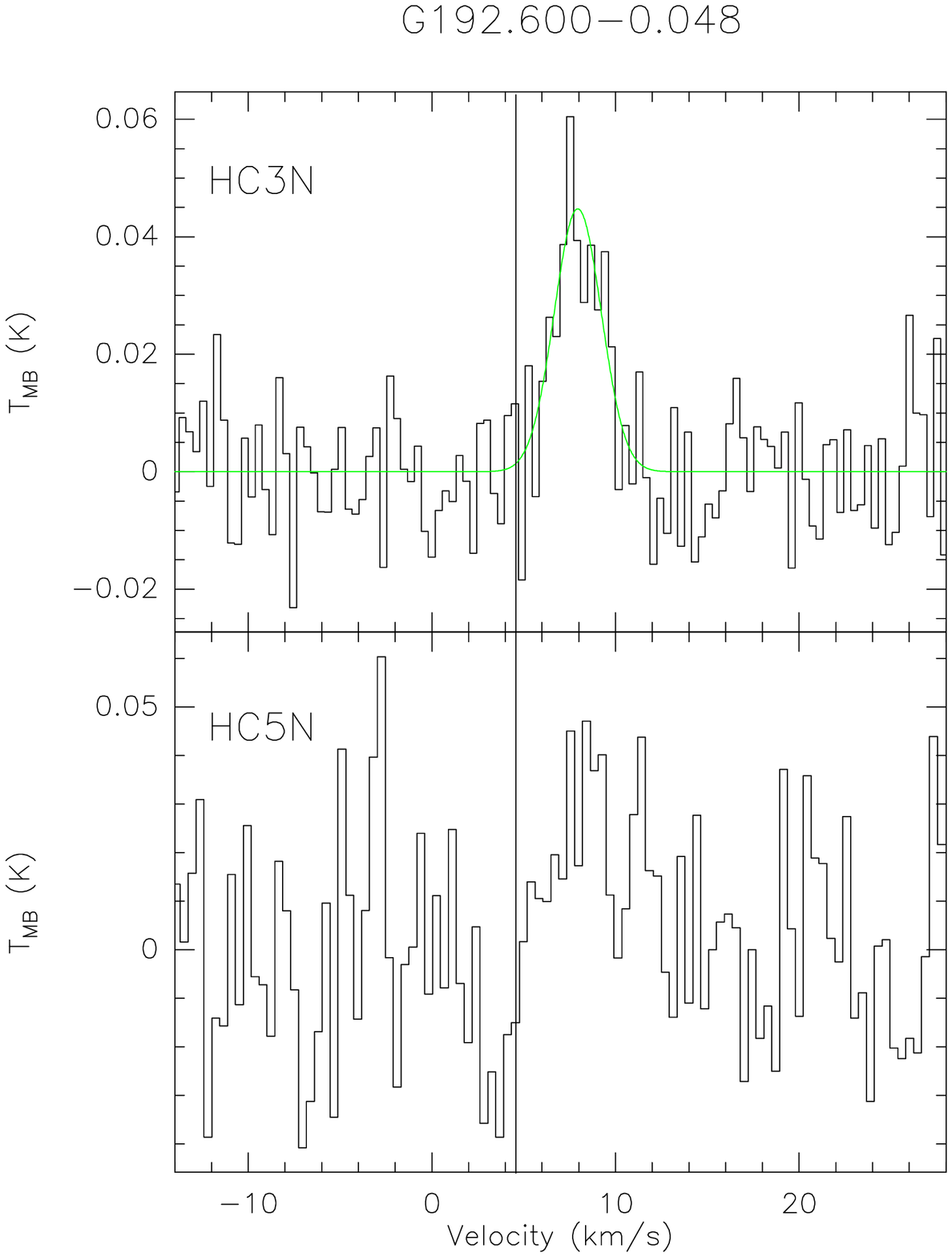}}
    \centering
    \subfigure{
    \centering
    \includegraphics[width=0.26\textwidth]{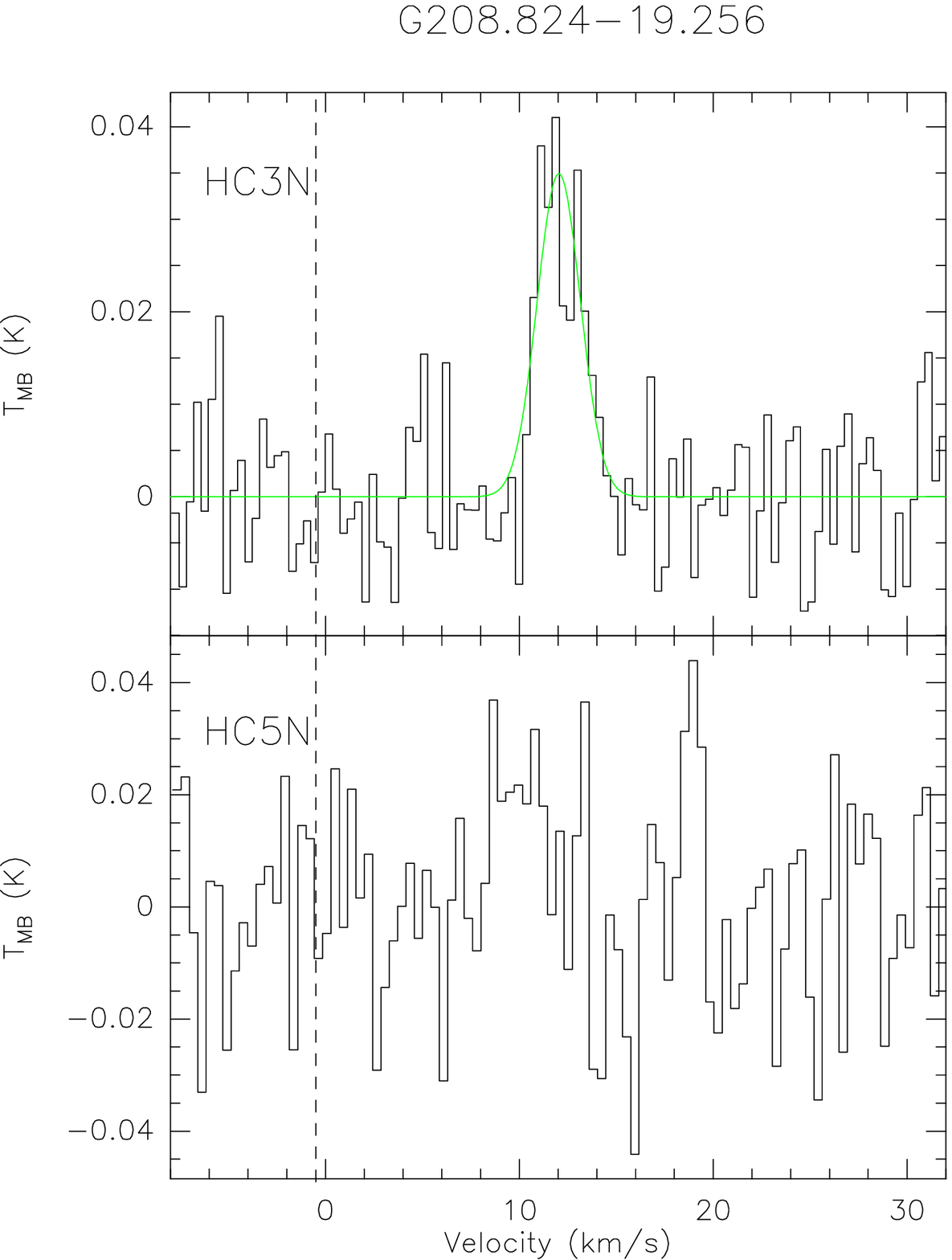}}
    \caption{Continued.}
    \label{figa1}
\end{figure}

\clearpage
\section{The observed properties of detected lines in the survey}
\begin{table}[h]
        \centering
        \caption{The HC$_{3}$N (J = 2-1) transitions detected with the TMRT.}
        \label{tab2}
    \setlength{\tabcolsep}{1.3mm}{
        \begin{tabular}{ccccccccccc} 
    \hline\hline
    Source Name         &       R.A.    &       DEC     &       V$_{LSR}$       &       $\Delta \nu$    &       $T_{\rm mb}$ &  rms             &       $\int$ $T_{\rm mb}$d$_{\nu }$ &    $D$  & $D_{\rm GC}$ &   Classifications \\
        &       (hh.mm.ss)      &       (hh.mm.ss)      &       (km s$^{-1}$)   &       (km s$^{-1}$) &     (mK) &  (mK) &  (K km s$^{-1}$) & (kpc) & (kpc) &               \\
(1)     &       (2)     &       (3)     &       (4)     &       (5)     &       (6)     &       (7)     &       (8)     &       (9)     &(10) & (11)  \\
\hline
G20.234+0.085  &  18:27:44.56  &  -11:14:54.2  &  71.03$\pm$0.08  &  3.49$\pm$0.24  &  80  &  8  &  0.30$\pm$0.02  &  4.41$^{\ast}$   &   4.27  &  Maser-only\\
G23.271-0.256  &  18:34:36.22  &  -08:42:26.6  &  81.88$\pm$0.13  &  3.08$\pm$0.33  &  68  &  10  &  0.22$\pm$0.02    &  5.92  &   3.56   &  Maser-RRL\\
G23.389+0.185  &  18:33:14.36  &  -08:23:57.6  &  75.96$\pm$0.18  &  3.33$\pm$0.40  &  52  &  12  &  0.19$\pm$0.02   &  4.81  &   4.17   &  Maser-RRL\\
G23.436-0.184  &  18:34:39.21  &  -08:31:40.4  &  101.58$\pm$0.11  &  5.65$\pm$0.25  &  101  &  12  &  0.61$\pm$0.02     &  5.88  &   3.59   &  Maser-RRL\\
G23.484+0.097  &  18:33:44.05  &  -08:21:20.6  &  85.36$\pm$0.16  &  4.51$\pm$0.43  &  62  &  11  &  0.30$\pm$0.02    &  4.93$^{\ast}$   &   4.10   &  Maser-only\\
G23.965-0.110  &  18:35:22.31  &  -08:01:28.0  &  73.06$\pm$0.34  &  6.03$\pm$0.80  &  65  &  21  &  0.42$\pm$0.05     &  5.07$^{\ast}$   &    4.05   &  Maser-RRL\\
G24.328+0.144  &  18:35:08.15  &  -07:35:04.4  &  113.81$\pm$0.07  &  4.11$\pm$0.17  &  115  &  9  &  0.50$\pm$0.02     &  7.23$^{\ast}$   &   3.35   &  Maser-RRL\\
G24.528+0.337  &  18:34:49.00  &  -07:19:05.9  &  107.36$\pm$0.20  &  3.59$\pm$0.50  &  57  &  13  &  0.22$\pm$0.03     &  6.00$^{\ast}$   &   3.65   &  Maser-RRL\\
G24.546-0.245  &  18:36:55.92  &  -07:34:13.9  &  100.11$\pm$0.14  &  3.05$\pm$0.37  &  57  &  10  &  0.19$\pm$0.02     &  5.70$^{\ast}$   &    3.77   &  RRL-only\\
G24.790+0.084  &  18:36:12.45  &  -07:12:10.7  &  110.29$\pm$0.02  &  3.79$\pm$0.04  &  186  &  4  &  0.75$\pm$0.01      &  6.67  &    3.48    &  Maser-RRL\\
G25.649+1.050  &  18:34:20.91  &  -05:59:42.5  &  43.19$\pm$0.18  &  4.08$\pm$0.57  &  57  &  12  &  0.25$\pm$0.03     &  3.80$^{\ast}$   &   4.98   &  Maser-only\\
G25.709+0.044  &  18:38:03.17  &  -06:24:15.5  &  99.47$\pm$0.17  &  4.32$\pm$0.45  &  63  &  12  &  0.29$\pm$0.02      &  5.87$^{\ast}$   &   3.81  &   Maser-RRL\\
G28.147-0.004  &  18:42:42.59  &  -04:15:35.0  &  99.42$\pm$0.20  &  3.25$\pm$0.41  &  38  &  10  &  0.13$\pm$0.02     &  6.33  &   3.92   &  Maser-RRL\\
G28.287-0.348  &  18:44:11.48  &  -04:17:32.1  &  49.11$\pm$0.19  &  3.52$\pm$0.50  &  48  &  11  &  0.18$\pm$0.02     &  4.52  &   4.66  &   Maser-RRL\\
G28.393+0.085  &  18:42:50.49  &  -04:00:00.0  &  79.06$\pm$0.04  &  3.68$\pm$0.09  &  133  &  6  &  0.52$\pm$0.01      &  4.33  &   4.78  &   Maser-RRL\\
G28.452+0.002  &  18:43:14.64  &  -03:59:07.9  &  98.84$\pm$0.12  &  0.88$\pm$0.22  &  43  &  10  &  0.04$\pm$0.01      &  8.17$^{\ast}$   &   8.09  &   RRL-only\\
  &    &    &  100.52$\pm$0.26  &  1.64$\pm$0.74  &  26  &    &  0.05$\pm$0.02  &    &   &     \\
G28.817+0.365  &  18:42:37.35  &  -03:29:41.0  &  87.21$\pm$0.26  &  3.09$\pm$0.52  &  62  &  21  &  0.20$\pm$0.03     &  4.51$^{\ast}$   &   4.70  &   Maser-only\\
G29.835-0.012  &  18:45:59.57  &  -02:45:04.4  &  100.46$\pm$0.10  &  0.76$\pm$0.18  &  76  &  20  &  0.06$\pm$0.02      &  7.64$^{\ast}$   &   4.09  &   Maser-RRL\\
G30.770-0.804  &  18:50:21.51  &  -02:17:27.5  &  79.21$\pm$0.13  &  2.79$\pm$0.29  &  42  &  8  &  0.13$\pm$0.01     &  3.71$^{\ast}$   &    5.29   &   Maser-only\\
G30.810-0.050  &  18:47:46.97  &  -01:54:26.4  &  64.02$\pm$0.43  &  7.97$\pm$1.12  &  24  &  8  &  0.20$\pm$0.03      &  3.12  &   5.68  &   Maser-RRL\\
  &    &    &  98.61$\pm$0.04  &  5.67$\pm$0.11  &  204  &    &  1.23$\pm$0.02  &    &    &      \\
  &    &    &  118.42$\pm$0.26  &  6.72$\pm$0.6  &  39  &    &  0.28$\pm$0.02  &    &    &     \\
G30.897+0.163  &  18:47:08.77  &  -01:44:12.7  &  106.17$\pm$0.19  &  2.94$\pm$0.37  &  56  &  14  &  0.17$\pm$0.02      &  7.07$^{\ast}$   &   4.17  &   Maser-RRL\\
G31.579+0.076  &  18:48:41.94  &  -01:10:02.5  &  96.65$\pm$0.14  &  3.95$\pm$0.34  &  56  &  10  &  0.24$\pm$0.02      &  5.46  &   4.50  &   Maser-RRL\\
G35.141-0.750  &  18:58:06.14  &  01:37:07.5  &  34.59$\pm$0.11  &  5.54$\pm$0.25  &  82  &  9  &  0.48$\pm$0.02     &  2.21$^{\ast}$   &   6.44  &   Maser-RRL\\
G35.194-1.725  &  19:01:42.33  &  01:13:32.3  &  43.19$\pm$0.21  &  3.42$\pm$0.42  &  41  &  10  &  0.15$\pm$0.02      &  2.43  &   6.30  &   Maser-RRL\\
G37.043-0.035  &  18:59:03.64  &  03:37:45.1  &  80.98$\pm$0.16  &  3.38$\pm$0.39  &  48  &  10  &  0.17$\pm$0.02      &  4.57$^{\ast}$   &   5.25  &   Maser-only\\
G49.028-0.217  &  19:22:12.74  &  14:10:58.3  &  65.58$\pm$0.20  &  1.27$\pm$0.41  &  47  &  15  &  0.06$\pm$0.02      &  5.49$^{\ast}$   &   6.13  &   RRL-only\\
G49.466-0.408  &  19:23:45.73  &  14:28:45.1  &  56.54$\pm$0.13  &  3.79$\pm$0.28  &  58  &  9  &  0.23$\pm$0.02     &  5.48$^{\ast}$  &   6.18  &   Maser-RRL\\
G69.543-0.973  &  20:10:09.00  &  31:31:35.9  &  11.36$\pm$0.21  &  4.66$\pm$0.51  &  125  &  28  &  0.62$\pm$0.06      &  2.46  &   7.62  &   Maser-only\\
G75.770+0.344  &  20:21:41.73  &  37:26:02.5  &  -1.08$\pm$0.34  &  4.25$\pm$0.88  &  29  &  11  &  0.13$\pm$0.02      &  3.83  &   8.09  &   Maser-RRL\\
G81.752+0.591  &  20:39:02.01  &  42:24:59.3  &  -3.18$\pm$0.07  &  2.87$\pm$0.16  &  134  &  12  &  0.41$\pm$0.02      &  1.50  &   8.05   &   Maser-RRL\\
G81.871+0.779  &  20:38:36.66  &  43:37:30.5  &  10.26$\pm$0.04  &  3.56$\pm$0.11  &  103  &  5  &  0.39$\pm$0.01      &  1.30  &   8.04  &   Maser-RRL\\
G84.638-1.140  &  20:56:18.81  &  43:34:25.1  &  6.93$\pm$0.06  &  0.49$\pm$0.11  &  72  &  10  &  0.04$\pm$0.01     &  1.34$^{\ast}$   &   8.08  &   RRL-only\\
  &    &    &  8.37$\pm$0.06  &  0.45$\pm$0.14  &  46  &    &  0.02$\pm$0.01  &    &    & \\
G90.921+1.487  &  21:09:12.87  &  50:01:02.6  &  -67.82$\pm$0.11  &  0.75$\pm$0.20  &  65  &  20  &  0.05$\pm$0.02      &  5.85  &   10.09  &   Maser-only\\
G92.670+3.072  &  21:09:21.22  &  52:22:35.7  &  -5.32$\pm$0.13  &  3.02$\pm$0.28  &  52  &  9  &  0.17$\pm$0.01     &  1.63  &   8.36  &   RRL-only\\
G111.532+0.759  &  23:13:43.87  &  61:26:55.8  &  -55.7$\pm$0.13  &  5.05$\pm$0.30  &  97  &  13  &  0.52$\pm$0.03      &  2.69$^{\ast}$   &   9.45  &   Maser-RRL\\
G174.205-0.069  &  05:30:49.01  &  33:47:46.7  &  -2.41$\pm$0.13  &  2.71$\pm$0.31  &  34  &  6  &  0.10$\pm$0.01     &  2.13$^{\ast}$   &   10.24  &   Maser-only\\
G192.600-0.048  &  06:12:53.99  &  17:59:23.7  &  7.95$\pm$0.17  &  3.02$\pm$0.37  &  45  &  10  &  0.14$\pm$0.02     &  1.66  &   9.76  &   Maser-only\\
G208.824-19.256  &  05:35:24.29  &  -05:08:30.7  &  12.04$\pm$0.14  &  2.63$\pm$0.28  &  35  &  7  &  0.10$\pm$0.01     &  0.42$^{\ast}$   &   8.10  &   RRL-only\\
       \hline 
        \end{tabular}}       
        \tablefoot{(1): Source name; (2): Right ascension (J2000); (3): Declination(J2000); (4): LSR velocity; (5): Full width at half maximum (FWHM) ; (6): Peak $T_{\rm mb}$ value; (7): The rms noise value; (8): Integrated line intensity; (9): The heliocentric distance \textit{D} is taken from the trigonometric parallax measurements \citep{2014ApJ...783..130R,2019ApJ...885..131R}. Others (with $^{\ast}$) are obtained from the latest Parallax-based Distance Calculator V2; (10): The galactocentric distance $D_{\rm GC}$; (11): Classifications.}
\end{table}

\begin{table}
        \centering
        \caption{The HC$_{5}$N (J = 3-2) and HC$_{7}$N(J = 15-14) transitions detected with TMRT.}
        \label{tab3}
    \begin{threeparttable}
    \setlength{\tabcolsep}{1.5mm}{
        \begin{tabular}{cccccccccc} 
    \hline\hline
    Molecule & Source Name      &       R.A.    &       DEC     &       $V_{\rm LSR}$   &       $\Delta \nu$ &  $T_{\rm mb}$ &  rms &   $\int T_{\rm mb}d_{\nu }$      &       Classifications \\
           &    &       (hh.mm.ss)      &       (hh.mm.ss)      &       (km s$^{-1}$)       &       (km s$^{-1}$) & (mK) &  (mK) &  (K km s$^{-1}$) &               \\
    (1) &       (2)     &       (3)     &       (4)     &       (5)     &       (6)     &       (7)     &       (8)     &       (9)     &       (10) \\
    \hline 
HC$_{5}$N       &       G20.234+0.085   &       18:27:44.56     &       -11:14:54.2     &       70.54$\pm$0.15  &       1.55$\pm$0.36   &       67         &       17      &       0.11$\pm$0.02   &       Maser-only      \\
        &       G23.436-0.184   &       18:34:39.21     &       -08:31:40.4     &       101.97$\pm$0.15 &       1.05$\pm$0.43   &       47         &       13      &       0.05$\pm$0.02   &       Maser-RRL       \\
        &       G24.328+0.144   &       18:35:08.15     &       -07:35:04.4     &       113.27$\pm$0.10 &       0.83$\pm$0.29   &       53         &       12      &       0.05$\pm$0.01   &       Maser-RRL       \\
        &       G24.790+0.084   &       18:36:12.45     &       -07:12:10.7     &       109.06$\pm$0.06 &       3.41$\pm$0.13   &       47         &       3       &       0.17$\pm$0.01   &       Maser-RRL       \\
        &       G28.393+0.085   &       18:42:50.49     &       -04:00:00.0     &       77.20$\pm$0.25  &       2.78$\pm$0.97   &       33         &       9       &       0.10$\pm$0.02   &       Maser-RRL       \\
        &       G30.770-0.804   &       18:50:21.51     &       -02:17:27.5     &       78.83$\pm$0.18  &       1.57$\pm$0.49   &       34         &       11      &       0.06$\pm$0.01   &       Maser-only      \\
        &       G30.810-0.050   &       18:47:46.97     &       -01:54:26.4     &       98.07$\pm$0.24  &       3.44$\pm$0.46   &       45         &       13      &       0.16$\pm$0.02   &       Maser-RRL       \\
        &       G30.897+0.163   &       18:47:08.77     &       -01:44:12.7     &       105.45$\pm$0.16 &       1.47$\pm$0.43   &       30         &       9       &       0.05$\pm$0.01   &       Maser-RRL       \\
        &       G35.141-0.750   &       18:58:06.14     &       01:37:07.5      &       34.97$\pm$0.17  &       1.19$\pm$0.49   &       60         &       19      &       0.08$\pm$0.02   &       Maser-RRL       \\
        &       G81.752+0.591   &       20:39:02.01     &       42:24:59.3      &       -3.48$\pm$0.18  &       2.77$\pm$0.40   &       79         &       19      &       0.23$\pm$0.03   &       Maser-RRL       \\
        &       G81.871+0.779   &       20:38:36.66     &       43:37:30.5      &       9.36$\pm$0.17   &       3.07$\pm$0.49   &       26         &       5       &       0.09$\pm$0.01   &       Maser-RRL       \\
HC$_{7}$N       &       G24.790+0.084   &       18:36:12.45     &       -07:12:10.7     &       109.98$\pm$0.35 &       3.55$\pm$1.05   &       8         &       3       &       0.03$\pm$0.01   &       Maser-RRL       \\
    \hline 
        \end{tabular}}
\tablefoot{(1): Molecules species; (2): Source name; (3): Right ascension (J2000); (4): Declination(J2000); (5): LSR velocity; (6): FWHM; (7): Peak $T_{\rm mb}$ value; (8): The rms noise value; (9): Integrated line intensity; (10): Classifications. }
\end{threeparttable}
\end{table}

\begin{landscape}
\begin{table}
        \centering
        \caption{The column density and the relative abundance for HC$_{3}$N, HC$_{5}$N, and HC$_{7}$N.}
    \setlength{\tabcolsep}{1mm}
    \begin{threeparttable}
        \begin{tabular}{ccccccccccc} 
    \hline\hline
{Classifications}  & {Source Name}& {$T_{\rm rot}$ }& $N(HC_{3}N)$ & $X(HC_{3}N)$ & $N(HC_{5}N)$ & $X(HC_{5}N)$ & $N(HC_{7}N)$ & $X(HC_{7}N)$ & $N(HC_{3}N)$/$N(HC_{5}N)$ & Reference \\
& &{(K)}& $(\times 10^{13} cm^{-2})$ & $(\times 10^{-10})$ & $(\times 10^{13} cm^{-2})$ & $(\times 10^{-10})$ & $(\times 10^{13} cm^{-2})$ & $(\times 10^{-10})$ & & \\
(1)& (2) & (3)& (4) & (5) & (6) & (7) & (8) & (9) & (10) & (11)\\
\hline
Maser-only      &       G20.234+0.085   &       15.11   &       1.96 $\pm$ 0.13    &        1.07$\pm$0.08  &       0.43 $\pm$ 0.08 &        0.23$\pm$0.05  &       ...     &       ...     & 4.55 & 1        \\
        &       G23.484+0.097   &       16.10   &       2.05 $\pm$ 0.14 &        2.86$\pm$0.21   &       ...     &       ...     &       ...     &       ...     & ... &   2       \\
        &       G25.649+1.050   &       20.60   &       2.06 $\pm$ 0.25 & 1.94$\pm$0.17   &       ...     &       ...     &       ...     &       ...     & ... &   3       \\
        &       G28.817+0.365   &       18.96   &       1.54 $\pm$ 0.23 &        1.04$\pm$0.15   &       ...     &       ...     &       ...     &       ...     & ... &   2       \\
        &       G30.770-0.804   &       17.56   &       0.95 $\pm$ 0.07 &        0.34$\pm$0.03   &       0.26 $\pm$ 0.04 &        0.09$\pm$0.02  &       ...     &       ...     & 3.64 &  2       \\
        &       G37.043-0.035   &       15.22   &       1.12 $\pm$ 0.13 &        0.02$\pm$0.01   &       ...     &       ...     &       ...     &       ...     & ... &   2       \\
        &       G69.543-0.973   &       21.20   &       5.22 $\pm$ 0.51 &       ...     &       ...     &       ...     &       ...     &       ...     & ... &   3       \\
        &       G90.921+1.487   &       18.90$^{\ast}$  &       0.39 $\pm$ 0.16    &       ...     &       ...     &       ...     &       ...     &       ...     & ... &   ...     \\
        &       G174.205-0.069  &       19.80   &       0.80 $\pm$ 0.08 &       ...     &       ...     &       ...     &       ...     &       ...     & ... &   3       \\
        &       G192.600-0.048  &       25.52   &       1.37 $\pm$ 0.20 &       ...     &       ...     &       ...     &       ...     &       ...     & ... &   1       \\
Maser-RRL       &       G23.271-0.256   &       18.62   &       1.70 $\pm$ 0.15    &        4.32$\pm$0.39  &       ...     &       ...     &       ...     &       ...     & ... &   2       \\
        &       G23.389+0.185   &       16.80   &       1.31 $\pm$ 0.14 &        1.02$\pm$0.11   &       ...     &       ...     &       ...     &       ...     & ... &   3       \\
        &       G23.436-0.184   &       17.27   &       4.37 $\pm$ 0.14 &        12.32$\pm$0.51  &       0.21 $\pm$ 0.08 &        0.58$\pm$0.23  &       ...     &       ...     & 20.81 & 4       \\
        &       G23.965-0.110   &       22.36   &       3.66 $\pm$ 0.44 &        8.55$\pm$1.03   &       ...     &       ...     &       ...     &       ...     & ... &   2       \\
        &       G24.328+0.144   &       13.31   &       3.02 $\pm$ 0.12 &        10.37$\pm$0.41  &       0.18 $\pm$ 0.04 &        0.61$\pm$0.13  &       ...     &       ...     & 16.78 & 5       \\
        &       G24.528+0.337   &       11.22   &       1.19 $\pm$ 0.16 &       ...     &       ...     &       ...     &       ...     &       ...     & ... &   6       \\
        &       G24.790+0.084   &       22.01   &       6.53 $\pm$ 0.09 &        12.22$\pm$0.16  &       0.84 $\pm$ 0.05 &        1.54$\pm$0.09  &       0.11 $\pm$ 0.04      &        0.27$\pm$0.09  & 7.77 &        1       \\
        &       G25.709+0.044   &       18.30   &       2.17 $\pm$ 0.15 &       2.13$\pm$0.02   &       ...     &       ...     &       ...     &       ...     & ... &   3       \\
        &       G28.147-0.004   &       18.67   &       1.01 $\pm$ 0.15 &       2.46$\pm$0.15   &       ...     &       ...     &       ...     &       ...     & ... &   2       \\
        &       G28.287-0.348   &       13.62   &       1.10 $\pm$ 0.12 &       ...     &       ...     &       ...     &       ...     &       ...     & ... &   5       \\
        &       G28.393+0.085   &       11.30   &       2.84 $\pm$ 0.05 &       9.16$\pm$0.18   &       0.34 $\pm$ 0.07      &       ...     &       ...     &       ...     & 8.35 &        7       \\
        &       G29.835-0.012   &       20.10   &       0.50 $\pm$ 0.16 &        0.56$\pm$0.18   &       ...     &       ...     &       ...     &       ...     & ... &   3       \\
        &       G30.810-0.050   &       22.00   &       10.13 $\pm$ 0.26        &       ...     &       0.79 $\pm$ 0.10      &       ...     &       ...     &       ...     & 12.82 &       2       \\
        &       G30.897+0.163   &       18.49   &       1.32 $\pm$ 0.15 &        1.91$\pm$0.22   &       0.22 $\pm$ 0.04 &        0.31$\pm$0.06  &       ...     &       ...     & 6.00 &  2       \\
        &       G31.579+0.076   &       23.80   &       2.17 $\pm$ 0.18 &        1.52$\pm$0.13   &       ...     &       ...     &       ...     &       ...     & ... &   3       \\
        &       G35.141-0.750   &       20.05   &       3.91 $\pm$ 0.16 &        3.65$\pm$0.16   &       0.37 $\pm$ 0.09 &       0.35$\pm$0.08   &       ...     &       ...     & 10.57 & 2       \\
        &       G35.194-1.725   &       19.09$^{\ast}$  &       1.18 $\pm$ 0.16    &       ...     &       ...     &       ...     &       ...     &       ...     & ... &   ...     \\
        &       G49.466-0.408   &       19.09$^{\ast}$  &       1.80 $\pm$ 0.16    &       ...     &       ...     &       ...     &       ...     &       ...     & ... &   ...     \\
        &       G75.770+0.344   &       32.60   &       1.56 $\pm$ 0.24 &       ...     &       ...     &       ...     &       ...     &       ...     & ... &   3       \\
        &       G81.752+0.591   &       18.60   &       3.12 $\pm$ 0.15 &       ...     &       1.02 $\pm$ 0.13      &       ...     &       ...     &       ...     & 3.06 &        3       \\
        &       G81.871+0.779   &       19.09$^{\ast}$  &       3.06 $\pm$ 0.08    &       ...     &       ...     &       ...     &       ...     &       ...     & ... &   ...     \\
        &       G111.532+0.759  &       23.60   &       4.78 $\pm$ 0.28 &       ...     &       ...     &       ...     &       ...     &       ...     & &       3       \\
RRL-only        &       G24.546-0.245   &       17.96   &       1.38 $\pm$ 0.15    &        3.11$\pm$0.36  &       ...     &       ...     &       ...     &       ...     & ... &   2       \\
        &       G28.452+0.002   &       17.96$^{\ast}$  &       0.67 $\pm$ 0.07    &        0.18$\pm$0.02  &       ...     &       ...     &       ...     &       ...     & ... &   ...     \\
        &       G49.028-0.217   &       17.96$^{\ast}$  &       0.47 $\pm$ 0.15    &       ...     &       ...     &       ...     &       ...     &       ...     & ... &   ...     \\
        &       G84.638-1.140   &       17.96$^{\ast}$  &       0.59 $\pm$ 0.07    &       ...     &       ...     &       ...     &       ...     &       ...     & ... &   ...     \\
        &       G92.670+3.072   &       17.96$^{\ast}$  &       1.23 $\pm$ 0.07    &       ...     &       ...     &       ...     &       ...     &       ...     & ... &   ...     \\
        &       G208.824-19.256 &       17.96$^{\ast}$  &       0.72 $\pm$ 0.07    &       ...     &       ...     &       ...     &       ...     &       ...     & ... &   ...     \\
        \hline 
        \end{tabular}
\tablefoot{(1): Classifications; (2): Source name; (3): The rotational temperature of NH$_{3}$. $T_{\rm rot}$ value with $^{\ast}$ shows the mean $T_{\rm rot}$ of each type; (4)-(9): The column density and relative abundance of HC$_{3}$N, HC$_{5}$N and HC$_{7}$N; (10): The abundance ration of $N(HC_{3}N)$/$N(HC_{5}N)$; (11): The rotational temperature values are taken from: 1 \citep{Yangtian2021}, 2 \citep{2012A&A...544A.146W}, 3 \citep{2011MNRAS.418.1689U}, 4 \citep{2016AJ....152...92L}, 5 \citep{2013ApJ...764...61C}, 6 \citep{2016ApJ...822...59S} and 7 \citep{2013A&A...552A..40C}.}
      \end{threeparttable}
      \label{tab4}
\end{table}
\end{landscape}

\begin{table}[htp]
        \centering
        \caption{The integrated line intensities of other molecules.}
    \setlength{\tabcolsep}{5mm}
    \begin{threeparttable}
        \begin{tabular}{cccccc} 
        \hline\hline
Classifications & Source name & $I_{\rm H_{2}CO}$ & $N_{\rm H_{2}CO}$ & $I_{\rm SiO}$ & $N_{\rm SiO}$ \\
 & &  (K km s$^{-1}$) & $(\times 10^{13} cm^{-2})$ & (K km s$^{-1}$) & $(\times 10^{13} cm^{-2})$ \\
 (1) & (2) & (3) & (4) & (5) & (6)\\
\hline
Maser-only      & G30.770-0.804 & 1.11$\pm$0.09 & 0.82$\pm$0.07 & ... & ... \\
    &   G37.043-0.035   &  ... &...     &       1.23$\pm$0.12 & 0.22    \\
    &   G69.543-0.973   &   3.51$\pm$0.07 & 8.75$\pm$0.17  &   ... & ... \\
    &   G174.205-0.069  &   1.18$\pm$0.09 & 0.77$\pm$0.06  &   ... & ... \\
Maser-RRL       &       G23.389+0.185   &       0.53$\pm$0.05 & 0.45$\pm$0.04 &       1.36$\pm$0.25 & 0.24 \\
        &       G23.436-0.184   &       13.75$\pm$0.37 & 23.04$\pm$0.62 &       14.07$\pm$0.37 & 2.50  \\
        &       G23.965-0.110   &       2.48$\pm$0.08 & 3.37$\pm$0.11   &       7.53$\pm$0.25 & 1.30  \\
        &       G24.328+0.144   &       3.29$\pm$0.11 & 4.76$\pm$0.16   &       ... & ...   \\
        &       G24.528+0.337   &       3.97$\pm$0.08 & 2.18$\pm$0.04   &       ... & ...   \\
        &       G24.790+0.084   &       3.57$\pm$0.07 & 13.73$\pm$0.27  &       17.41$\pm$0.37 & 3.10  \\
        &       G28.147-0.004   &       2.27$\pm$0.17 & 1.55$\pm$0.12   &       ... & ...   \\
        &       G28.287-0.348   &       2.35$\pm$0.05 & 1.47$\pm$0.03   &       ... & ...   \\
        &       G28.393+0.085   &       8.39$\pm$0.07 & 10.85$\pm$0.09  &       9.28$\pm$0.22 &  1.70 \\
        &       G30.897+0.163   & ...   & ... & 3.19$\pm$0.21 & 0.57    \\
        &       G31.579+0.076   &       ...     & ... & 3.83$\pm$0.25 & 0.68    \\
        &       G35.141-0.750   &       ...     & ... & 12.47$\pm$0.37 & 2.20    \\
        &       G49.466-0.408   &       58.82$\pm$0.62 & 61.49$\pm$0.65 &       ... & ...   \\
        &       G81.752+0.591   &       2.71$\pm$0.09 & 6.46$\pm$0.21   &       ...     & ... \\
        &       G81.871+0.779   &       1.98$\pm$0.08 & 3.75$\pm$0.15   &       ...     & ...\\
RRL-only        &       G24.546-0.245   &  1.43$\pm$0.08 & 1.42$\pm$0.08        &       ... & ...   \\
        &   G28.452+0.002       &   2.55$\pm$0.11 & 1.48$\pm$0.06       &       ... & ...   \\
        &       G49.028-0.217   &   1.01$\pm$0.07 & 0.82$\pm$0.06       &       ... & ...   \\
        &       G92.670+3.072   &       0.72$\pm$0.05 & 0.64$\pm$0.04   &       ... & ...   \\
    \hline 
        \end{tabular}
      \end{threeparttable}
      \label{tab6}
      \tablefoot{(1): Classification; (2): Source name; (3): The integrated line intensity of H$_{2}$CO (taken from TMRT C band survey); (4): The column density of H$_{2}$CO, which is derived as done for cyanopolyynes in Sect. 3.1; (5-6): The integrated line intensity and column density of SiO, taken from \cite{2016A&A...586A.149C}.}
\end{table}

\section{Source catalogue without cyanopolyyne detection}
\FloatBarrier
\begin{center}
\begin{longtable}{ccccc}
\caption{Sources without HC$_{3}$N (J = 2-1) detection.}
\label{appendix} \\
\hline\hline
    Source Name        &       R.A.       &       DEC       &       Classifications  &  rms    \\
     &       (hh.mm.ss)       &       (hh.mm.ss)       &              &  (mK)   \\
     (1)       &       (2)       &       (3)       &       (4)       &       (5)     \\
     \hline
     \endfirsthead 
     \hline\hline
     Source Name        &       R.A.       &       DEC       &       Classifications  &  rms    \\
     &       (hh.mm.ss)       &       (hh.mm.ss)       &              &  (mK)   \\
     (1)       &       (2)       &       (3)       &       (4)       &       (5)     \\
     \hline
     \endhead
     \hline
     \endfoot
G20.749-0.112        &       18:29:21.28       &       -10:52:38.5       &       RRL-only       &       12       \\
G23.009-0.379       &       18:34:33.43       &       -08:59:46.9       &       Maser-RRL       &       13       \\
G23.010-0.410       &       18:34:40.25       &       -09:00:38.2       &       Maser-RRL       &       16       \\
G23.338-0.213       &       18:34:34.42       &       -08:37:41.3       &       RRL-only       &       10       \\
G23.428-0.231       &       18:34:48.38       &       -08:33:22.0       &       Maser-RRL       &       6       \\
G23.653-0.143       &       18:34:54.65       &       -08:18:59.2       &       Maser-RRL       &       6       \\
G23.680-0.189       &       18:35:07.46       &       -08:18:47.9       &       Maser-RRL       &       7       \\
G23.899+0.065       &       18:34:37.20       &       -08:00:09.7       &       Maser-RRL       &       6       \\
G24.010+0.503       &       18:33:15.42       &       -07:42:04.7       &       RRL-only       &       11       \\
G24.148-0.009       &       18:35:21.02       &       -07:48:54.1       &       Maser-only       &       6       \\
G24.313-0.154       &       18:36:10.51       &       -07:44:06.1       &       Maser-RRL       &       6       \\
G24.362-0.146       &       18:36:14.34       &       -07:41:19.3       &       Maser-RRL       &       5       \\
G24.485+0.180       &       18:35:17.92       &       -07:25:44.4       &       Maser-RRL       &       5       \\
G24.564-0.308       &       18:37:11.55       &       -07:35:01.3       &       RRL-only       &       6       \\
G24.633+0.153       &       18:35:40.11       &       -07:18:34.8       &       Maser-RRL       &       11       \\
G24.818-0.108       &       18:36:56.84       &       -07:15:59.3       &       RRL-only       &       12       \\
G24.943+0.074       &       18:36:31.49       &       -07:04:16.9       &       Maser-RRL       &       11       \\
G25.177+0.211       &       18:36:28.20       &       -06:48:00.7       &       Maser-RRL       &       11       \\
G25.346-0.189       &       18:38:12.83       &       -06:50:00.8       &       Maser-RRL       &       11       \\
G25.395+0.033       &       18:37:30.44       &       -06:41:18.3       &       Maser-RRL       &       14       \\
G26.545+0.423       &       18:38:14.50       &       -05:29:17.2       &       Maser-RRL       &       39       \\
G26.579-0.120       &       18:40:14.60       &       -05:42:23.2       &       RRL-only       &       8       \\
G26.645+0.021       &       18:39:31.66       &       -05:35:00.1       &       Maser-only      &       11       \\
G27.220+0.261       &       18:40:03.69       &       -04:57:42.4       &       Maser-only       &       16       \\
G27.287+0.154       &       18:40:34.11       &       -04:57:05.4       &       Maser-only       &       13       \\
G28.832-0.250        &       18:44:50.45       &       -03:45:44.7       &       Maser-RRL       &       11       \\
G28.855-0.219       &       18:44:46.41       &       -03:43:41.5       &       RRL-only       &       10       \\
G29.320-0.162        &       18:45:25.16       &       -03:17:16.9       &       Maser-RRL       &       15       \\
G30.392+0.121       &       18:46:22.34       &       -02:12:18.7       &       RRL-only       &       9       \\
G30.902-0.035       &       18:47:51.61       &       -01:49:21.0       &       RRL-only       &       9       \\
G31.101+0.265       &       18:47:09.28       &       -01:30:30.7       &       RRL-only       &       6       \\
G33.641-0.228       &       18:53:32.57       &       00:31:39.2       &       Maser-only       &       6       \\
G34.267-0.210       &       18:54:37.25       &       01:05:33.7       &       Maser-only       &       12       \\
G35.067-1.569       &       19:00:55.11       &       01:11:02.9       &       RRL-only       &       9       \\
G35.197-0.729       &       18:58:10.11       &       01:40:58.4       &       Maser-only       &       6       \\
G37.430+1.517       &       18:54:14.38       &       04:41:40.3       &       Maser-only       &       8       \\
G37.669-0.093       &       19:00:25.60       &       04:10:19.9       &       RRL-only       &       8       \\
G37.873-0.399       &       19:01:53.53       &       04:12:50.3       &       RRL-only       &       6       \\
G38.258-0.074       &       19:01:26.23       &       04:42:17.3       &       Maser-RRL       &       11       \\
G43.076-0.078       &       19:10:22.05       &       08:58:51.5       &       Maser-RRL       &       10       \\
G43.148+0.013       &       19:10:10.98       &       09:05:18.1       &       Maser-RRL       &       10       \\
G45.070+0.124       &       19:13:23.60       &       10:50:34.7       &       Maser-only       &       8       \\
G45.124+0.136       &       19:13:27.22       &       10:53:49.0       &       RRL-only       &       9       \\
G45.493+0.126       &       19:14:11.36       &       11:13:06.4       &       Maser-only       &       10       \\
G45.541-0.016       &       19:14:47.48       &       11:11:43.3       &       RRL-only       &       13      \\
G48.946-0.331       &       19:22:27.89       &       14:03:24.8       &       RRL-only       &       8       \\
G49.072-0.327       &       19:22:41.83       &       14:10:11.9       &       RRL-only       &       7       \\
G49.224-0.334       &       19:23:01.07       &       14:18:00.5       &       RRL-only       &       10       \\
G49.265+0.311       &       19:20:44.86       &       14:38:26.9       &       Maser-only       &       20       \\
G49.391-0.235       &       19:22:59.18       &       14:29:41.1       &       RRL-only       &       7       \\
G49.416+0.326       &       19:20:59.21       &       14:46:49.7       &       Maser-only       &       5       \\
G51.341+0.065       &       19:25:43.12       &       16:21:14.1       &       RRL-only       &       11       \\
G58.775+0.647       &       19:38:49.13       &       23:08:40.2       &       Maser-only       &       10       \\
G59.634-0.192       &       19:43:50.00       &       23:28:38.8       &       Maser-only       &       15       \\
G59.785+0.068       &       19:43:10.72       &       23:44:14.9       &       Maser-only       &       6      \\
G75.841+0.425       &       20:21:33.82       &       37:32:21.4       &       RRL-only       &       6       \\
G76.659+1.922       &       20:17:37.28       &       39:03:35.3       &       RRL-only      &       15       \\
G78.881+1.427       &       20:26:19.99       &       40:36:11.0       &       RRL-only       &       7       \\
G79.024+2.449        &       20:22:20.27       &       41:18:23.4       &       RRL-only       &       11       \\
G79.170+0.396        &       20:31:36.70       &       40:13:59.5       &       RRL-only       &       11       \\
G79.877+2.476       &       20:24:49.89       &       42:01:18.3       &       RRL-only       &       6       \\
G80.862+0.383       &       20:37:01.02       &       41:34:56.9       &       Maser-RRL       &       11       \\
G80.865+0.342       &       20:37:12.22       &       41:33:34.8       &       Maser-RRL       &       12       \\
G80.939-0.127       &       20:39:25.98       &       41:20:01.5       &       RRL-only       &       7       \\
G81.250+1.123       &       20:35:05.61       &       42:20:16.3       &       RRL-only       &       10       \\
G81.252+0.982       &       20:35:42.50       &       42:15:16.9       &       RRL-only      &       14       \\
G81.337+0.824       &       20:36:40.29       &       42:13:37.9       &       RRL-only       &       7       \\
G81.683+0.541       &       20:39:01.04       &       42:19:52.7       &       RRL-only       &       8       \\
G82.308+0.729       &       20:40:16.72       &       42:56:28.6       &       Maser-only       &       8       \\
G84.649-1.089       &       20:56:08.33       &       43:36:54.5       &       RRL-only       &       7       \\
G84.941-1.162       &       20:57:29.82       &       43:47:21.5       &       RRL-only      &       9       \\
G84.951-0.691       &       20:55:32.50       &       44:06:10.2       &       Maser-RRL       &       13       \\
G97.527+3.184       &       21:32:11.30       &       55:53:39.5       &       Maser-RRL       &       4       \\
G110.196+2.476       &       22:57:29.78       &       62:29:45.1       &       Maser-only       &       5       \\
G111.526+0.803       &       23:13:33.10       &       61:29:15.2       &       Maser-RRL       &       9       \\
G123.035-6.355       &       00:52:10.97       &       56:30:58.8       &       Maser-only       &       20       \\
G123.050-6.310       &       00:52:17.18       &       56:33:42.6       &       Maser-only       &       9       \\
G126.645-0.786       &       01:22:59.95       &       61:51:30.3       &       RRL-only       &       12       \\
G133.690+1.113       &       02:25:10.17       &       62:00:39.9       &       RRL-only       &       8       \\
G133.716+1.207       &       02:25:39.71       &       62:05:24.0       &       RRL-only       &      8       \\
G134.004+1.144       &       02:27:45.35       &       61:55:40.2       &       RRL-only      &       16       \\
G150.525-0.930       &       04:03:01.58       &       51:22:31.7       &       RRL-only      &       5       \\
G173.596+2.823       &       05:41:05.42       &       35:52:02.4       &       Maser-RRL       &       18       \\
G213.706-12.602       &       06:07:46.82       &       -06:23:08.3       &       Maser-RRL       &       10       \\
G213.752-12.615       &       06:07:48.79       &       -06:25:55.3       &       Maser-RRL       &       6       \\
\hline
\end{longtable}
\tablefoot{(1): Source name; (2): Right ascension (J2000); (3): Declination(J2000); (4): Classifications; (5): The rms noise value.}
    \label{tabc3}
\end{center}

\end{appendix}
\end{document}